\pdfoutput=1
\newcommand*{\ATLASLATEXPATH}{latex/}
\documentclass[cernpreprint,texlive=2016,txfonts,UKenglish]{latex/atlasdoc}
\usepackage{\ATLASLATEXPATH atlaspackage}
\newcommand{\AtlasCoordFootnote}{%
ATLAS uses a right-handed coordinate system with its origin at the nominal interaction point (IP)
in the centre of the detector and the $z$-axis along the beam pipe.
The $x$-axis points from the IP to the centre of the LHC ring,
and the $y$-axis points upwards.
Cylindrical coordinates $(r,\phi)$ are used in the transverse plane, 
$\phi$ being the azimuthal angle around the $z$-axis.
The pseudorapidity is defined in terms of the polar angle $\theta$ as $\eta = -\ln \tan(\theta/2)$.
Angular distance is measured in units of $\Delta R \equiv \sqrt{(\Delta\eta)^{2} + (\Delta\phi)^{2}}$.
Rapidity is defined as $y \equiv 0.5 \ln{[(E+p_{\text{z}})/(E-p_{\text{z}})]}$ where $E$ denotes the 
energy and $p_{\text{z}}$ is the momentum component along the beam direction.}
\usepackage{\ATLASLATEXPATH atlasbiblatex}
\usepackage{\ATLASLATEXPATH atlascontribute}
\usepackage{\ATLASLATEXPATH atlasphysics}
\usepackage{morefloats}
\usepackage{adjustbox}
\usepackage{diagbox}
\usepackage{hhline}
\usepackage{bm}
\usepackage{tikz}
\usepackage{hyperref}
\usepackage{multirow}
\usepackage{verbatim}
\usepackage{lscape}
\usepackage{float}
\usepackage[capitalize]{cleveref}
\usepackage{graphicx}
\usepackage{pifont}
\usepackage[utf8]{inputenc}
\usepackage[T1]{fontenc}
\usepackage{a4wide}
\graphicspath{{logos/}{figures/}}
\usepackage{EPS2017DCHPaper-defs}
\AtlasTitle{Search for doubly charged Higgs boson production in multi-lepton 
final states with the ATLAS detector using proton--proton collisions at $\sqrt{s}=\SI{13}{\TeV}$}
\author{The ATLAS Collaboration}
\usepackage{authblk}
\makeatletter
\renewcommand\AB@affilsepx{, \protect\Affilfont}
\makeatother
\renewcommand\Authands{, }
\renewcommand\Affilfont{\itshape\small}
\AtlasRefCode{EXOT-2016-07}
\PreprintIdNumber{CERN-EP-2017-198}
\AtlasJournal{EPJC}
\AtlasDOI{10.1140/EPJC/S10052-018-5661-Z}
\AtlasJournalRef{Eur. Phys. J. C78 (2018) 199}
\AtlasAbstract{%
A search for doubly charged Higgs bosons with pairs of prompt, isolated, 
highly energetic leptons with the same electric charge is presented.
The search uses a proton--proton collision data sample at a centre-of-mass energy of 
\SI{13}{\TeV} corresponding to \SI{36.1}{\ifb} of integrated 
luminosity recorded in 2015 and 2016  by the ATLAS detector at the LHC. 
This analysis focuses on the decays $H^{\pm\pm}\rightarrow e^{\pm}e^{\pm}$, 
$H^{\pm\pm}\rightarrow e^{\pm}\mu^{\pm}$ and $H^{\pm\pm}\rightarrow \mu^{\pm}\mu^{\pm}$, 
fitting the dilepton mass spectra in several exclusive signal regions. No 
significant evidence of a signal is observed and corresponding limits on 
the production cross-section and consequently a lower limit on 
$m(H^{\pm \pm})$ are derived at 95\% confidence level. With 
$\ell^{\pm}\ell^{\pm}=e^{\pm}e^{\pm}/\mu^{\pm}\mu^{\pm}/e^{\pm}\mu^{\pm}$,
the observed lower limit on the mass of a doubly charged Higgs boson only 
coupling to left-handed leptons varies from \SI{770}{\GeV} to \SI{870}{\GeV} (\SI{850}{\GeV} expected) 
for $B(H^{\pm\pm}\rightarrow \ell^{\pm}\ell^{\pm})=100\%$ and both the expected and
observed mass limits are above \SI{450}{\GeV} for 
$B(H^{\pm\pm}\rightarrow \ell^{\pm}\ell^{\pm})=10\%$ and any combination of 
partial branching ratios.
}
\hypersetup{pdftitle={ATLAS document},pdfauthor={The ATLAS Collaboration}}
\begin{document}
\maketitle
\section{Introduction}
\label{sec:Introduction}

Events with two prompt, isolated, highly energetic leptons with the same electric 
charge (same-charge leptons) are produced very rarely in a proton--proton collision 
according to the predictions of the Standard Model (SM), but may occur with higher 
rate in various theories Beyond the Standard Model (BSM). This analysis focuses on 
BSM theories that contain a doubly charged Higgs particle $H^{\pm\pm}$ using the 
observed invariant mass of same-charge lepton pairs. In the absence of evidence for 
a signal, lower limits on the mass of the $H^{\pm \pm}$ particle are set at the $95\%$ 
confidence level.

Doubly charged Higgs bosons can arise in a large variety of BSM theories, namely 
in left-right symmetric (LRS) 
models~\cite{Pati:1974yy,Mohapatra:1974hk,Senjanovic:1975rk,Dev:2016vle,Borah:2016hqn}, 
Higgs triplet models~\cite{montalvo2006searching,gunion1990higgs}, the little Higgs 
model~\cite{arkani2002minimal}, type-II see-saw 
models~\cite{muhlleitner2003note,akeroyd2005single,hektor2007testing,perez2008testing,chao2008correlative}, 
the Georgi--Machacek model~\cite{Georgi:1985nv}, scalar singlet 
dark matter~\cite{Bhattacharya:2016qsg}, and the Zee--Babu neutrino mass 
model~\cite{zee1986quantum,babu1988model,nebot2008prospects}. Theoretical 
studies~\cite{gunion1989higgs,deshpande1991left,Babu:2016rcr} indicate that the 
doubly charged Higgs bosons would be predominantly pair-produced via the 
Drell--Yan process at the LHC. For this search, the cross-sections utilised to
set the final exclusion limits are computed according to the model in 
Ref.~\cite{muhlleitner2003note}.

\begin{figure}[t]
\centering
\resizebox{0.4\textwidth}{!}{
  \begin{tikzpicture}[line width=1.25 pt, scale=1]
  \draw[antifermion] (0,1) -- (1,0) ;
  \draw[fermion] (0,-1) -- (1,0) ;
  \node at (0,1.2) {$\bar{q}$};
  \node at (0,-1.2) {$q$};

  \draw[vector] (1,0) -- (2,0);
  \node at (1.5,-0.35) {$Z/ \gamma^*$};

  \draw[scalar] (2,0) -- (3,1);
  \draw[scalar] (2,0) -- (3,-1);
  \node at (2.95,0.4) {$H^{++}$};
  \node at (2.95,-0.4) {$H^{--}$};

  \draw[antifermion] (3,1) -- (3.5,1.5) ;
  \draw[antifermion] (3,1) -- (3.5,0.5) ;
  \node at (3.75,1.5) {$\ell^{+}$};
  \node at (3.75,0.5) {$\ell^{+}$};

  \draw[fermion] (3,-1) -- (3.5,-1.5) ;
  \draw[fermion] (3,-1) -- (3.5,-0.5) ;
  \node at (3.75,-1.5) {$\ell^{-}$};
  \node at (3.75,-0.5) {$\ell^{-}$};

  \end{tikzpicture}
}
\caption{ Feynman diagram of the pair production process $pp\rightarrow H^{++}H^{--}$. 
The analysis studies only the electron and muon channels, where at least one of the 
lepton pairs is $e^{\pm}e^{\pm}$, $e^{\pm}\mu^{\pm}$, or $\mu^{\pm}\mu^{\pm}$. }
\label{fig:DCHfey}
\end{figure}
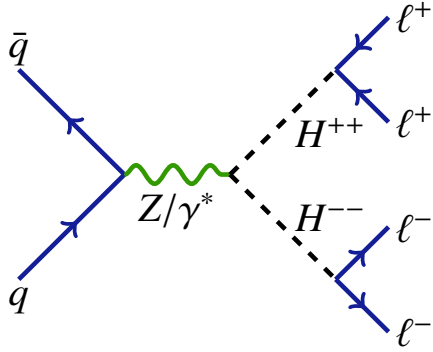

Doubly charged Higgs particles can couple to either left-handed or right-handed 
leptons. In LRS models, two cases are distinguished and denoted $H_{L}^{\pm \pm}$ 
and $H_{R}^{\pm \pm}$. The cross-section for $H_{L}^{++}H_{L}^{--}$ 
production is about $2.3$ times larger than for $H_{R}^{++}H_{R}^{--}$ 
due to the different couplings to the $Z$ boson~\cite{huitu1997doubly}. Besides 
the leptonic decay, the $H^{\pm \pm}$ particle can decay into a pair of $W$ bosons 
as well. For low values of the Higgs triplet vacuum expectation value $v_{\Delta}$, 
it decays almost exclusively to leptons while for high values of $v_{\Delta}$ the 
decay is mostly to a pair of $W$ bosons~\cite{perez2008testing,muhlleitner2003note}. 
In this analysis, the coupling to $W$ bosons is assumed to be negligible and only 
pair production via the Drell--Yan process is considered. The Feynman diagram of 
the production mechanism is presented in Figure~\ref{fig:DCHfey}.

The analysis targets only decays of the $H^{\pm\pm}$ particle into electrons 
and muons, denoted by $\ell$. Other final states $X$ that are not directly
selected in this analysis are taken into account by reducing the lepton multiplicity 
of the final state. These states $X$ would include, for instance, 
$\tau$ leptons or $W$ bosons, as well as particles which escape detection.
The total assumed branching ratio of $H^{\pm \pm}$ is therefore 
$B(H^{\pm \pm} \rightarrow e^{\pm}e^{\pm})+B(H^{\pm \pm} \rightarrow e^{\pm}\mu^{\pm}) + B(H^{\pm \pm} \rightarrow \mu^{\pm}\mu^{\pm}) + B(H^{\pm \pm} \rightarrow X) = B(H^{\pm \pm} \rightarrow \ell^{\pm} \ell^{\pm}) + B(H^{\pm \pm} \rightarrow X) = 100\%$. 
Moreover, the decay width is assumed to be negligible compared to the detector 
resolution, which is compatible with theoretical predictions. Two-, three-, and 
four-lepton signal regions are defined to select the majority of such events. 
These regions are further divided into unique flavour categories ($e$ or $\mu$) 
to increase the sensitivity.  The partial decay width of $H^{\pm \pm}$ to leptons is given by:

\begin{linenomath}
\begin{equation*}
\Gamma (H^{\pm \pm} \rightarrow \ell^{\pm} \ell'^{\pm}) = k\frac{h^{2}_{\ell \ell'}}{16 \pi} m(H^{\pm \pm}),
\label{eq:decayWidth}
\end{equation*}
\end{linenomath}

with $k=2$ if both leptons have the same flavour ($\ell=\ell'$) and $k=1$ for different 
flavours. The factor $h_{\ell \ell'}$ has an upper bound that depends on the 
flavour combination~\cite{Dutta:2014dba,Rentala:2011mr}. In this analysis, only prompt 
decays of the $H^{\pm \pm}$ bosons ($c\tau < \SI{10}\mathrm{\mu}$m) are considered, 
corresponding to $h_{\ell \ell'} \gtrsim 1.5\times10^{-6}$ for $m(H^{\pm \pm})=\SI{200}{\GeV}$. 
In general, there is no preference for decays into $\tau$ leptons, as the coupling 
is not proportional to the lepton mass like it is for the SM Higgs boson.

Additional motivation to study cases with 
$B(H^{\pm \pm} \rightarrow \ell^{\pm} \ell^{\pm}) < 100\%$ is given by type-II 
see-saw models with specific neutrino mass hypotheses resulting in a fixed branching 
ratio combination~\cite{kadastik2008direct,chao2008correlative,perez2008neutrino} 
which does not necessarily correspond to $B(H^{\pm \pm} \rightarrow \ell^{\pm} \ell^{\pm}) = 100\%$.

The ATLAS Collaboration previously analysed data corresponding to \SI{20.3}{\ifb} 
of integrated luminosity which were recorded in 2012 at a centre-of-mass energy of 
\SI{8}{\TeV}~\cite{EXOT-2012-25}. This study resulted in the most stringent lower 
limits on the mass of a potential $H^{\pm\pm}_L$ particle. Depending on the 
flavour of the final-state leptons, the observed limits vary between \SI{465}{\GeV} 
and \SI{550}{\GeV} assuming $B(H^{\pm\pm}_L\to \ell^\pm\ell^\pm)=100\%$. The analysis 
presented in this paper extends the one described in Ref.~\cite{EXOT-2012-25} and 
is based on \SI{36.1}{\ifb} of integrated luminosity collected in 2015 and 2016 
at a centre-of-mass energy of \SI{13}{\TeV}. A similar search has also been performed 
by the CMS Collaboration~\cite{CMS-HIG-12-005}.

\FloatBarrier

\section{ATLAS detector}
\label{sec:atlas1}

The ATLAS detector~\cite{PERF-2007-01} at the LHC is a multi-purpose particle detector
with a forward-backward symmetric cylindrical geometry and an almost $4\pi$ coverage 
in solid angle.\footnote{\AtlasCoordFootnote} It consists of an inner tracking 
detector (ID) surrounded by a thin superconducting solenoid providing a \SI{2}{\tesla} 
axial magnetic field, electromagnetic (EM) and hadronic calorimeters, and a muon spectrometer.
The inner tracking detector covers the pseudorapidity range $|\eta| < 2.5$.
It is composed of silicon pixel, silicon micro-strip, and transition radiation tracking detectors.
A new innermost layer of pixel detectors~\cite{ATLAS-TDR-19} was installed prior
to the start of data taking in 2015.
Lead/liquid-argon (LAr) sampling calorimeters provide electromagnetic energy measurements
with high granularity.
A hadronic (steel/scintillator-tile) calorimeter covers the central pseudorapidity range ($|\eta| < 1.7$).
The end-cap and forward regions are instrumented with LAr calorimeters
for both EM and hadronic energy measurements up to $|\eta| = 4.9$.
The muon spectrometer surrounds the calorimeters and features three large air-core toroidal
superconducting magnets with eight coils each.
The field integral of the toroids ranges between \num{2.0} to \SI{6.0}{\tesla\metre}
across most of the detector.
The muon system includes precision tracking chambers and fast detectors for triggering.
A two-level trigger system is used to select events~\cite{TRIG-2016-01} that are 
interesting for physics analyses. The first-level trigger is implemented as part of 
the hardware. Subsequently a software-based high-level trigger executes algorithms 
similar to those used in the offline reconstruction software, reducing the event 
rate to about 1~kHz.

\section{Dataset and simulated event samples}
\label{sec:monteCarloProduction}

The data used in this analysis were collected at centre-of-mass energy of \SI{13}{\TeV}
during 2015 and 2016, and correspond to an integrated luminosity of \SI{3.2}{\ifb} in 2015
and \SI{32.9}{\ifb} in 2016. The average number of $pp$ interactions per bunch crossing in the
dataset is 24. Interactions other than the hard-scattering one are referred to as pile-up.
The uncertainty on the combined 2015 and 2016 integrated luminosity is $3.2\%$. 
Following a methodology similar to the one described in Ref.~\cite{DAPR-2013-01}, 
this uncertainty is derived from a preliminary calibration of the luminosity scale 
using $x$--$y$ beam-separation scans performed in August 2015 and May 2016.

Signal candidate events in the electron channel are required to pass a dielectron
trigger with a threshold of \SI{17}{\GeV} on the transverse energy ($E_\text{T}$) of each of the electrons.
Candidate events in the muon channel are selected using a combination of two single-muon
triggers with transverse momentum (\pt) thresholds of \SI{26}{\GeV} and \SI{50}{\GeV}. 
The single-muon trigger with the lower \pt threshold also requires track-based 
isolation of the muon according to the isolation criteria described in Ref.~\cite{PERF-2015-10}.
Events containing both electrons and muons (mixed channel) are required to pass either the 
combined electron--muon trigger or any of the triggers used for the muon channel or the electron channel. 
The combined trigger has an $E_\text{T}$ threshold of 17 GeV for the electron and a $p_\text{T}$ 
threshold of 14 GeV for the muon. Events with four leptons are selected using a combination of 
dilepton triggers. In general, single-lepton triggers are more efficient than dilepton triggers. 
However, single-electron triggers impose stringent electron identification criteria, which 
interfere with the data-driven background estimation.

An irreducible background originates from SM processes resulting in same-charge 
leptons, hereafter referred to as prompt background. Prompt background and signal 
model predictions were obtained from Monte Carlo (MC) simulated event samples which are 
summarised in Table~\ref{tab:MC}. Prompt background events mainly originate from 
diboson ($W^\pm W^\pm$ / $ZZ$ / $WZ$) and $t\bar{t}\,X$ processes 
($t\bar{t}\,W$, $t\bar{t}\,Z$, and $t\bar{t}\,H$). They also provide a source of reducible 
background due to charge misidentification in channels that contain electrons.\footnote{The 
probability of muon charge misidentification is negligible because muon tracks are measured 
both in the inner detector and in the muon spectrometer which provides a much larger 
lever arm for the curvature measurement.} As described in Section~\ref{sec:backgroundEstimation},
the modelling of charge misidentification in simulation deviates from data and consequently charge
reconstruction scale factors are derived in a data-driven way and applied to the simulated events
to compensate for the differences. The highest-yield process which enters the analysis through charge
misidentification is Drell--Yan ($q\bar{q} \rightarrow Z/\gamma^{*} \rightarrow \ell^{+}\ell^{-}$)
followed by $t\bar{t}$ production. MC samples are in general normalised using theoretical cross-sections 
referenced in Table~\ref{tab:MC}. However, yields of some MC samples are considered as 
free parameters in the likelihood fit, as described in Section~\ref{sec:results}.

\begin{table*}[ht]
\caption{Simulated signal and background event samples: the corresponding event generator, parton shower, cross-section normalisation, PDF set used for the matrix element and 
set of tuned parameters are shown for each sample. The cross-section in the event generator that produces the sample is used where not specifically stated otherwise.}
\label{tab:MC}
\begin{center}
\scriptsize
\resizebox{\textwidth}{!}
{
\begin{tabular}{  l  c  c  c  c  c  }
\toprule
Physics process    & Event generator & ME PDF set    & Cross-section & Parton shower & Parton shower \\
                   &                 &               & normalisation &               &         tune  \\
\midrule
Signal             &                 &               &               &               &               \\
$H^{\pm \pm}$                         &
\PYTHIA 8.186~\cite{Sjostrand:2007gs} &
NNPDF2.3NLO~\cite{Ball:2012cx}        &
NLO (see Table~\ref{tab:DCHNLOxsec})  &
\PYTHIA 8.186                         &
A14~\cite{ATL-PHYS-PUB-2014-021}      \\
\hline
Drell--Yan                                       &&&&& \\
$Z/\gamma^{*}\rightarrow ee/\tau\tau$                  &
\POWHEGBOX v2~\cite{alioli2010general,powheg1,powheg2} &
CT10~\cite{Lai:2010vv}                                 &
NNLO~\cite{dulat2015ct14}                              &
\PYTHIA 8.186                                          & 
AZNLO~\cite{STDM-2012-23}                             \\
\hline
Top                               &&&&& \\
$t\bar{t}$                               &
\POWHEGBOX v2                            &
NNPDF3.0NLO~\cite{ball2015parton}        &
NNLO~\cite{Czakon:2011xx}                &
\PYTHIA 8.186                            &
A14                                     \\
Single top                               &
\POWHEGBOX v2                            &
CT10                                     &
NLO~\cite{Kidonakis:2015nna}             &
\PYTHIA 6.428~\cite{sjostrand2006pythia} &
Perugia 2012~\cite{Skands:2010ak}       \\
$\ttbar W$, $\ttbar Z/\gamma^{*}$        &
\AMCATNLO 2.2.2~\cite{Alwall:2014hca}    &
NNPDF2.3NLO                              &
NLO~\cite{YR4}                           &
\PYTHIA 8.186                            &
A14                                     \\
$\ttbar H$                               &
\AMCATNLO 2.3.2                          &
NNPDF2.3NLO                              &
NLO~\cite{YR4}                           &
\PYTHIA 8.186                            &
A14                                     \\
\hline
Diboson                           &&&&& \\
$ZZ$, $WZ$                               &
\SHERPA 2.2.1~\cite{Gleisberg:2008ta}    &
NNPDF3.0NLO                              &
NLO                                      &
\SHERPA                                  &
\SHERPA default                         \\
Other (inc. $W^{\pm}W^{\pm}$)            &
\SHERPA 2.1.1                            &
CT10                                     &
NLO                                      &
\SHERPA                                  &
\SHERPA default                         \\
\hline
Diboson Sys.                      &&&&& \\
$ZZ$, $WZ$                               &
\POWHEGBOX v2                            &
CT10NLO                                  &
NLO                                      &
\PYTHIA 8.186                            &
AZNLO                                   \\
\bottomrule
\end{tabular}
}
\end{center}
\end{table*}

Another source of reducible background arises from events with non-prompt electrons
or muons or with other physics objects misidentified as electrons or muons,
collectively called `fakes'. For both, electrons and muons, this contribution 
originates within jets, from decays of light-flavour or heavy-flavour hadrons into light 
leptons. For electrons, a significant component of fakes arises from jets which satisfy 
the electron reconstruction criteria and from photon conversions. MC samples are not used 
to estimate this background because the simulation of jets and hadronisation has large 
uncertainties. Instead, a data-driven approach is used to assess this contribution from 
production of $W$+jets, $t\bar{t}$ and multi-jet events. The method is validated in 
specialised validation regions.

The SM Drell--Yan process was modelled using \POWHEGBOX v2~\cite{alioli2010general,powheg1,powheg2} 
interfaced to \PYTHIA 8.186~\cite{Sjostrand:2007gs} for parton showering. The CT10 
set of parton distribution functions (PDF)~\cite{Lai:2010vv} was used to calculate the hard scattering process. 
A set of tuned parameters called the AZNLO tune~\cite{STDM-2012-23} was used in combination with the 
CTEQ6L1 PDF set~\cite{pumplin2002new} to model non-perturbative effects.  \textsc{Photos}++ 
version 3.52~\cite{Davidson:2010ew} was used for photon emissions from electroweak vertices 
and charged leptons. The generation of the process was divided into 19 samples with 
subsequent invariant mass intervals to guarantee a good statistical coverage over the 
entire mass range.

Higher-order corrections were applied to the Drell--Yan simulated events to scale the 
mass-dependent cross-section computed at next-to-leading order (NLO) in the strong coupling 
constant with the CT10 PDF set to next-to-next-to-leading order (NNLO) in the strong coupling constant 
with the CT14NNLO PDF set~\cite{dulat2015ct14}. The corrections were calculated with 
VRAP~\cite{anastasiou2004high} for QCD effects and \textsc{Mcsanc}~\cite{bondarenko2013nlo} for 
electroweak effects. The latter are corrected from leading-order (LO) to NLO. 

A sample of $Z\to ee$ events was generated with \SHERPA 2.2.1~\cite{Gleisberg:2008ta},
in addition to the \POWHEG prediction, to measure the probability of electron charge misidentification, 
as explained in Section~\ref{sec:backgroundEstimation}. The electron $p_\text{T}$ spectrum 
is a crucial ingredient for the estimate of this probability and was found to be better 
described by \SHERPA than by \POWHEG, especially for invariant masses of the electron pair 
close to the $Z$ boson mass. \SHERPA uses Comix~\cite{Gleisberg:2008fv} and 
OpenLoops~\cite{Cascioli:2011va} to calculate the matrix elements up to two partons at
NLO and up to four partons at LO in the strong coupling constant. The merging with the \SHERPA parton 
shower~\cite{Schumann:2007mg} follows the ME+PS@NLO prescription in~\cite{Hoeche:2012yf}.

The $t\bar{t}$ process was generated with the NLO QCD event generator \POWHEGBOX v2 which was 
interfaced to \PYTHIA 8.186 for parton showering. The A14 parameter set~\cite{ATL-PHYS-PUB-2014-021} 
was used together with the NNPDF2.3~\cite{Ball:2012cx} PDF set for tuning the shower. Furthermore,
the PDF set used for generation was NNPDF3.0~\cite{ball2015parton}. 
Additionally, top-quark spin correlations were preserved through the use of 
\textsc{MadSpin}~\cite{Artoisenet:2012st}. The predicted $t\bar{t}$ production cross-section 
is $832^{+20}_{-30}$ (scale) $\pm 35$ (PDF + $\alpha_{\text{S}}$) pb as calculated 
with \textsc{Top}++2.0~\cite{Czakon:2011xx} to NNLO in perturbative QCD, including soft-gluon 
resummation to next-to-next-to-leading-log order. The top-quark mass was assumed to be 
\SI{172.5}{\GeV}. The scale uncertainty results from independent variations of the 
factorisation and renormalisation scales, while the second uncertainty is associated with 
variations of the PDF set and $\alpha_{\text{S}}$, following the PDF4LHC\cite{Botje:2011sn} 
prescription using the MSTW2008 $68\%$ CL NNLO\cite{Martin:2009bu}, 
CT10 NNLO~\cite{gao2014ct10}, and NNPDF2.3 PDF sets.

Single-top-quark events produced in $Wt$ final states were generated by \POWHEGBOX v2 with 
the CT10 PDF set used in the matrix element calculations. Single-top-quark 
events in other final states were generated by \POWHEGBOX v1. This event generator 
uses the four-flavour scheme for the NLO QCD matrix element calculations together with the fixed 
four-flavour PDF set CT10f4. The parton shower, hadronisation, and underlying event were simulated 
with \PYTHIA 6.428~\cite{sjostrand2006pythia} using the CTEQ6L1 PDF set and the corresponding 
Perugia 2012 tune (P2012)~\cite{Skands:2010ak}. The top-quark mass was set to \SI{172.5}{\GeV}. 
The NLO cross-sections used to normalise these MC samples are summarised in Ref.~\cite{Kidonakis:2015nna}.

The $t\bar{t}\,W$, $t\bar{t}\,Z$, and $t\bar{t}\,H$ processes were generated at LO with 
\MADGRAPH~v2.2.2~\cite{alwall2011madgraph} and \MADGRAPH~v2.3.2 using the NNPDF2.3 PDF set. 
\PYTHIA 8.186 was applied for shower modelling configured with the A14 
tune~\cite{ATL-PHYS-PUB-2014-021}, as explained in more detail in 
Ref.~\cite{ATL-PHYS-PUB-2016-005}. They were normalised using theoretical cross-sections 
summarised in Ref.~\cite{YR4}.

Diboson processes with four charged leptons, three charged leptons and one neutrino, or two charged 
leptons and two neutrinos were generated with \SHERPA 2.2.1, using matrix elements containing all 
diagrams with four electroweak vertices.
They were calculated for up to three partons at LO accuracy and up to one (4$\ell$, 2$\ell$+2$\nu$)
or zero partons (3$\ell$+1$\nu$) at NLO QCD using Comix and OpenLoops. The merging with the \SHERPA 
parton shower~\cite{Schumann:2007mg} follows the ME+PS@NLO prescription. The NNPDF3.0NNLO~\cite{ball2015parton} 
PDF set was used in conjunction with dedicated parton shower tuning by the \SHERPA authors.

Diboson processes with one boson decaying hadronically and the other one decaying leptonically 
were predicted by \SHERPA 2.1.1~\cite{Gleisberg:2008ta}. They were calculated 
for up to three additional partons at 
LO accuracy and up to one ($ZZ$) or zero ($WW$, $WZ$) additional partons at
NLO using Comix and OpenLoops matrix element 
generators. The merging with the \SHERPA parton shower~\cite{Schumann:2007mg}
follows the ME+PS@NLO prescription. The CT10 PDF set was used in conjunction 
with a dedicated parton shower tuning. The \SHERPA 2.1.1 diboson prediction was
scaled by $0.91$ to account for differences between the internal electroweak scheme used in this \SHERPA
version and the $G_\mu$ scheme which is the common default.
Similarly, loop-induced diboson production with both gauge bosons decaying fully leptonically
was simulated with \SHERPA 2.1.1. The prediction is at LO accuracy while up to one additional
jet is merged with the matrix element.

Additional diboson samples for $WZ$ and $ZZ$ production were generated with \POWHEGBOX v2
to estimate theoretical uncertainties. \PYTHIA 8.186 provided the parton shower. The CT10 PDF set was used
for the matrix element calculation while the parton shower was configured with the CTEQL1 PDF set. 
The non-perturbative effects were modelled using the AZNLO~\cite{STDM-2012-23} tune. 

Signal samples were generated at LO using the LRS package of \PYTHIA 8.186 which implements
the $H^{\pm \pm}$ scenario described in 
Ref.~\cite{huitu1997doubly}. The program was configured to use the NNPDF23LO PDF set.
The $h_{\ell \ell'}$ couplings of lepton pairs were assumed to be the same for $H_R^{\pm\pm}$ and 
$H_L^{\pm\pm}$ particles. This choice resulted in a good statistical coverage 
for all possible decay channels. The production of the $H^{\pm\pm}$ was implemented only 
via the Drell--Yan process. Originally, the cross-section at $\sqrt{s} = \SI{14}{\TeV}$ was calculated
with NLO accuracy by the authors of Ref.~\cite{muhlleitner2003note}. Subsequently, a rescaling to 
$\sqrt{s}=\SI{13}{\TeV}$ with the CTEQ6 PDF~\cite{nadolsky2008implications} set was provided.
The cross-sections and corresponding $K$-factors are summarised in Table~\ref{tab:DCHNLOxsec}.

Since this analysis exclusively targets the leptonic decays of the $H^{\pm \pm}$ bosons, the vacuum 
expectation value of the neutral component of the left-handed Higgs triplet ($v^L_{\Delta}$) 
was set to zero in order to exclude $H^{\pm\pm}\rightarrow WW$ decays. The decay width of the $H^{\pm \pm}$ 
particle to leptons depends on the $h_{\ell \ell'}$ couplings. These were set to the value 
$h_{\ell\ell'}=$ $0.02$ in all \PYTHIA 8.186 samples. This setting corresponds to a decay width
that is negligible compared to the detector resolution. The $h_{\ell \tau}$ and $h_{\tau\tau}$ couplings were fixed at zero.
There are 23 MC samples with different $H^{\pm \pm}$ particle masses, starting from \SI{200}{\GeV}
up to 1300\,{\GeV} in steps of \SI{50}{\GeV}. The ATLAS detector is expected to have the best $H^{\pm \pm}$ 
mass resolution in the electron--electron final states. Resolutions around \SI{30}{\GeV}
for masses of 200-\SI{500}{\GeV} and \SI{50}{\GeV} to \SI{100}{\GeV} for higher masses can be
achieved with the event selection defined in Section~\ref{sec:eventSelection}.
Furthermore, the $H^{\pm \pm}$ mass resolution in electron--muon final states varies from \SI{50}{\GeV} to \SI{150}{\GeV}
and from \SI{50}{\GeV} to \SI{200}{\GeV} in muon--muon final states.

For all simulated samples except those obtained with \SHERPA, the EvtGen v1.2.0 
program~\cite{EvtGen} was used to model bottom and charm hadron decays.
The effect of the pile-up was included by overlaying minimum-bias collisions,
simulated with \PYTHIA~8.186, on each generated signal and background event.
The number of overlaid collisions is such that the distribution of the average number
of interactions per $pp$ bunch crossing in the simulation matches the pile-up
conditions observed in the data. The pile-up simulation is described in more detail in Ref.~\cite{PileUpSimulation}.

\begin{table}[t]
\caption{NLO cross-sections for the pair production of $H_{L}^{++}H_{L}^{--}$ and $H_{R}^{++}H_{R}^{--}$ 
in $pp$ collisions at $\sqrt{s}=\SI{13}{\TeV}$, together with the correction factors ($K=\sigma_{\mathrm{NLO}}/\sigma_{\mathrm{LO}}$) used to
obtain those values from the LO prediction. These $K$-factors are calculated by the authors of Ref.~\cite{muhlleitner2003note} 
using the CTEQ6 PDF~\cite{nadolsky2008implications}.}
\label{tab:DCHNLOxsec}
\begin{center}
\begin{tabular}{ r S S S S }
\toprule
{$m(H^{\pm\pm})\;[{\GeV}]$} & {$\sigma(H^{\pm\pm}_{L})\;[\mathrm{fb}]$} & {$K$-factor~$(H^{\pm\pm}_{L})$} & {$\sigma(H^{\pm\pm}_{R})\;[\mathrm{fb}]$} & {$K$-factor~$(H^{\pm\pm}_{R})$} \\
\noalign{\vskip 0.3ex} 
\midrule
\noalign{\vskip 0.2ex} 
300  &  13      &   1.25  &  5.6    & 1.25    \\
350  &  7.0     &   1.25  &  3.0    & 1.25    \\
400  &  3.9     &   1.24  &  1.7    & 1.24    \\
450  &  2.3     &   1.24  &  0.99   & 1.24    \\
500  &  1.4     &   1.24  &  0.61   & 1.24    \\
600  &  0.58    &   1.23  &  0.25   & 1.24    \\
700  &  0.26    &   1.23  &  0.11   & 1.23    \\
800  &  0.12    &   1.22  &  0.054  & 1.23    \\
900  &  0.062   &   1.22  &  0.027  & 1.23    \\ 
1000 &  0.032   &   1.22  &  0.014  & 1.24    \\
1100 &  0.017   &   1.23  &  0.0076 & 1.24    \\
1200 &  0.0094  &   1.23  &  0.0042 & 1.25    \\
1300 &  0.0052  &   1.24  &  0.0023 & 1.26    \\ 
\bottomrule
\end{tabular}
\end{center}
\end{table}

The response of the ATLAS detector was simulated using the \GEANT4 toolkit~\cite{agostinelli2003geant4}. 
Data and simulated events were reconstructed with the default ATLAS software~\cite{SOFT-2010-01} while 
simulated events were corrected with calibration factors to better match the performance measured in data. 

\FloatBarrier

\section{Event reconstruction and selection}
\label{sec:eventSelection}

Events are required to have at least one reconstructed primary vertex with at least 
two associated tracks with \pt > 400 MeV. Among all the vertices in the event the 
one with the highest sum of squared transverse momenta of the associated tracks 
is chosen as the primary vertex.

\subsection{Event reconstruction}
\label{sec:ObjSelection}

This analysis classifies leptons in two exclusive categories called \textit{tight} 
and \textit{loose}, defined specifically for each lepton flavour as described below. 
Leptons selected in the tight category feature a predominant component of prompt
leptons, while loose leptons are mostly fakes, which are used for the fake-background
estimation. All tracks associated with lepton candidates must have a longitudinal 
impact parameter with respect to the primary vertex of less than $0.5$ mm.

Electron candidates are reconstructed using information from the EM calorimeter and 
ID by matching an isolated calorimeter energy deposit to an ID track. They are 
required to have $|\eta|<2.47$, $\pt>\SI{30}{\GeV}$, and to pass at least the 
\texttt{LHLoose} identification level based on a multivariate likelihood 
discriminant~\cite{ATLAS-CONF-2016-024, ATL-PHYS-PUB-2015-041}. The 
likelihood discriminant is based on track and calorimeter cluster information. 
Electron candidates within the transition region between the barrel and endcap 
electromagnetic calorimeters ($1.37<|\eta|<1.52$) are vetoed due to limitations
in their reconstruction quality. The track associated with the electron candidate 
must have an impact parameter evaluated at the point of closest approach
between the track and the beam axis in the transverse plane ($d_0$) that satisfies
$|d_0|/\sigma(d_0)<5$, where $\sigma(d_0)$ is the uncertainty on $d_0$. In addition
to this, electron candidates are classified as tight if they satisfy the \texttt{LHMedium} 
working point of the likelihood discriminant and the isolation criteria described
in Ref.~\cite{ATLAS-CONF-2016-024}. This is based on calorimeter cluster 
and track isolation, which vary to obtain a fixed efficiency for selecting prompt 
electrons of $99\%$ across \pt and $\eta$. Electrons are classified as loose if 
they fail to satisfy either of the identification or the isolation criteria.

Muon candidates are selected by combining information from the muon spectrometer and 
the ID. They satisfy the \texttt{medium} quality criteria described in Ref.~\cite{PERF-2015-10}
and are required to have $\pt>\SI{30}{\GeV}$, $|\eta|<2.5$ and $|d_0|/\sigma(d_0)<10$.
Muon candidates are classified as tight if their impact parameter satisfies $|d_0|/\sigma(d_0)<3.0$
and they satisfy the most stringent isolation working point of the \textit{cut-based}
track isolation~\cite{ATLAS-CONF-2016-024}. Muons are loose if they fail the isolation 
requirement.

Jets or particles originating from the hadronisation of partons are reconstructed
by clustering energy deposits in the calorimeter calibrated at the EM scale. The anti-$k_{t}$ 
algorithm~\cite{cacciari2008anti} is used with a radius parameter of 0.4, which is implemented 
with the \textsc{FastJet}~\cite{Cacciari:2011ma} package. The majority 
of pile-up jets are rejected using the \textit{jet-vertex-tagger}~\cite{ATLAS-CONF-2014-018}, 
which is a combination of track-based variables providing discrimination against pile-up jets. 
For all jets the expected average transverse energy contribution from pile-up is subtracted
using an area-based \pt density subtraction method and a residual correction derived from the
MC simulation, both detailed in Refs.~\cite{Cacciari:2007fd,PERF-2016-04}.
In this analysis, events containing jets identified as originating from $b$-quarks 
are vetoed. They are identified with a multivariate discriminant~\cite{PERF-2016-04}
that has a $b$-jet efficiency of $77\%$ in simulated $t\bar{t}$ events and a rejection factor 
of $\approx40$ ($\approx20$) for jets originating from gluons and light quarks ($c$-quarks).

After electron and muon identification, jet calibration, and pile-up jet removal, overlaps 
between reconstructed particles or jets are resolved. First, electrons are removed if they share a track with 
a muon. Secondly, ambiguities between electrons and jets are resolved. If a jet is
closer than $\sqrt{(\Delta y)^{2} + (\Delta\phi)^{2}} = 0.2$ the jet is rejected.
If $0.2 < \sqrt{(\Delta y)^{2} + (\Delta\phi)^{2}} < 0.4$ the electron is removed. 
Finally, if a muon and a jet are closer than $\sqrt{(\Delta y)^{2} + (\Delta\phi)^{2}} = 0.4$,
and the jet features less than 3 tracks, the jet is removed. Otherwise the muon is discarded.

\subsection{Event selection}
\label{sec:ExclusiveRegions}

In this search, events are classified in independent categories, called \textit{analysis regions}, 
which serve different purposes. The so-called \textit{control regions} are used to constrain free 
background parameters in the statistical analysis detailed in Section~\ref{sec:results}.
The background model is validated against data in \textit{validation regions}. Both the control 
and validation regions are designed to reject signal events. A dedicated selection
targeting signal events is utilised to define the \textit{signal regions}. The selection
criteria utilised for each region are summarised in Table~\ref{tab:AllRegions}. The main
variable that defines the type of the region is the invariant mass of same-charge
lepton pairs. Invariant masses are required to be above \SI{200}{\GeV}
in signal regions and below \SI{200}{\GeV} in most control and validation regions.

The lepton multiplicity in the event is used to define the analysis regions. Events with
two or three leptons are required to contain exactly one same-charge lepton pair, while 
four-lepton events are required to feature two same-charge pairs where the sum of all lepton
charges has to be zero. An exception is the \textit{opposite-charge control region} (OCCR) 
where exactly two electrons with opposite 
charge are required. In all regions, events with at least one $b$-tagged jet are vetoed, in order 
to suppress background events arising from top-quark decays. In regions with more than two leptons, 
events are rejected if any opposite-charge same-flavour lepton pair is within \SI{10}{\GeV} 
of the $Z$ boson mass ($\SI{81.2}{\GeV} < m(\ell^{+}\ell^{-}) < \SI{101.2}{\GeV}$). This 
requirement is applied to reject diboson events featuring a $Z$ boson in the final state, 
and is inverted in \textit{diboson control regions}, where at least one $Z$ boson is present. 
Furthermore, the $Z$ boson veto is not applied in four-lepton control and validation regions 
to increase the available number of simulated diboson events.

The invariant mass of the same-charge lepton pair is used in the final fit of the analysis
for the two- and three-lepton regions. In the OCCR, the invariant mass of the opposite-charge 
lepton pair is used. A lower bound of \SI{60}{\GeV} on the invariant mass is imposed in all 
regions to discard low-mass events which would potentially bias the background estimation of 
the analysis while maximising the available number of events.

In the electron and mixed channels the lower bound is increased 
to \SI{90}{\GeV} in the three-lepton regions and to \SI{130}{\GeV} in the two-lepton regions. 
The motivation for increasing the lower mass bound in regions containing electrons is the 
data-driven charge misidentification background correction, where the $Z \rightarrow ee$ 
peak is used to measure the charge misidentification rates (described in 
Section~\ref{sec:backgroundEstimation}). Differences between data and MC simulation in the dielectron 
same-charge $Z\rightarrow ee$ peak~(see~Figure~\ref{fig:ZPeak}) were minimised by construction 
following the methodology described in Section~\ref{sec:backgroundEstimation}, and the 
$Z\rightarrow ee$ peak was therefore not used in the fit. In the two-lepton regions, this bound 
is set to \SI{130}{\GeV} to completely remove the $Z$ peak region. In the three-lepton regions, 
where this effect is not as strong, the bound is relaxed to \SI{90}{\GeV} to reduce the
statistical uncertainty of the sample. As the charge misidentification background is not present 
in the muon channel, there is no need to increase the lower mass bound there.

In the mixed channel, events are further divided into two categories, where the same-charge
pair features different-flavour leptons or not, indicated by $e^{\pm} \mu^{\pm} \ell^{\mp}$
and $e^{\pm} e^{\pm} \mu^{\mp}$ or $\mu^{\pm} \mu^{\pm} e^{\mp}$, respectively.

In order to maximise the sensitivity in two-lepton and three-lepton signal regions (SR1P2L and SR1P3L), 
additional requirements are imposed on same-charge lepton pairs, regardless of the flavour. These 
exploit both the boosted decay topology of the $H^{\pm \pm}$ resonance and the high energy of the decay products.
The same-charge lepton separation is required to be $\Delta R(\mathrm{\ell^{\pm},\ell^{\pm}})<3.5$.
Their combined transverse momentum has to be $\pt(\mathrm{\ell^{\pm}\ell^{\pm}})>\SI{100}{\GeV}$.
\footnote{The variable $\pt(\mathrm{\ell^{\pm}\ell^{\pm}})$ is the vector sum of the leptons' transverse momenta.} 
Finally, the scalar sum of the leptons' transverse momenta is required to be above \SI{300}{\GeV} 
in the signal regions. In SR1P2L and SR1P3L, the signal selection efficiency combined with the
detector acceptance varies greatly with the assumed branching ratio into light leptons.
It is the highest for $B(H^{\pm\pm}\rightarrow \ell^{\pm} \ell^{\pm}) \approx 60\%$ where about 40\%
of signal events are selected either in SR1P2L or SR1P3L. For $B(H^{\pm\pm}\rightarrow \ell^{\pm} \ell^{\pm})=100\%$,
about 25\% of signal events are selected in either of the regions.

In the four-lepton signal region (SR2P4L), the fit variable is the average invariant mass of the 
two same-charge lepton pairs $\bar{M}\equiv (m^{++}+m^{--})/2$. A selection on the variable 
$\Delta M/\bar{M} \equiv |m^{++} - m^{--}| / \bar{M} $ is applied to reject background where the 
two same-charge pairs have inconsistent invariant masses. The $\Delta M/\bar{M}$ requirement 
is optimised for different flavour combinations which generally feature different mass resolutions. 
This selection corresponds to $\Delta M$ values which are required to be below 15-\SI{50}{\GeV} for 
$\bar{M}=\SI{200}{\GeV}$, 30-\SI{160}{\GeV} for $\bar{M}=\SI{500}{\GeV}$, and 
50-\SI{500}{\GeV} for $\bar{M}=\SI{1000}{\GeV}$. In the 2P4L signal region, the fraction of signal
events that are selected is approximately 50\% for the $B(H^{\pm\pm}\rightarrow \ell^{\pm} \ell^{\pm})=100\%$
case and lower for branching ratios into light leptons below 100\%.

The \textit{same-charge validation region} (SCVR) is used to validate the data-driven 
fake-background estimation and the charge misidentification effect in the electron channel.
The \textit{three-lepton validation region} (3LVR) is used to validate the SM diboson background 
and fake events with three reconstructed leptons with different proportions across 
channels. The \textit{four-lepton validation region} (4LVR) is used to validate the diboson 
modelling in the four-lepton region. Furthermore, the \textit{diboson control region} (DBCR) 
is used to constrain the diboson background yield in each channel while the 
opposite-charge control region is used to constrain the Drell--Yan contribution 
in the electron channel only. The \textit{four-lepton control region} (4LCR) is used to constrain 
the yield of the diboson background in four-lepton regions.

\begin{table}[t]
\caption{Summary of all regions used in the analysis. The table is split into three blocks: the upper block
indicates the final states for each region, the middle block indicates the mass range of the corresponding final state, and the
lower block indicates the event selection criteria for the region.
The application of a selection requirement 
is indicated by a check-mark (\ding{51}). The 2P4L regions include all 
lepton flavour combinations. In the three lepton regions, $\ell^{\pm}\ell^{\pm}\ell'^{\mp}$ 
indicates that same-charge leptons have the same flavour, while the opposite-sign lepton has a 
different flavour.}
\label{tab:AllRegions}
\renewcommand{\arraystretch}{1.3}
\centering
\footnotesize
\setlength\tabcolsep{2.4pt}
\begin{tabular}{l|ccc|ccc|ccc}
\hhline{==========}
\multirow{2}{*}{\backslashbox{Channel}{Region}}
 & \multicolumn{3}{c}{Control Regions}
 & \multicolumn{3}{c}{Validation Regions}
 & \multicolumn{3}{c}{Signal Regions}\tabularnewline
\hhline{~---------}

& OCCR 
& DBCR
& 4LCR
& SCVR 
& 3LVR
& 4LVR
& 1P2L 
& 1P3L 
& 2P4L

\tabularnewline

\hline
  
 Electron channel
 & $e^{\pm}e^{\mp}$ 
 & $e^{\pm}e^{\pm}e^{\mp}$ 
 & \multirow{4}{*}{$\ell^{\pm}\ell^{\pm} \ell^{\mp}\ell^{\mp}$ }
 & $e^{\pm}e^{\pm}$
 & $e^{\pm}e^{\pm}e^{\mp}$
 & \multirow{4}{*}{$\ell^{\pm}\ell^{\pm} \ell^{\mp}\ell^{\mp}$ }
 & $e^{\pm}e^{\pm}$ 
 & $e^{\pm}e^{\pm}e^{\mp}$
 & \multirow{4}{*}{$\ell^{\pm}\ell^{\pm} \ell^{\mp}\ell^{\mp}$ }
\tabularnewline

 Mixed channel
 & -
 & $e^{\pm}\mu^{\pm}\ell^{\mp}$
 &
 & $e^{\pm}\mu^{\pm}$
 & 
 \begin{tabular}{c}
 $e^{\pm}\mu^{\pm}\ell^{\mp}$\\
 $\ell^{\pm}\ell^{\pm}\ell'^{\mp}$
 \end{tabular}
 & 
 & $e^{\pm}\mu^{\pm}$
 & 
 \begin{tabular}{c}
 $e^{\pm}\mu^{\pm}\ell^{\mp}$\\
 $\ell^{\pm}\ell^{\pm}\ell'^{\mp}$
 \end{tabular}
 &\tabularnewline

 Muon channel
 & -
 & $\mu^{\pm}\mu^{\pm}\mu^{\mp}$
 &
 & $\mu^{\pm}\mu^{\pm}$
 & $\mu^{\pm}\mu^{\pm}\mu^{\mp}$
 & 
 & $\mu^{\pm}\mu^{\pm}$
 & $\mu^{\pm}\mu^{\pm}\mu^{\mp}$
 &
 \tabularnewline

 \hline
 $m(e^{\pm}e^{\pm})$ [GeV]
 & $[130,2000]$
 & $[90,200)$
 & \multirow{3}{*}{$[60,150)$}
 & $[130,200)$
 & $[90,200)$
 & \multirow{3}{*}{$[150,200)$}
 & $[200,\infty)$
 & $[200,\infty)$
 & \multirow{3}{*}{$[200,\infty)$}
 \tabularnewline

 $m(\ell^{\pm}\ell^{\pm})$ [GeV]
 & -
 & $[90,200)$
 &
 & $[130,200)$
 & $[90,200)$
 & 
 & $[200,\infty)$
 & $[200,\infty)$
 &
 \tabularnewline

 $m(\mu^{\pm}\mu^{\pm})$ [GeV]
 & -
 & $[60,200)$
 &
 & $[60,200)$
 & $[60,200)$
 & 
 & $[200,\infty)$
 & $[200,\infty)$
 &
 \tabularnewline
 
 \hline
 $b$-jet veto
 & \ding{51}
 & \ding{51}
 & \ding{51}
 & \ding{51}
 & \ding{51}
 & \ding{51}
 & \ding{51}
 & \ding{51}
 & \ding{51}
 \tabularnewline
 $Z$ veto
 & -
 & inverted
 & -
 & -
 & \ding{51}
 & -
 & -
 & \ding{51}
 & \ding{51}
 \tabularnewline
  $\Delta R(\ell^{\pm},\ell^{\pm})<3.5$
 & -
 & -
 & -
 & -
 & -
 & -
 & \ding{51}
 & \ding{51}
 & -
 \tabularnewline
$p_{\mathrm{T}}(\ell^{\pm}\ell^{\pm})>100\;\mathrm{GeV}$
 & -
 & -
 & -
 & -
 & -
 & -
 & \ding{51}
 & \ding{51}
 & -
 \tabularnewline
$\sum|p_{\mathrm{T}}(\ell)|>300\;\mathrm{GeV}$
 & -
 & -
 & -
 & -
 & -
 & -
 & \ding{51}
 & \ding{51}
 & -
 \tabularnewline

$\Delta M/\bar{M}$ requirement
 & -
 & -
 & -
 & -
 & -
 & -
 & -
 & -
 & \ding{51}
 \tabularnewline
\hline
\hline
\end{tabular}
\end{table}

\FloatBarrier

\section{Background composition and estimation}
\label{sec:backgroundEstimation}

Prompt SM backgrounds in all regions are estimated using the simulated samples listed in 
Section~\ref{sec:monteCarloProduction}. Prompt light leptons are defined as leptons 
originating from $Z$, $W$, and $H$ boson decays or leptons from $\tau$ decays if
the $\tau$ has a prompt source (e.g. $Z\rightarrow\tau\tau$). 
MC events containing at least one non-prompt or fake selected tight or loose lepton are discarded 
to avoid an overlap with the data-driven fake-background estimation. Prompt electrons 
in the remaining simulated events are corrected to account for  
different charge misidentification probabilities in data and simulation.

\begin{figure}[htbp]
\centering
\includegraphics[width=0.6\textwidth]{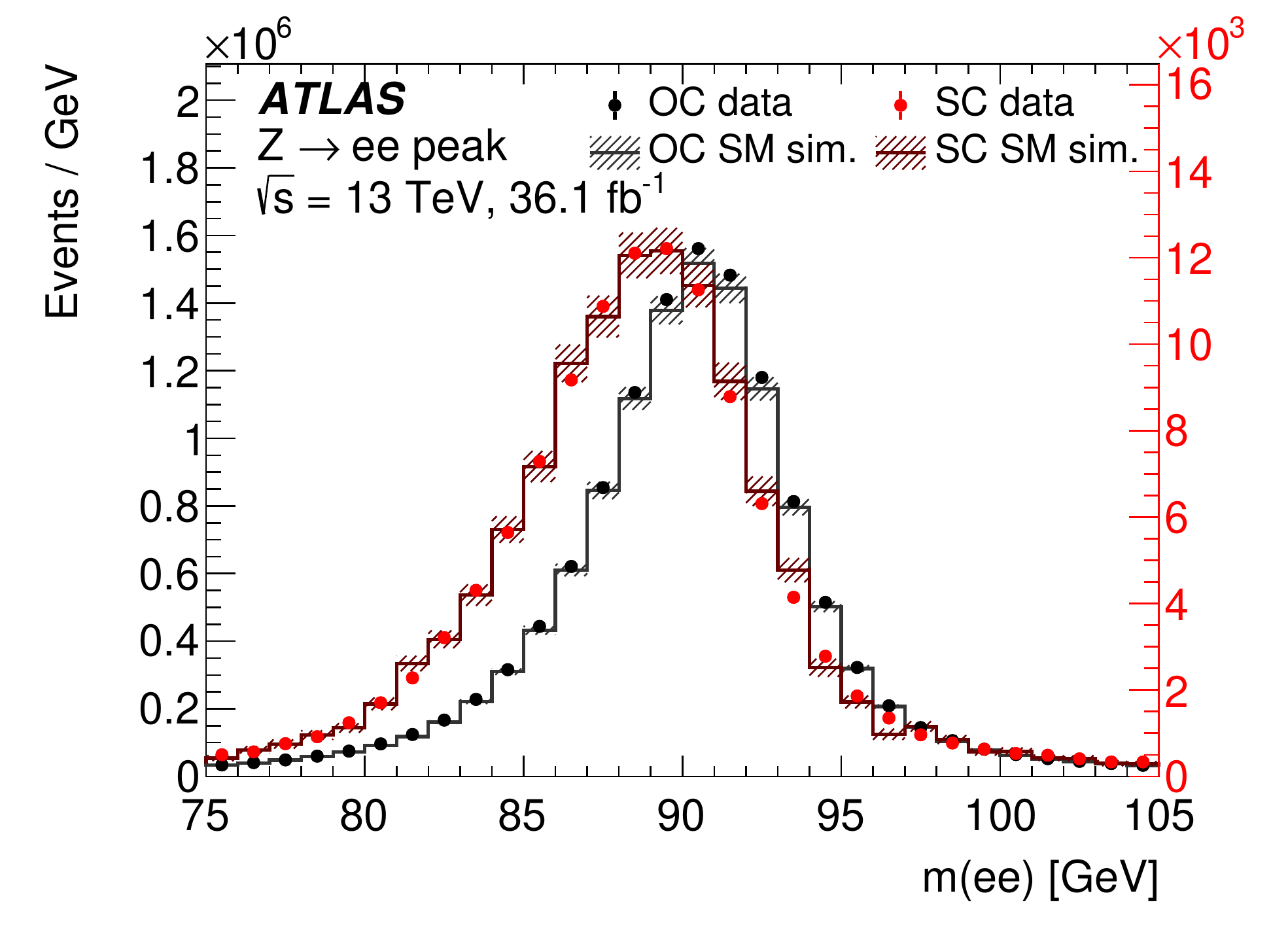}
\caption{Dielectron mass distributions for opposite-charge (black) and same-charge (red) pairs
for data (filled circles) and MC simulation (continuous line). The latter includes a correction
for charge misidentification. The hatched band indicates the statistical error
and the luminosity uncertainty summed in quadrature applied to MC simulated events.}
\label{fig:ZPeak}
\end{figure}

Electron charge misidentification is caused predominantly by bremsstrahlung. The emitted 
photon can either convert to an electron--positron pair, which happens in most of the cases, 
or traverse the inner detector without creating any track. In the first case, the cluster 
corresponding to the initial electron can be matched to the wrong-charge track, or most of the energy is 
transferred from one track to the other because of the photon. In case of photon emission without
subsequent pair production, the electron track has usually very few hits only in the 
silicon pixel layers, and thus a short lever arm on its curvature. Because the electron
charge is derived from the track curvature, it could be incorrectly determined while the electron energy 
is likely appropriate as the emitted photon deposits all of its energy in the 
EM calorimeter as well. For a similar reason high-energy electrons are more often affected by 
charge misidentification, as their tracks are approximately straight and therefore challenging for the 
curvature measurement. The modelling of charge misidentification in simulation deviates from 
data due to the complex processes involved, which particularly rely on a very precise description 
of the detector material. A correction is obtained by comparing the charge misidentification probability 
measured in data to the one in simulation. The charge misidentification probability is 
extracted by performing a likelihood fit on a dedicated $Z \rightarrow ee$ data sample 
(see Figure~\ref{fig:ZPeak}). Electron pairs are selected around the $Z$ boson peak and
categorised in opposite-charge (OC) and same-charge (SC) selections with the 
invariant mass requirements $|m_{\text{OC}}(ee)-m(Z)|<14\;\GeV$ and $|m_{\text{SC}}(ee)-m(Z)|<15.8\;\GeV$, 
respectively. Events from contributions other than $Z \rightarrow ee$ are subtracted from 
the peak regions. They are modelled with simulation and their normalisation is determined 
from data in mass windows around the $Z$ peak defined as $14\;\GeV\; < |m_{\text{OC}}(ee)-m(Z)| < 18\;\GeV\;$ 
for OC and $15.8\;\GeV\; < |m_{\text{OC}}(ee)-m(Z)| < 31.6\;\GeV\;$ for SC.
The number of OS and SC electron pairs in the two regions ($N^{ij}=N^{ij}_{\text{SC}} + N^{ij}_{\text{OC}}$) 
are then used as inputs of the likelihood fit.

The probability to observe $N^{ij}_{\text{SC}}$ same-charge pairs is the Poisson probability:

\begin{linenomath}
\begin{equation*}
f(N^{ij}_{\text{SC}};\lambda)=\frac{\lambda^{N^{ij}_{\text{SC}}} \mathrm{e}^{-\lambda}}{N^{ij}_{\text{SC}}!},
\label{eq:thePoisson}
\end{equation*}
\end{linenomath}

with $\lambda = N^{ij}(P_i(1-P_j) + P_j(1-P_i))$ denoting the expected number of same-charge 
pairs in bin $(i,j)$, where $i$ and $j$ indicate the kinematic configuration of the two electrons 
in the pair, given the charge misidentification probabilities $P_i$ and $P_j$. $N^{ij}_{\text{SC}}$ 
is the measured number of same-charge pairs. The formula for the negative log likelihood used in the
likelihood fit is given in Eq.~\ref{eq:logLikelihood}:

\begin{linenomath}
\begin{equation}
-\log L(\bm{P}|\bm{N_{\text{SC}}},\bm{N}) = \sum\limits_{i,j}\log(N^{ij}(P_i(1-P_j) + P_j(1-P_i)))N^{ij}_{\text{SC}}-N^{ij}(P_i(1-P_j) + P_j(1-P_i)).
\label{eq:logLikelihood}
\end{equation}
\end{linenomath}

\begin{figure}[htbp]
\centering
\subfloat[]{
\includegraphics[width= 0.48\textwidth]{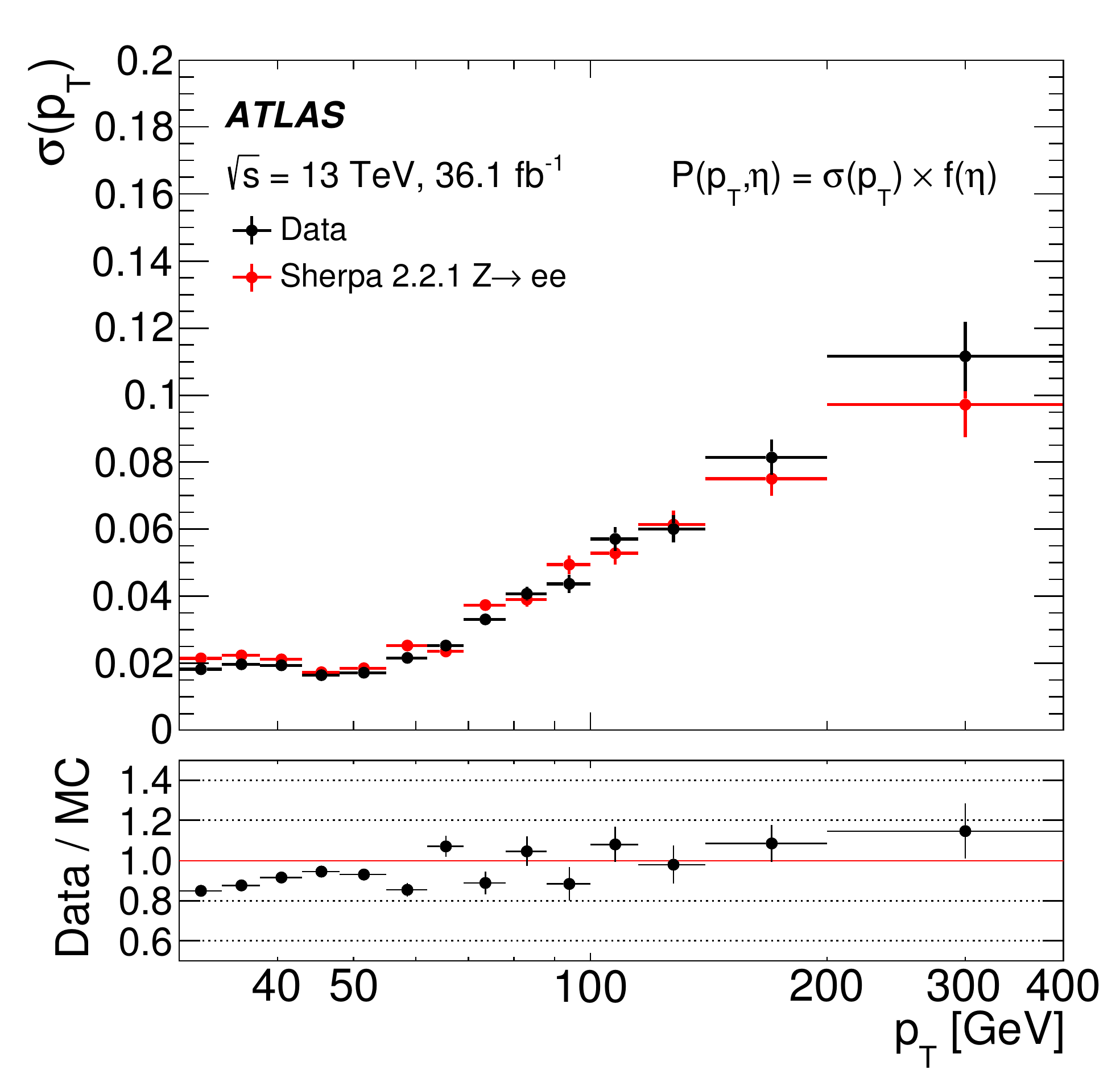}
\label{fig:chargeFlipPt}
}
\subfloat[]{
\includegraphics[width=0.48\textwidth]{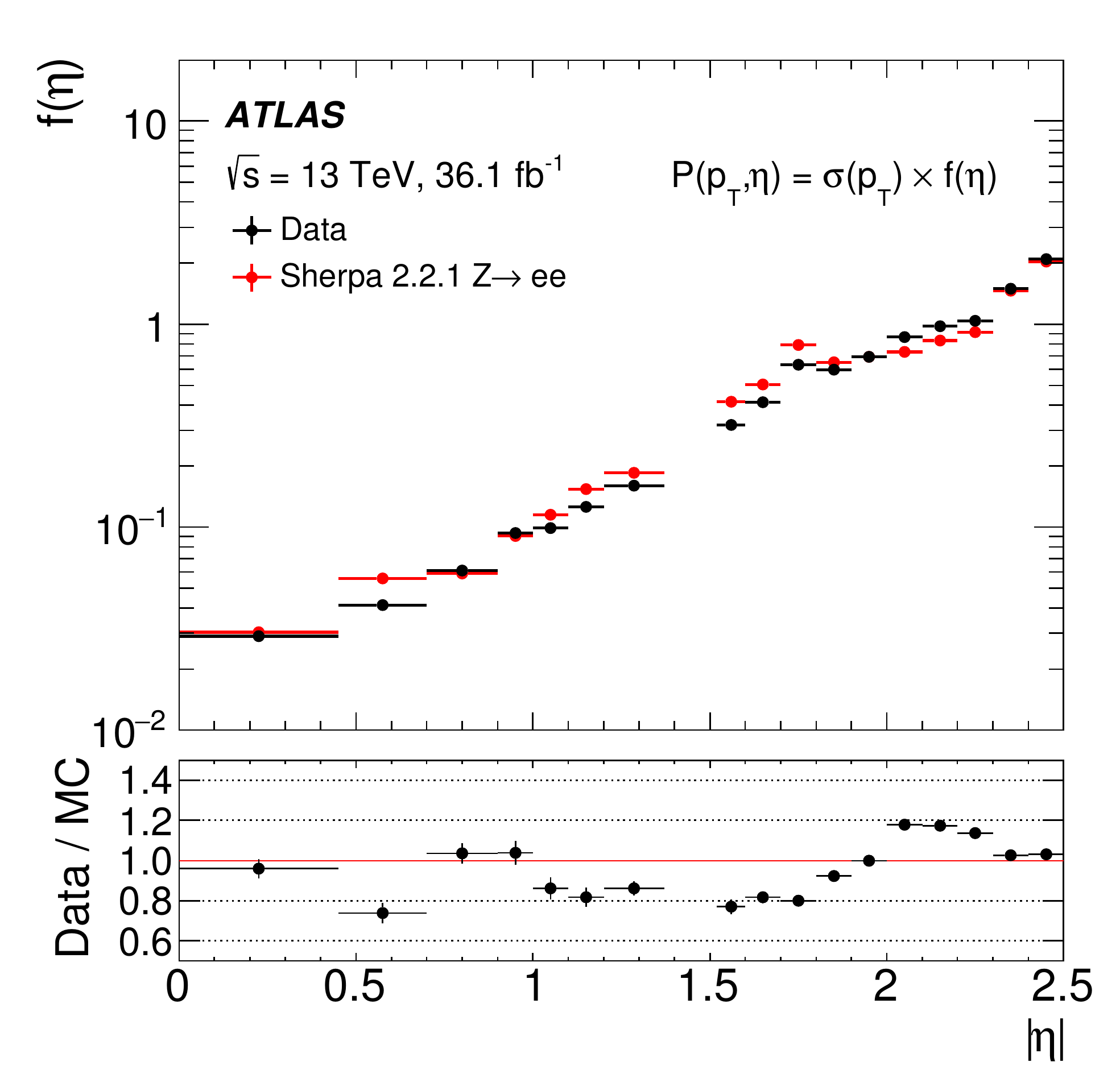}
\label{fig:chargeFlipEta}
}
\caption{Comparison of the factors composing the charge misidentification probability 
$P(\pt,\eta)=\sigma(\pt) \times f(\eta)$ measured in data and in
simulation using the likelihood fit in the $Z/\gamma^* \rightarrow ee$ region.
The area of the distribution describing $f(\eta)$ was set to unity (see text for details).
Error bars correspond to the statistical uncertainties estimated with the likelihood fit.
Plot \protect\subref{fig:chargeFlipPt} shows the charge misidentification probability 
component as a function of \pt and plot \protect\subref{fig:chargeFlipEta} shows the 
component as a function of $|\eta|$.
}
\label{fig:1DchargeFlipRates}
\end{figure}

The charge misidentification probability is parameterised as a function of electron \pt 
and $\eta$, $P(\pt,\eta)=\sigma(\pt) \times f(\eta)$. The binned values, $\sigma(\pt)$ and $f(\eta)$,
are free parameters in the likelihood fit. 
To ensure the proper normalisation of $P(\pt,\eta)$, the area of the distribution describing 
$f(\eta)$ was set to unity. The charge 
misidentification probability is measured with the same method in a simulated
$Z/\gamma^* \rightarrow ee$ sample and in data. The comparison of the result is shown in 
Figure~\ref{fig:1DchargeFlipRates}. All prompt electrons in 
simulated events are corrected with charge reconstruction scale factors. The scale factors are defined as
$P(\pt,\eta;\text{data})/P(\pt,\eta;\text{MC})$ if the charge is wrongly
reconstructed and $\left(1-P(\pt,\eta;\text{data})\right)/\left(1-P(\pt,\eta;\text{MC})\right)$ 
if the charge is properly reconstructed.

The fake-lepton background is estimated with a data-driven approach, the so-called
`fake factor' method, as described in Ref.~\cite{EXOT-2012-25}. 
The $b$-jet veto significantly reduces fake leptons from heavy-flavour decays. 
Most of the fake leptons still passing the analysis selection originate
from in-flight decays of mesons inside jets, jets misreconstructed as 
electrons, and conversions of initial- and final-state radiation photons. The fake factor method provides an estimation of events with 
fake leptons in analysis regions by extrapolating the yields from the so-called 
`side-band regions'. For each analysis region a corresponding side-band region is defined. 
It requires exactly the same selection and lepton multiplicity except that at least one 
lepton must fail to satisfy the tight identification criteria. The ratio of tight to loose leptons
is measured in dedicated `fake-enriched regions'. It is determined as a function of lepton 
flavour, $\pt$, and $\eta$, and referred to as the `fake factor' ($F(\pt,\eta,\text{flavour})$). 
It describes the probability 
for a fake lepton to be identified as a tight lepton.
The definitions of the fake-enriched regions for the electron and muon channels are
reported in Table~\ref{tab:FakeEnrichedRegions}. In the measurement of the fake factor, a requirement on  
the unbalanced momentum in the transverse plane of the event, $E^{\mathrm{miss}}_{\mathrm{T}}$,
is imposed to reject $W+{\mathrm{jets}}$ events and to further enrich the regions 
with fake leptons. The fake factor method relies on the assumption that no prompt leptons appear in the fake-enriched 
samples. This assumption is not fully correct with the imposed selection. Therefore, the 
number of residual prompt leptons in the fake-enriched regions is estimated using simulation and 
subtracted from the numbers of tight and loose leptons used to measure the fake factors. 

The number of events in the analysis regions containing at least one fake lepton, 
$N^{\mathrm{fake}}$, is estimated from the side-bands. Data are weighted with 
fake factors according to the loose lepton multiplicity of the region:

\begin{linenomath}
\begin{equation*}
N^{\mathrm{fake}} = \sum_{i=1}^{N^{\mathrm{data}}_{\mathrm{SB}} }(-1)^{N_{L,i}+1}\prod_{l=1}^{N_{L,i}}F_{l}
\; - \; \sum_{i=1}^{N^{\mathrm{MC}}_{\mathrm{SB}} }(-1)^{N_{L,i}+1}\prod_{l=1}^{N_{L,i}}F_{l},
\label{eq:FFformula}
\end{equation*}
\end{linenomath}

with $N^{\mathrm{data}}_{\mathrm{SB}}$ denoting the number of data events in the side-band,
$N_{L,i}$ is the loose lepton multiplicity in the $i$-th event of the side-band region and $l$ 
indicates the loose lepton. The contamination of prompt leptons in the side-band region is 
subtracted using simulated events, denoted by $N^{\mathrm{MC}}_{\mathrm{SB}}$.

Dedicated two-lepton and three-lepton validation regions, defined in Table~\ref{tab:AllRegions}, 
are used to verify the data-driven fake-lepton estimation in regions as similar to the signal 
regions as possible. They are designed to contain only a negligible number of signal events.
Orthogonality between signal and validation regions is ensured by requiring the invariant mass of the same-charge lepton pair 
$m(\ell^{\pm}\ell^{\pm})$ to be less than 200~GeV in the validation regions. Furthermore, 
diboson modelling and the electron charge misidentification backgrounds are tested. 
Each background estimation is validated in the corresponding regions, defined to be enriched 
in the given contribution.

\begin{table}[t]
\caption{Selection criteria defining the fake-enriched regions used to measure the 
ratio of the numbers of tight and loose leptons, the so-called fake factor, for the electron and muon channels.}
\label{tab:FakeEnrichedRegions}
\renewcommand{\arraystretch}{1.3}
\centering
\small
\begin{tabular}{c|c}
\hline
\hline
\multicolumn{2}{c}{Selection for fake-enriched regions}\tabularnewline
\multicolumn{1}{c}{Muon channel } & Electron channel\tabularnewline
\hline
Single-muon trigger & Single-electron trigger\tabularnewline
$b$-jet veto & $b$-jet veto\tabularnewline
One muon and one jet & One electron\tabularnewline
$p_{\mathrm{T}}(\mathrm{jet})>35\;\mathrm{GeV}$ & Number of tight electrons \textless{} 2\tabularnewline
$\Delta\phi(\mu,\mathrm{jet})>2.7$ &  $m(ee)\notin[71.2,111.2]\;\mathrm{GeV}$ \tabularnewline
$E_{\mathrm{T}}^{\mathrm{miss}}<40\;\mathrm{GeV}$ & $E_{\mathrm{T}}^{\mathrm{miss}}<25\;\mathrm{GeV}$\tabularnewline
\hline
\hline
\end{tabular}
\end{table}

\begin{figure}[htbp]
\centering
\subfloat[]{
\includegraphics[width=0.45\textwidth]{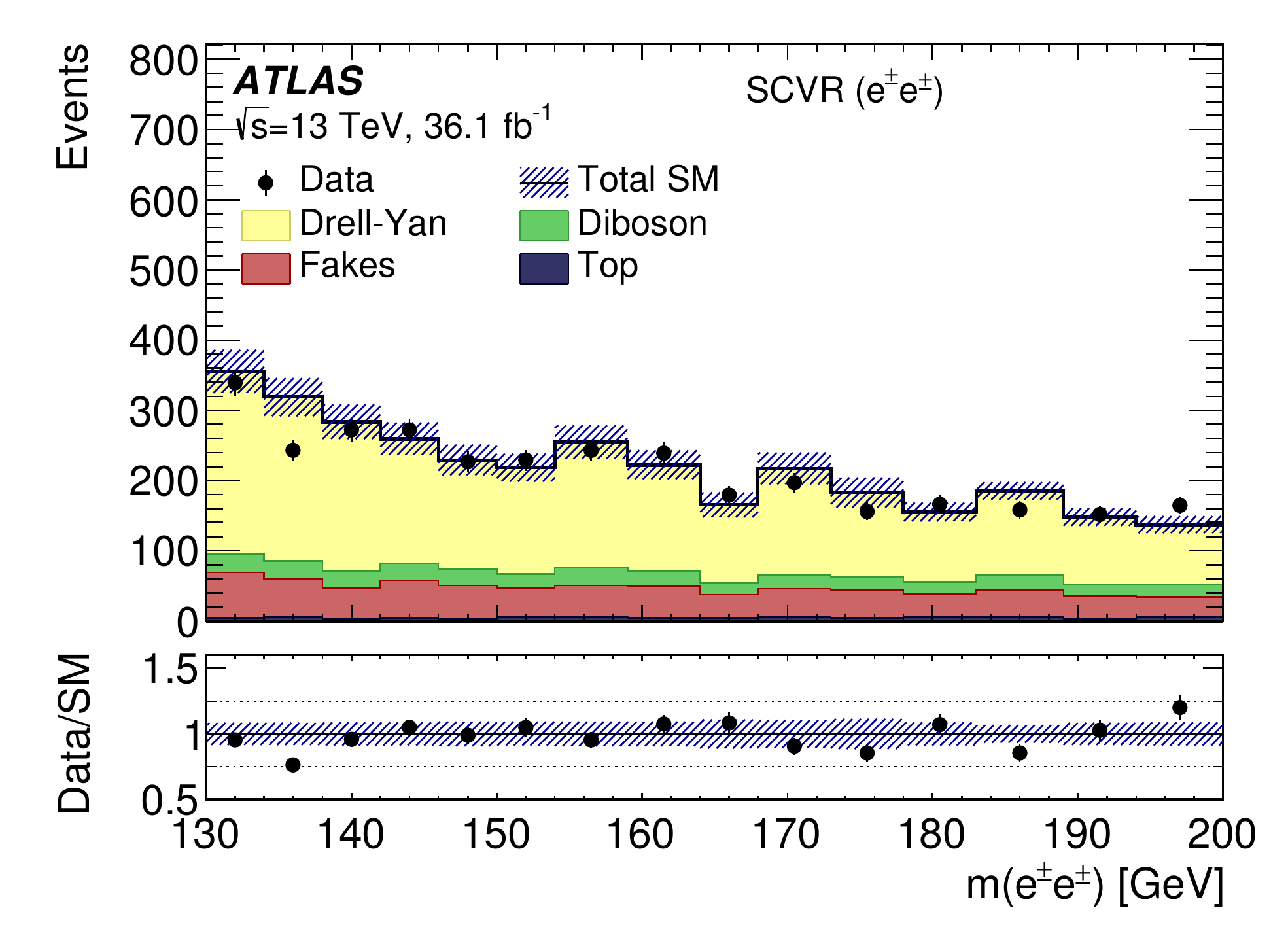}
\label{fig:histo_SSVRee}
}
\subfloat[]{
\includegraphics[width=0.45\textwidth]{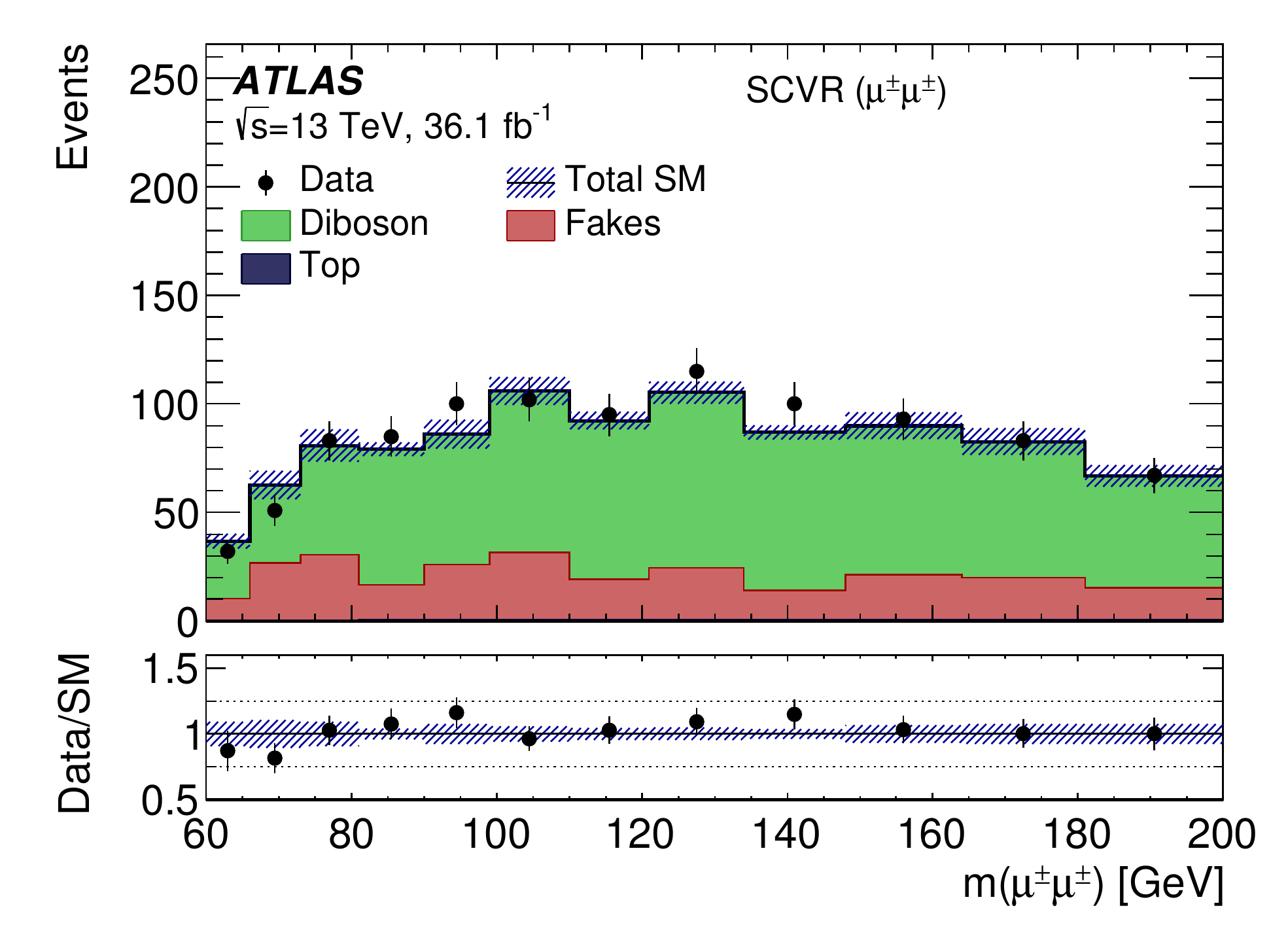}
\label{fig:histo_SSVRmm}
}
\\
\subfloat[]{
\includegraphics[width= 0.45\textwidth]{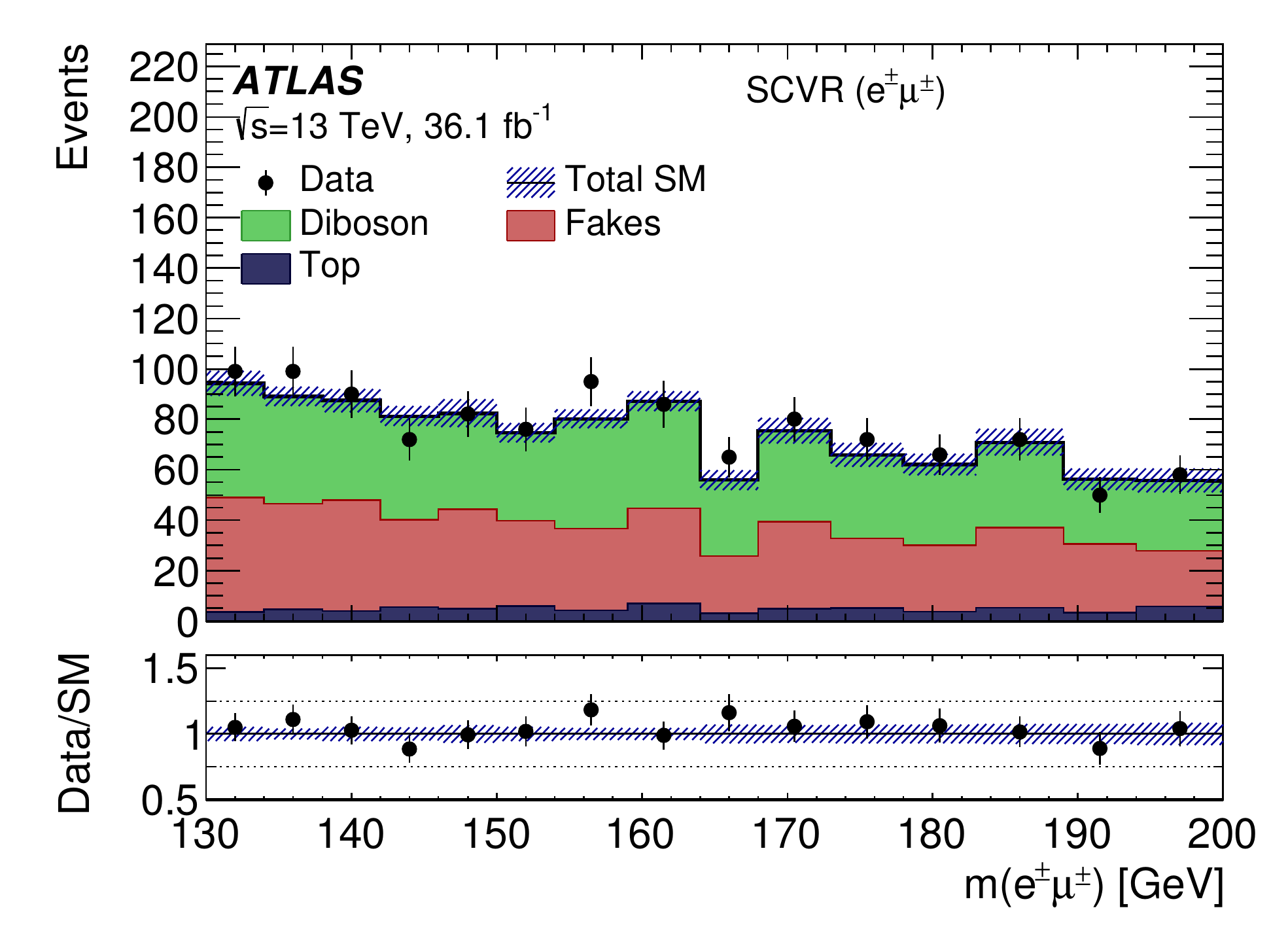}
\label{fig:histo_SSVRem}
}
\subfloat[]{
\includegraphics[width= 0.45\textwidth]{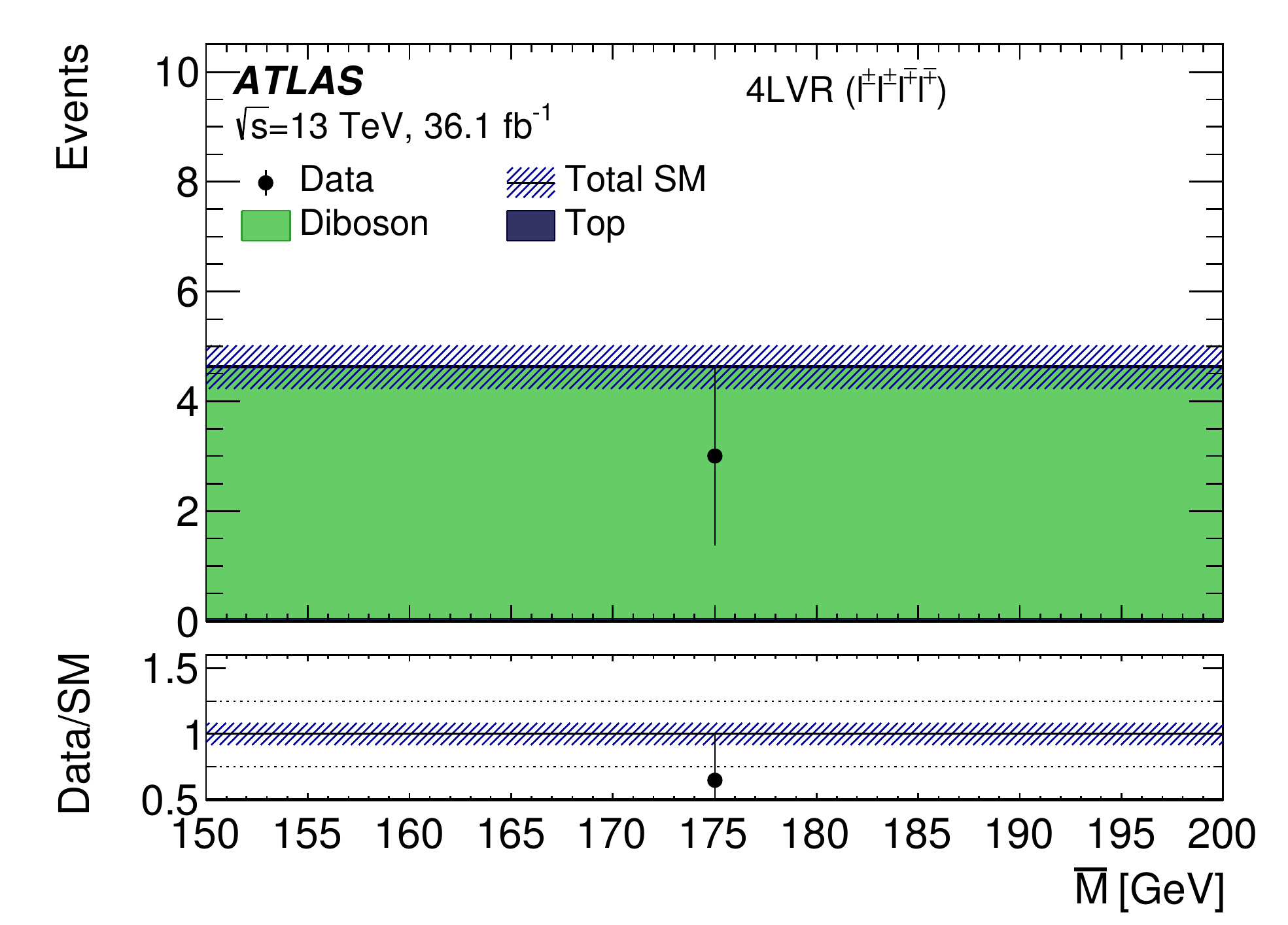}
\label{fig:histo_4LVRrelaxed}
}
\caption{Distributions of dilepton mass for data and SM background predictions in two- and four-lepton 
validation regions:
\protect\subref{fig:histo_SSVRee} the electron--electron,
\protect\subref{fig:histo_SSVRmm} the muon--muon, and
\protect\subref{fig:histo_SSVRem} the electron--muon two-lepton validation regions, as well as
\protect\subref{fig:histo_4LVRrelaxed} the four-lepton validation region.
The hatched bands include all systematic uncertainties post-fit, with the correlations 
between various sources taken into account.}
\label{fig:representativeVRs}
\end{figure}

\begin{figure}[htbp]
\centering
\subfloat[]{
\includegraphics[width= 0.45\textwidth]{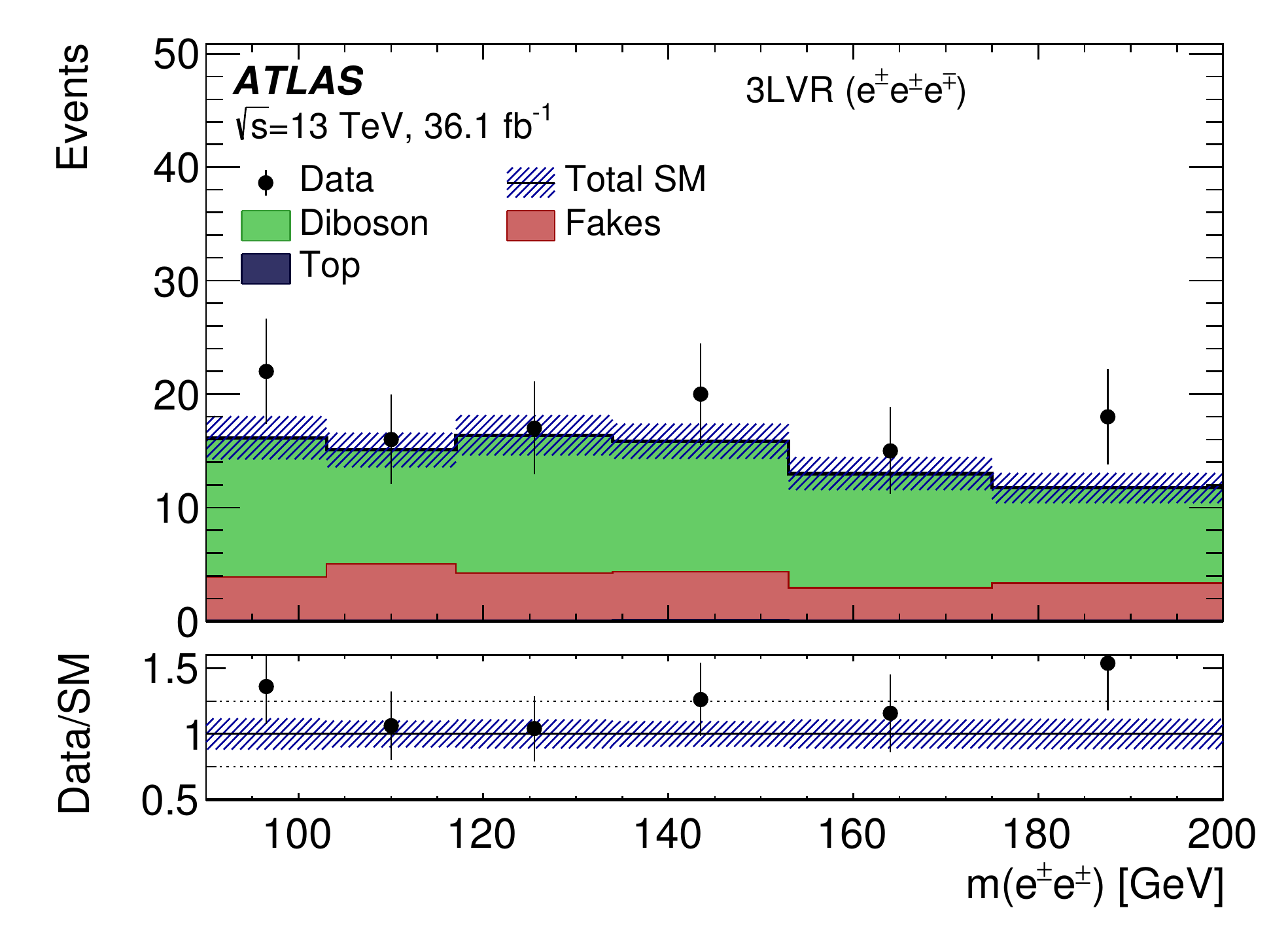}
\label{fig:histo_3LVRee}
}
\subfloat[]{
\includegraphics[width= 0.45\textwidth]{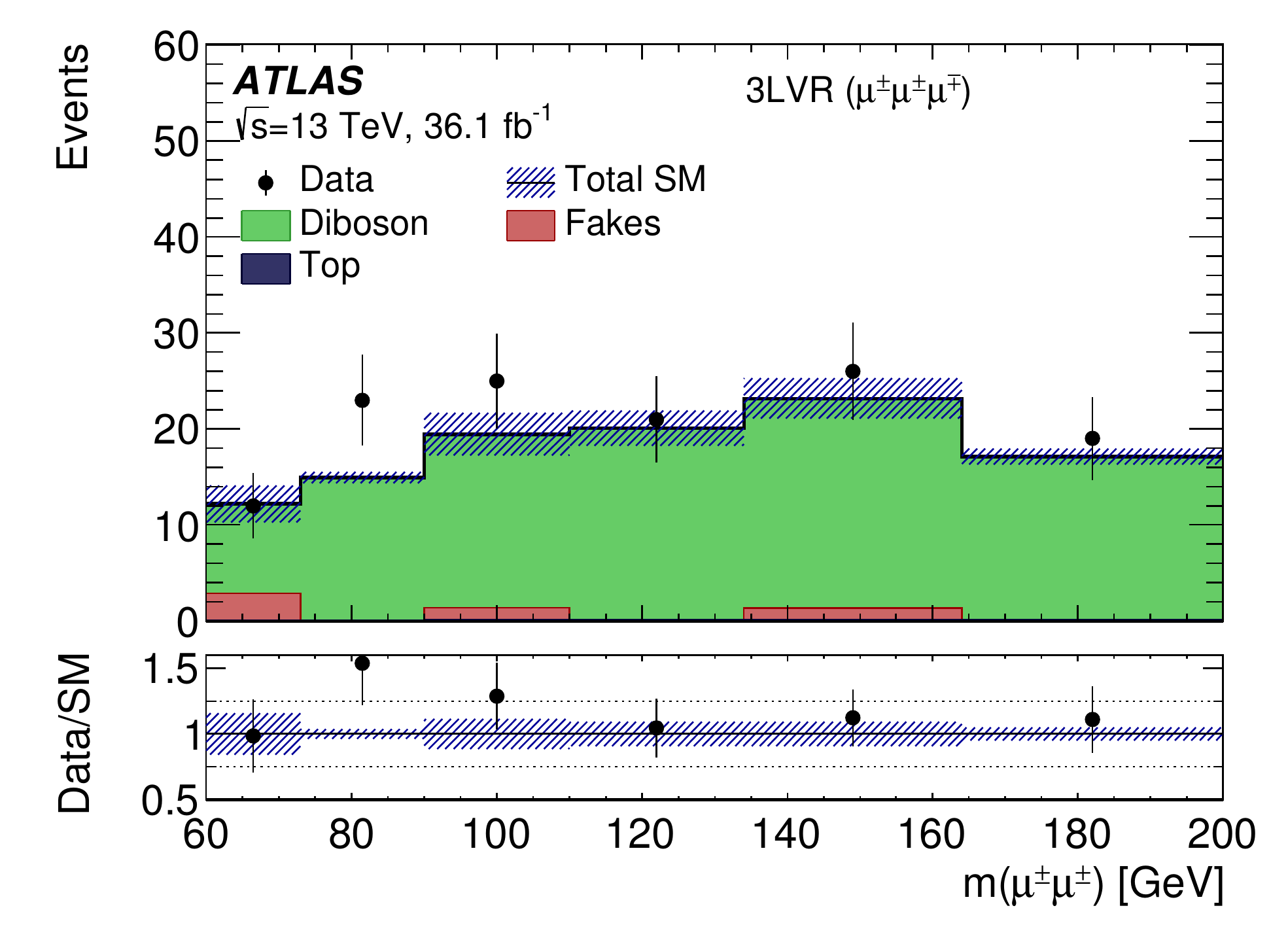}
\label{fig:histo_3LVRmm}
}
\\
\subfloat[]{
\includegraphics[width= 0.45\textwidth]{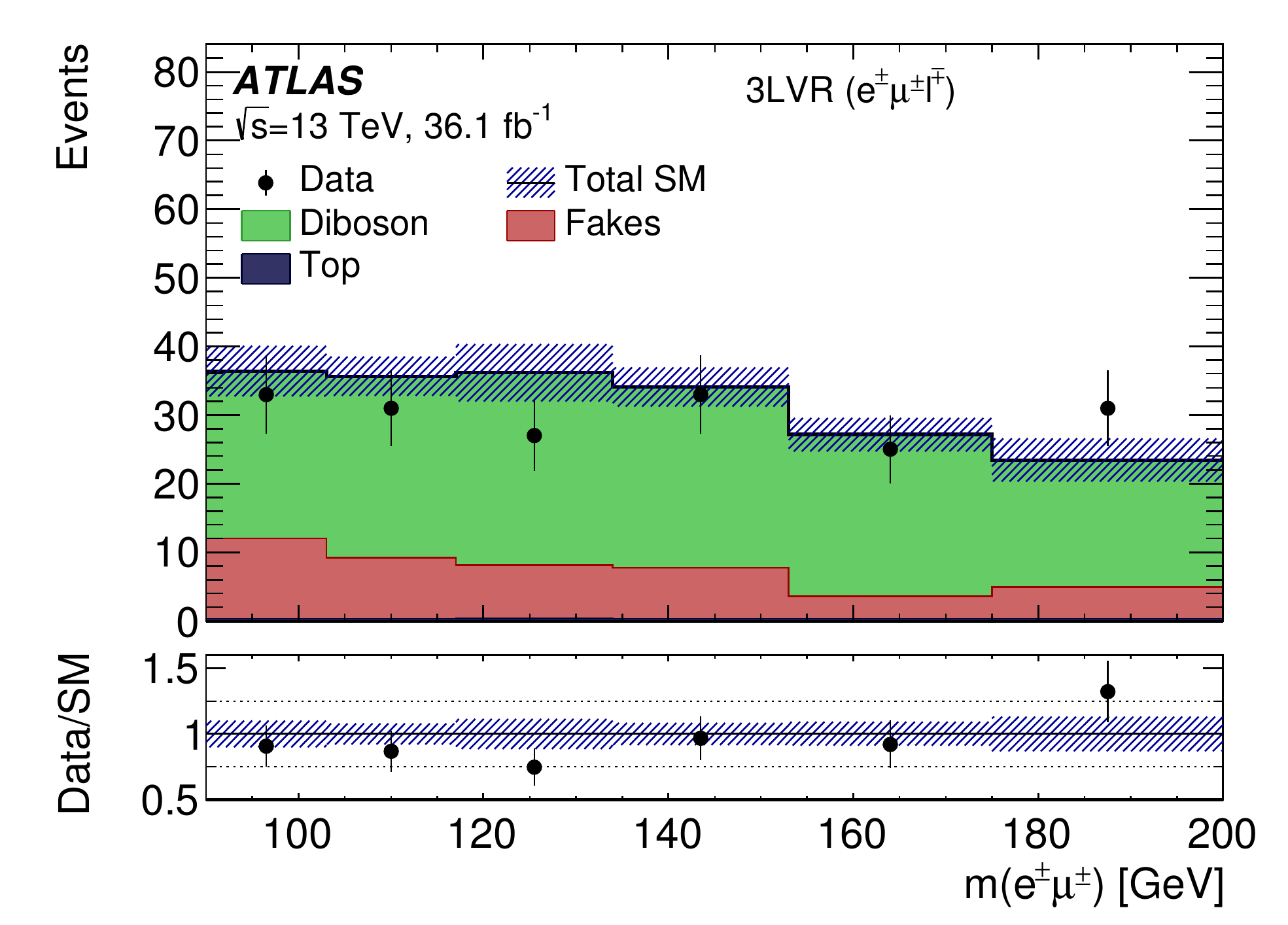}
\label{fig:histo_3LVRem}
}
\subfloat[]{
\includegraphics[width= 0.45\textwidth]{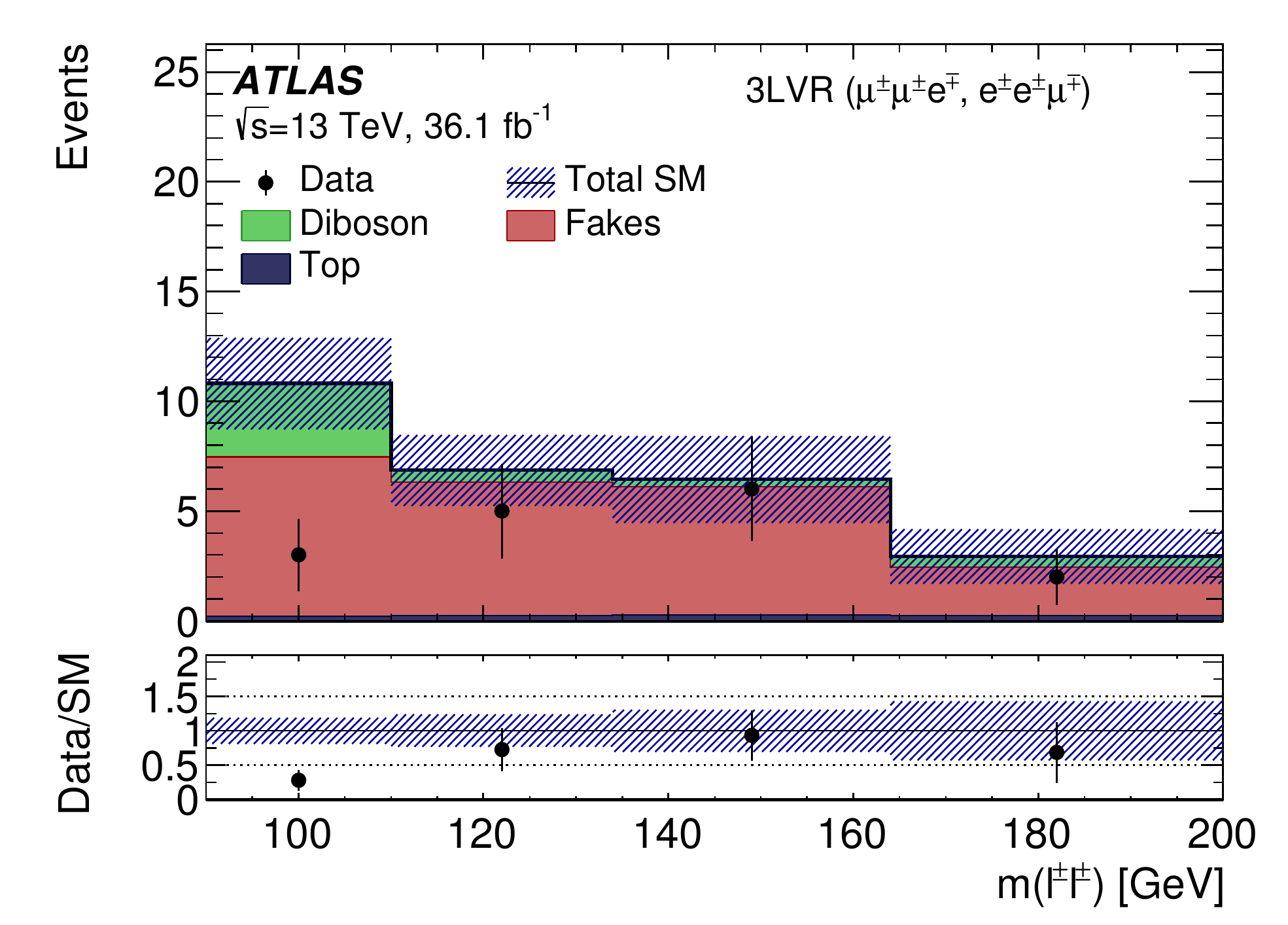}
\label{fig:histo_3LVR_mix}
}
\caption{Distribution of dilepton mass for data and SM background predictions in three-lepton validation 
regions:
\protect\subref{fig:histo_3LVRee}   the three-electron validation region,
\protect\subref{fig:histo_3LVRmm}   the three-muon validation region,
\protect\subref{fig:histo_3LVRem}   the 3LVR with an electron--muon same-charge pair ($e^{\pm}\mu^{\pm}\ell^{\mp}$), and
\protect\subref{fig:histo_3LVR_mix} the 3LVR with a same-flavour same-charge pair ($e^{\pm}e^{\pm}\mu^{\mp}$ or $\mu^{\pm}\mu^{\pm}e^{\mp}$).
The hatched bands include all systematic uncertainties post-fit, with the correlations 
between various sources taken into account.}
\label{fig:representativeVRs2}
\end{figure}

Figures~\ref{fig:representativeVRs} and~\ref{fig:representativeVRs2} present all validation 
regions sensitive to different background sources: same-charge two-lepton validation regions 
(SCVR) for testing the charge misidentification background modelling and fake-background 
predictions, and three-lepton and four-lepton validation regions (3LVR and 4LVR) for testing 
the diboson modelling. Good background modelling is observed in all these regions.

\FloatBarrier

\section{Systematic uncertainties}
\label{sec:SysUnc}

Several sources of systematic uncertainty are accounted for in the analysis.
These correspond to experimental and theoretical sources affecting both background and signal predictions.
All considered sources of systematic uncertainty affect the total event yield, and all except
the uncertainties on the luminosity and cross section also affect the distributions of the
variables used in the fit (Section~\ref{sec:results}).

The cross-sections used to normalise the simulated samples are varied to account
for the scale and PDF uncertainties in the cross-section calculation. The variation is $6\%$
for diboson production~\cite{ATL-PHYS-PUB-2017-005}, $13\%$ for $t\bar{t}\,W$ production,
$12\%$ for $t\bar{t}\,Z$ production, and $8\%$ for $t\bar{t}\,H$ production~\cite{YR4}.
The theoretical uncertainty in the Drell--Yan background is estimated by PDF eigenvector
variations of the nominal PDF set, variations of PDF scale, $\alpha_{\text{S}}$,
electroweak corrections, and photon-induced corrections. The effect of the PDF 
choice is considered by comparing the nominal PDF set to several others, namely 
CT10NNLO~\cite{gao2014ct10}, MMHT14~\cite{motylinski2014updates}, 
NNPDF3.0~\cite{ball2015parton}, ABM12~\cite{alekhin2014abm}, HERAPDF2.0~\cite{CooperSarkar:2011aa,zhang2015hera}, 
and JR14~\cite{jimenez2014delineating}. An envelope is constructed by taking into account the largest deviations
from the nominal choice. The predominant prompt background, arising 
from diboson production, is assigned an additional theoretical uncertainty by comparing 
the nominal \SHERPA 2.2.1 prediction with the \POWHEG prediction. This uncertainty varies from $5\%$ to $10\%$.
Furthermore, the theoretical uncertainty in the NLO cross-section for $pp\rightarrow H^{++}H^{--}$ is
reported to be about $15\%$~\cite{muhlleitner2003note}. It includes the renormalization and factorization
scale dependence and the uncertainty in the parton densities. Lastly, the theoretical uncertainty
in the simulated $pp\rightarrow H^{++}H^{--}$ events is assessed by varying the A14 parameter set in
\PYTHIA 8.186 and choosing alternative PDFs CTEQ6L1 and CT09MC1~\cite{Lai:2009ne}. The impact on the signal
acceptance is found to be negligible. 

A significant contribution arises from the statistical uncertainty in the MC
samples and data sideband regions. Analysis regions have a very restrictive selection and only a small
fraction of the initially generated MC events remains after applying all requirements. The statistical
uncertainty varies from $5\%$ to $40\%$ depending on the signal region.

Experimental systematic uncertainties due to different reconstruction, identification, 
isolation, and trigger efficiencies of leptons in data compared to simulation 
are estimated by varying the corresponding scale-factors. They are at most 3\%
and less significant than the other systematic uncertainties and MC statistical uncertainties.
The same is true for lepton energy or momentum calibration.

The experimental uncertainty related to the charge misidentification probability of 
electrons arises from the statistical uncertainty of both the data and the simulated 
sample of $Z/\gamma^* \rightarrow ee$ events used to measure this probability. The 
uncertainty ranges between $10\%$ and $20\%$ as a function of the electron \pt and $\eta$.
Possible systematic effects were investigated by altering the selection
requirements imposed on the invariant mass used to select $Z/\gamma^* \rightarrow ee$
events analysed to measure the misidentification probability. The effects estimated with 
this method are found to be negligible compared to the statistical uncertainty.

The experimental systematic uncertainty in the data-driven estimate of the fake-lepton 
background is evaluated by varying the nominal fake factor to account for different
effects. The $E^{\mathrm{miss}}_{\mathrm{T}}$ requirement is altered to consider
variations in the $W+\mathrm{jets}$ composition. The flavour composition of the
fakes is investigated by requiring an additional recoiling jet in the electron 
channel and changing the definition of the recoiling jet in the muon channel. Furthermore, the transverse
impact parameter criterion for tight muons (defined in Section~\ref{sec:ObjSelection}) is varied by one
standard deviation. Finally, in the fake-enriched regions, the 
normalisation of the subtracted simulated samples, to remove the prompt lepton component, 
is altered within its uncertainties. This accounts for uncertainties related to the luminosity, 
the cross-section, and the corrections applied to simulation-based predictions.
The statistical uncertainty in the fake factors is added in quadrature to the total 
systematic error. The uncertainty ranges between $10\%$ 
and $20\%$ across all \pt and $\eta$ bins.

\begin{figure}[htpb]
\centering
\includegraphics[width=1.03\textwidth]{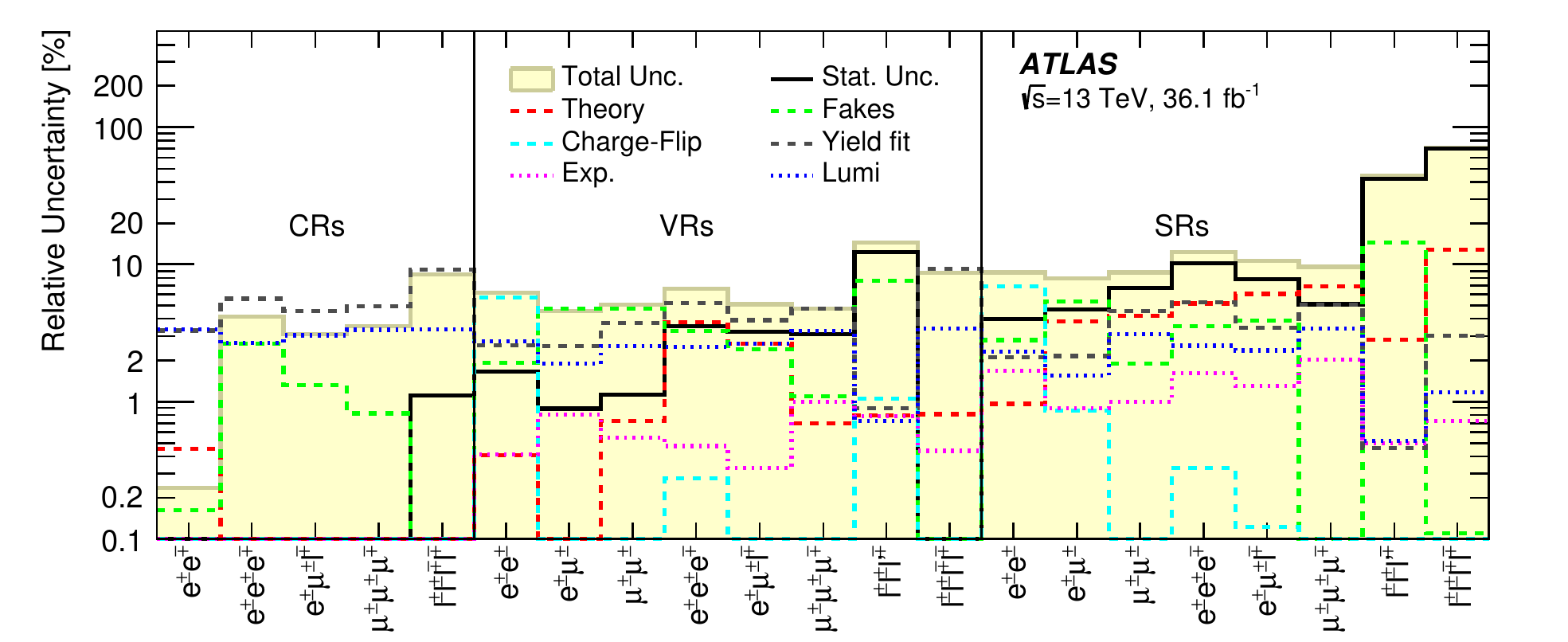}
\caption{Relative uncertainties in the total background yield estimation after 
the fit. `Stat. Unc.' corresponds to reducible and irreducible 
background statistical uncertainties. `Yield fit' corresponds to the uncertainty arising 
from fitting the yield of diboson and Drell--Yan backgrounds. `Lumi' corresponds to 
the uncertainty in the luminosity. `Theory' indicates the theoretical uncertainty in the 
physics model used for simulation (e.g. cross-sections). `Exp.' indicates the 
uncertainty in the simulation of electron and muon efficiencies (e.g. trigger, identification).
`Fakes' is the uncertainty associated with the model of the fake background.
Individual uncertainties can be correlated, and do not necessarily add in quadrature to 
the total background uncertainty, which is indicated by `Total Unc.'.}
\label{fig:allAnalysisRegionsSys}
\end{figure}

The total relative systematic uncertainty after the fit (Section~\ref{sec:results}), 
and its breakdown into components, is presented in Figure~\ref{fig:allAnalysisRegionsSys}.
All experimental systematic uncertainties discussed here affect the signal samples as 
well as the background.

\FloatBarrier

\section{Statistical analysis and results}
\label{sec:results}

The statistical analysis package \textsc{HistFitter}~\cite{baak2015histfitter} was used to 
implement a maximum-likelihood fit of the dilepton invariant mass distribution 
in all control and signal regions, and the $\bar{M}$ distribution in four-lepton regions to obtain
the numbers of signal and background events. The likelihood is the product of a Poisson 
probability density function describing the observed number of events and Gaussian distributions to constrain
the nuisance parameters associated with the systematic uncertainties. The widths of the Gaussian distributions 
correspond to the magnitudes of these uncertainties, whereas Poisson distributions are used 
for MC simulation statistical uncertainties. 
Furthermore, additional free parameters are introduced for the Drell--Yan and the diboson 
background contributions, to fit their yields in the analysis regions. Fitting the yields
of the largest backgrounds reduces the systematic uncertainty in the predicted yield from SM sources.
The fitted normalisations are compatible with their SM predictions within the uncertainties.
The diboson yield is described by four free parameters, each corresponding to a different
diboson region: electron channel, muon channel, mixed channel, and the four-lepton channel.
After the fit, the compatibility between the data and the expected background was 
assessed. For various branching ratio assumptions, $95\%$~CL upper limits were set 
on the $pp\rightarrow H^{++}H^{--}$ cross-section using the $\text{CL}_{\text{s}}$ 
method~\cite{read2002presentation}.

\subsection{Fit results}
\label{sec:FitResults}

\begin{figure}[htpb]
\centering
\includegraphics[width=1.03\textwidth]{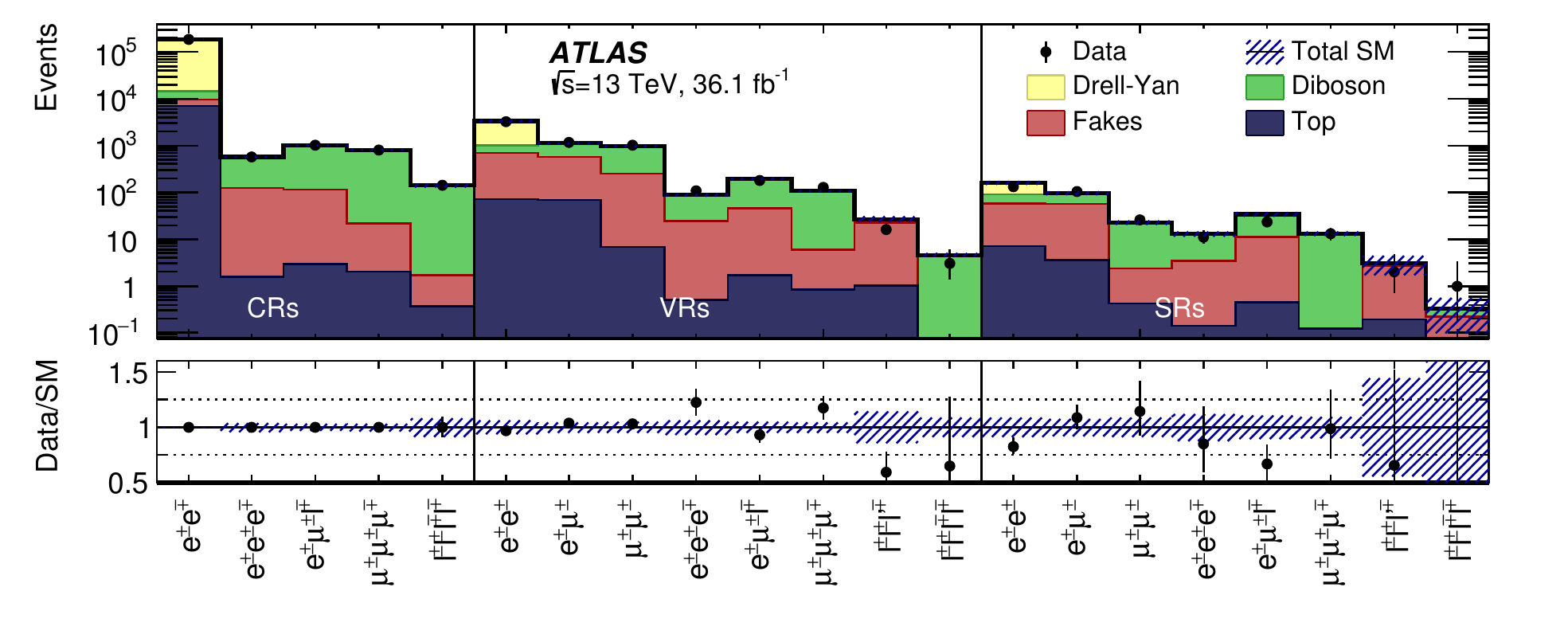}
\caption{Number of observed and expected events in the control, validation, and
signal regions for all channels considered. The background expectation is the result
of the fit described in the text. The hatched bands 
include all systematic uncertainties post-fit with the correlations between 
various sources taken into account. The notation $\ell^{\pm}\ell'^{\pm}\ell^{\mp}$ indicates  
that the same-charge leptons have different flavours and $\ell^{\pm}\ell^{\pm}\ell'^{\mp}$ 
indicates that same-charge leptons have the same flavour, while the opposite-charge 
lepton has a different flavour. }
\label{fig:allAnalysisRegions}
\end{figure}

\begin{figure}[t]
\centering
\subfloat[]{
\includegraphics[width=0.45\textwidth]{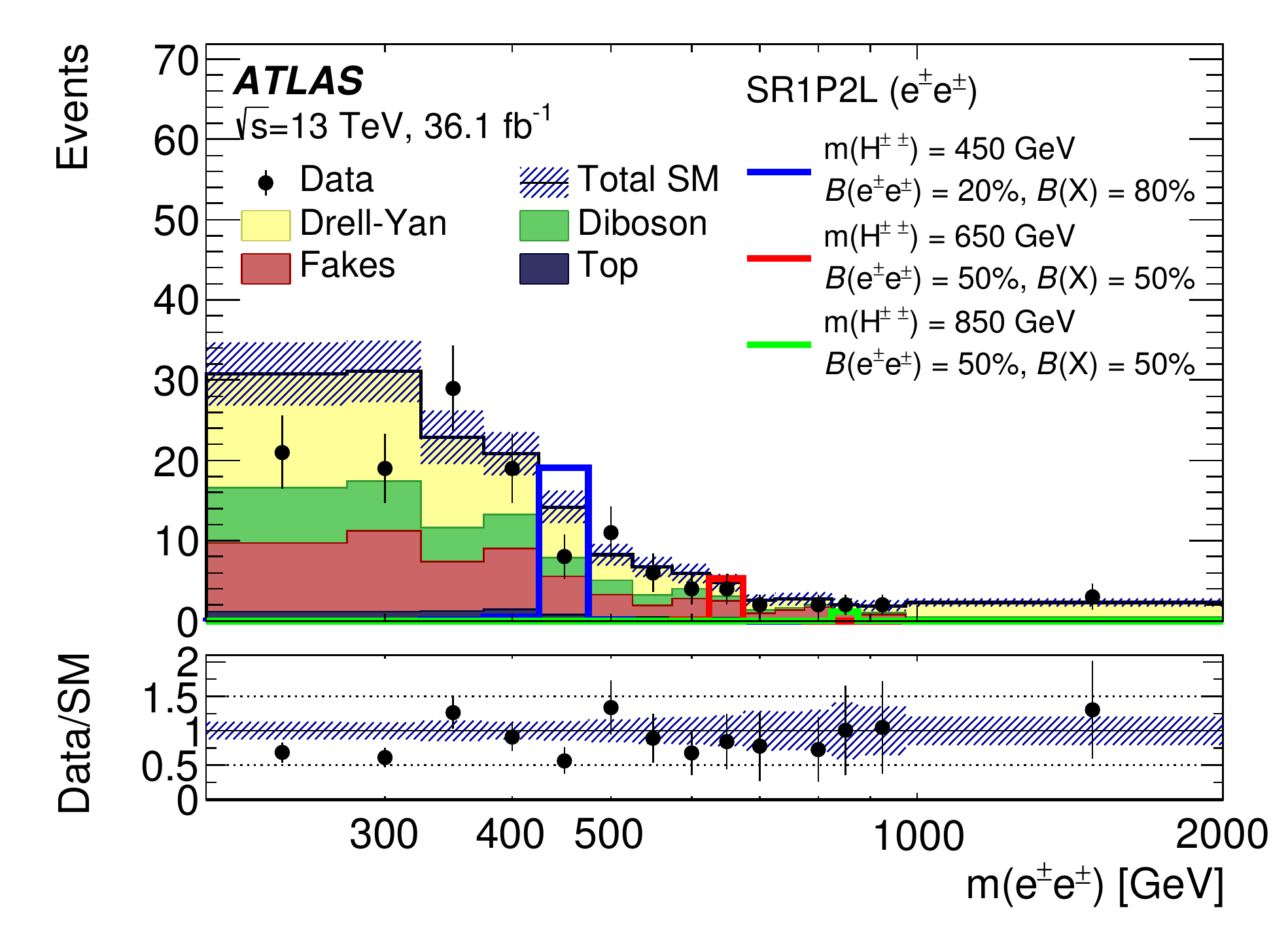}
\label{fig:histo_SR1P2Lee}
}
\subfloat[]{
\includegraphics[width=0.45\textwidth]{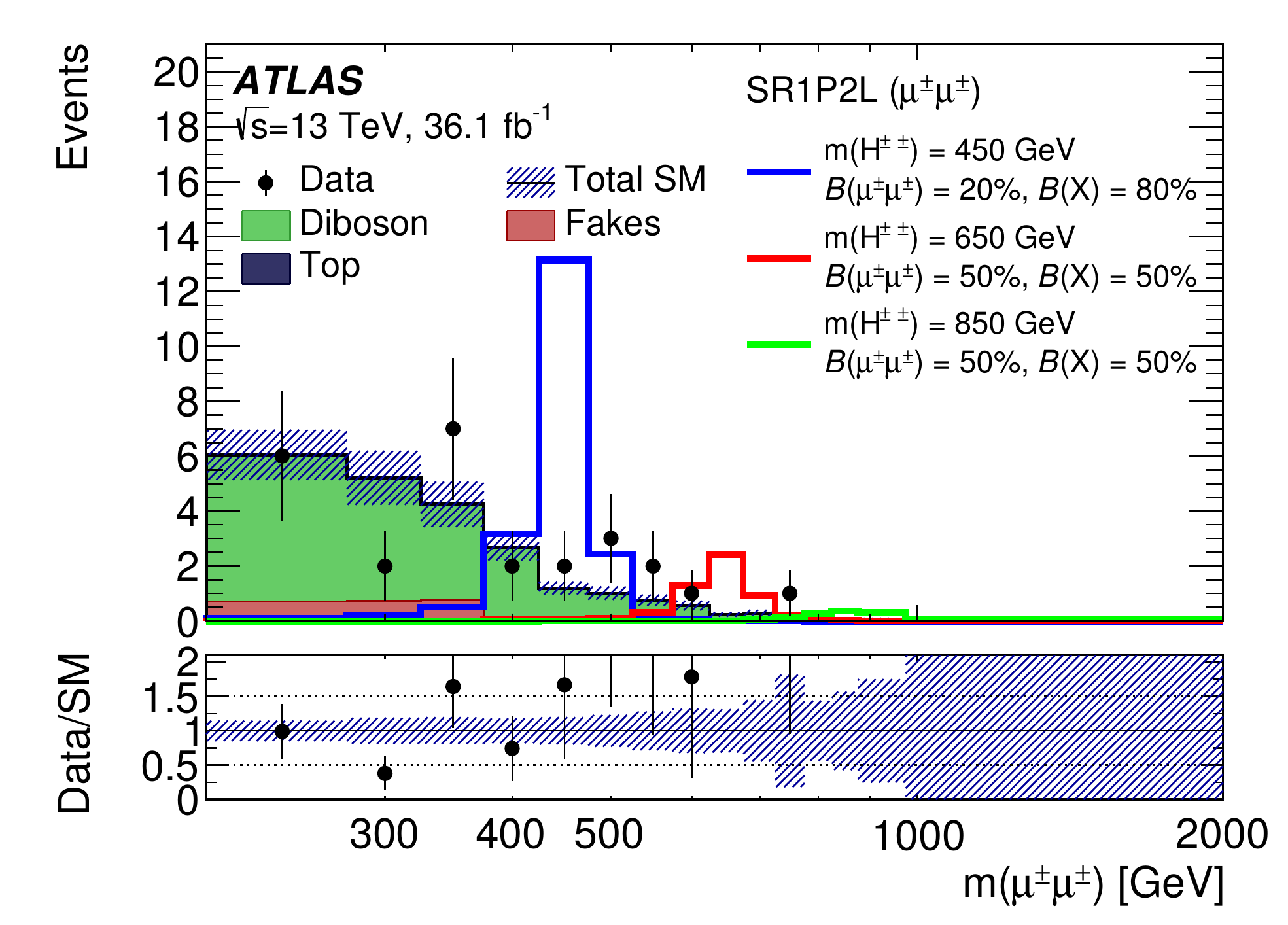}
\label{fig:histo_SR1P2Lmm}
}
\\
\subfloat[]{
\includegraphics[width= 0.45\textwidth]{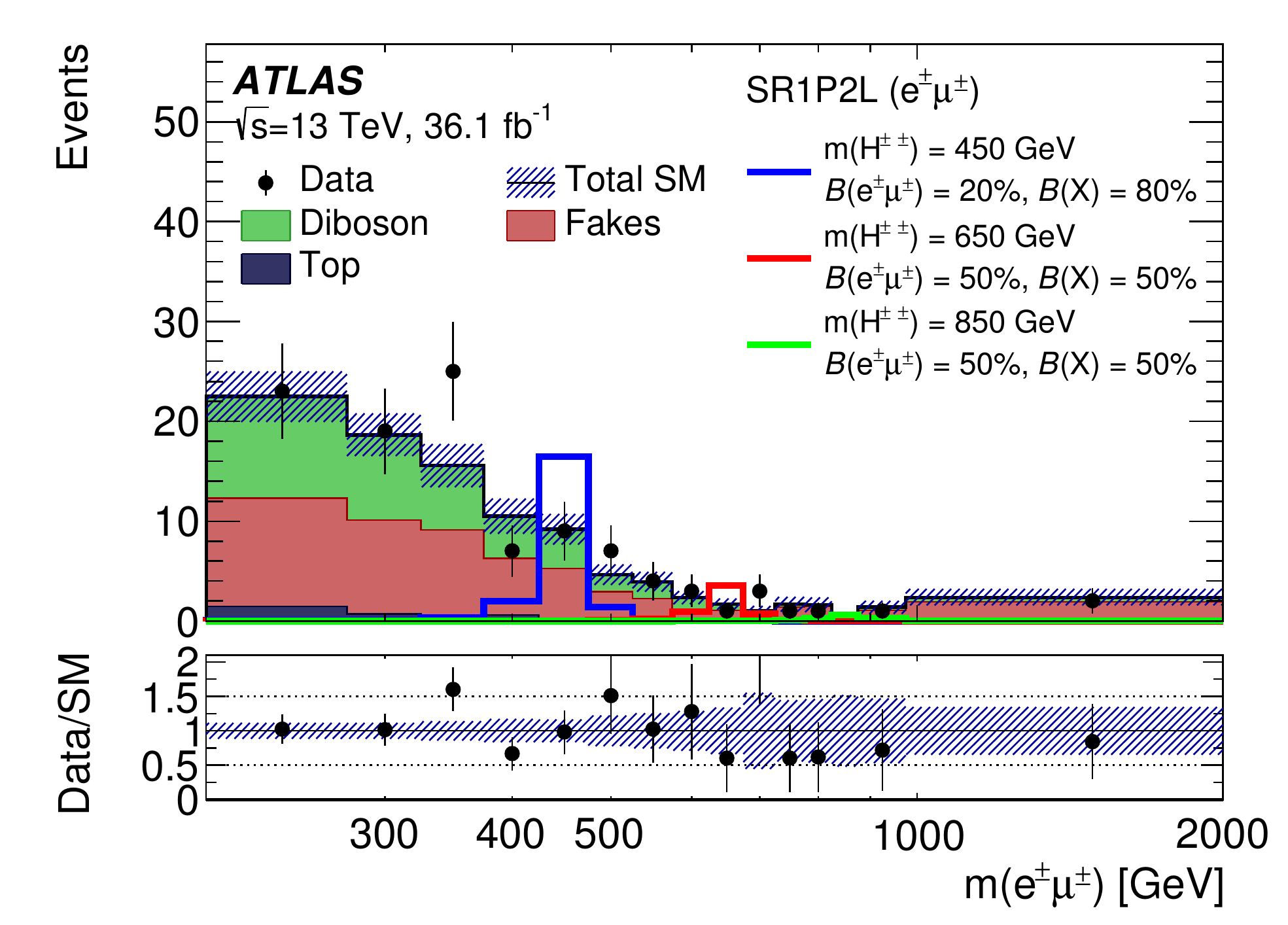}
\label{fig:histo_SR1P2Lem}
}
\subfloat[]{
\includegraphics[width= 0.45\textwidth]{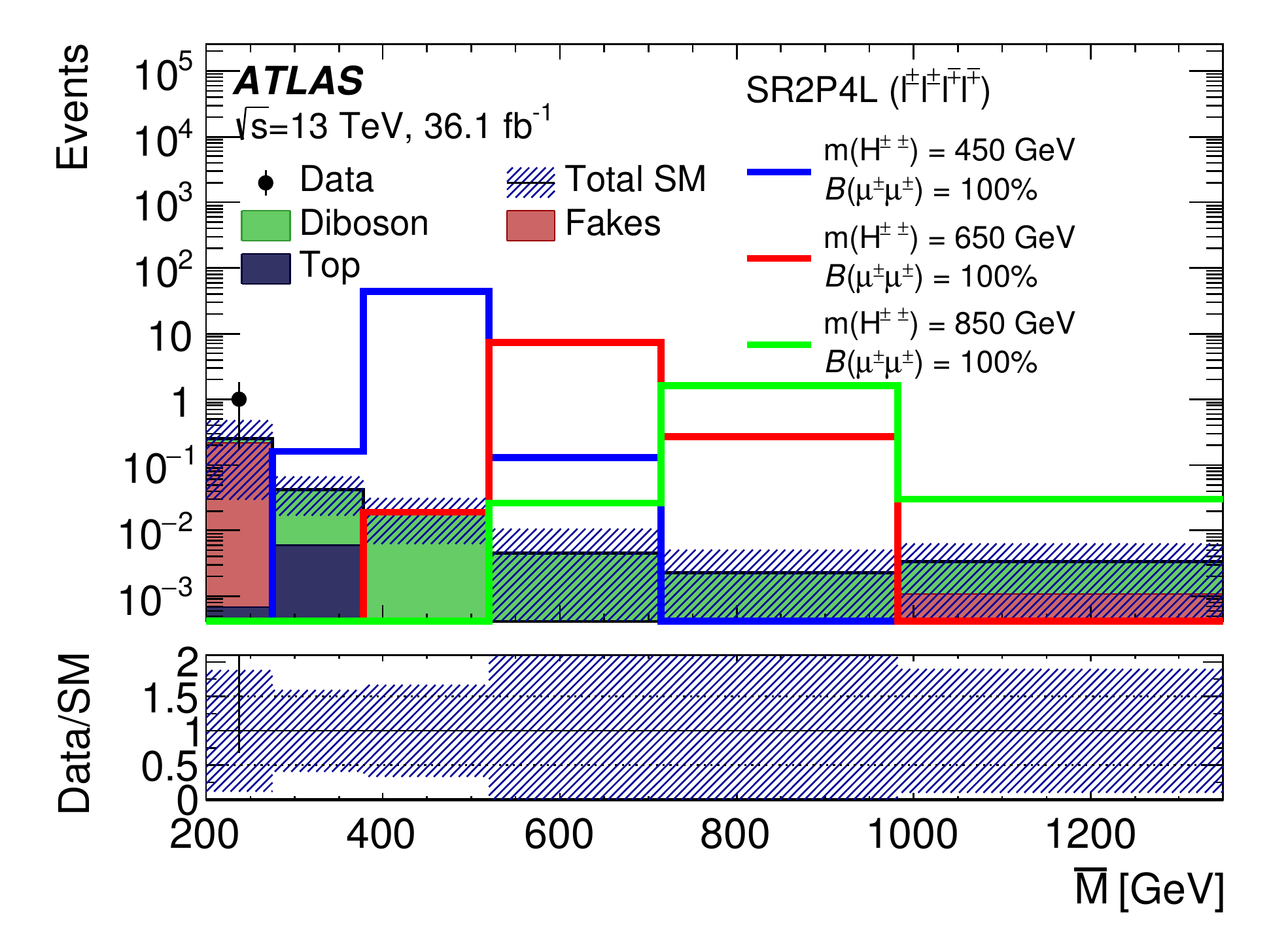}
\label{fig:histo_SR2}
}
\caption{Distributions of $m(\ell^{\pm}\ell^{\pm})$ in representative signal regions, namely 
\protect\subref{fig:histo_SR1P2Lee} the electron--electron two-lepton signal region (SR1P2L),
\protect\subref{fig:histo_SR1P2Lmm} the muon--muon two-lepton signal region (SR1P2L),
\protect\subref{fig:histo_SR1P2Lem} the electron--muon two-lepton signal region (SR1P2L), and
\protect\subref{fig:histo_SR2} the four-lepton signal region (SR2P4L). The hatched bands include 
all systematic uncertainties post-fit with the correlations between various sources 
taken into account. The solid coloured lines correspond to signal samples, normalised using the 
theory cross-section, with the $H^{\pm \pm}$ mass and decay modes marked in the legend.
}
\label{fig:representativeSRs}
\end{figure}

\begin{figure}[t]
\centering
\subfloat[]{
\includegraphics[width=0.45\textwidth]{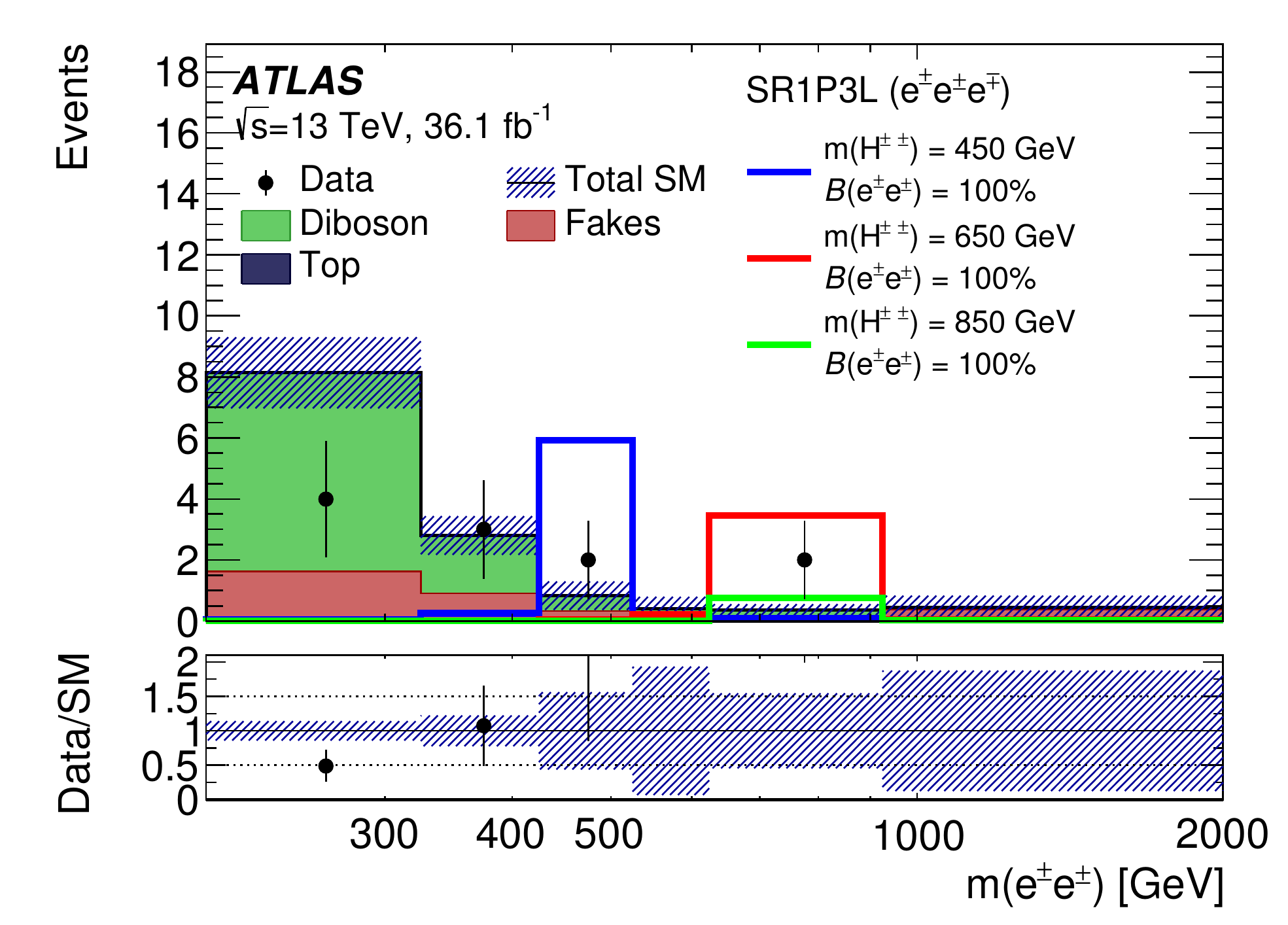}
\label{fig:histo_SR1P3Lee}
}
\subfloat[]{
\includegraphics[width=0.45\textwidth]{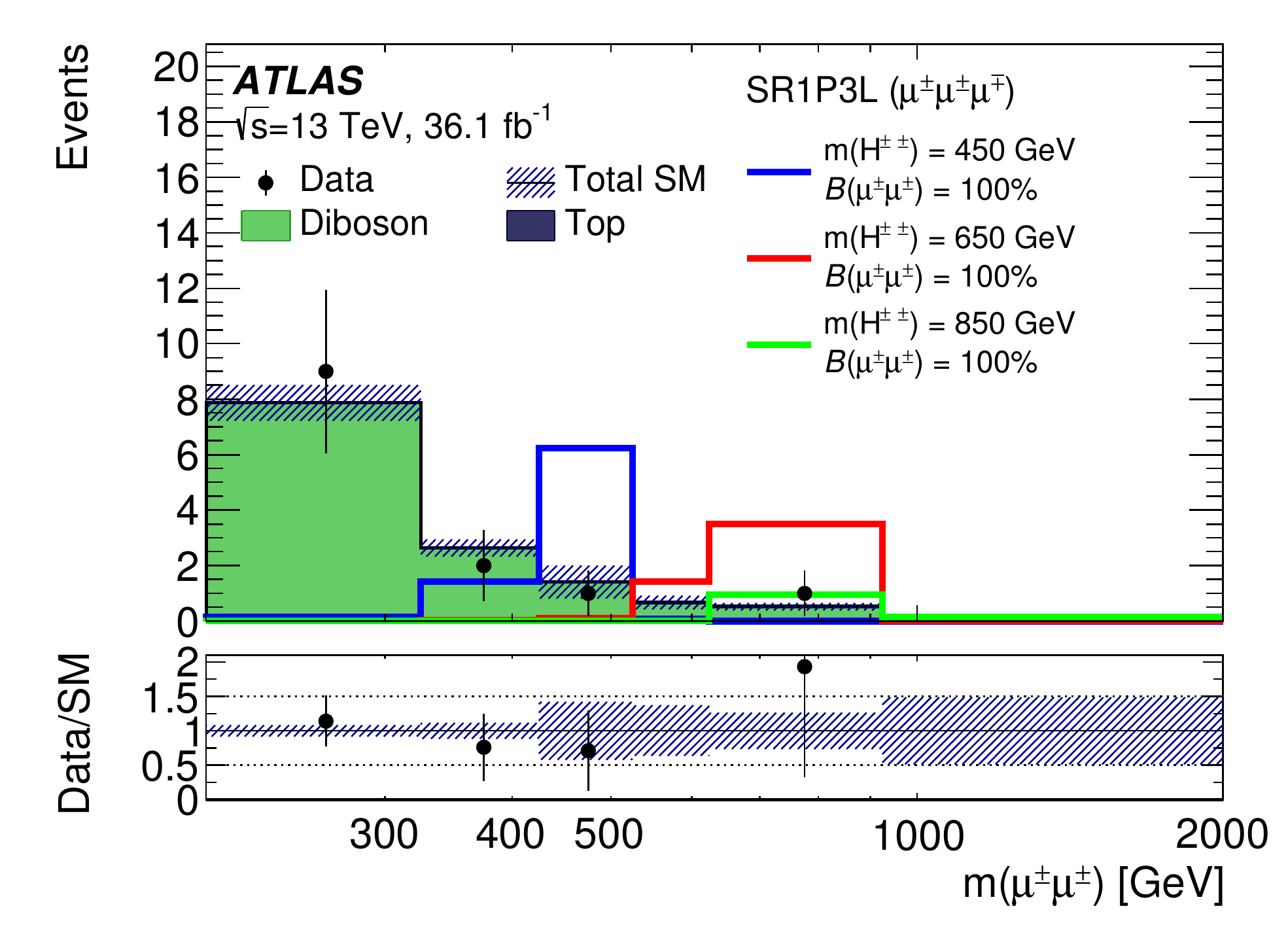}
\label{fig:histo_SR1P3Lmm}
}
\\
\subfloat[]{
\includegraphics[width= 0.45\textwidth]{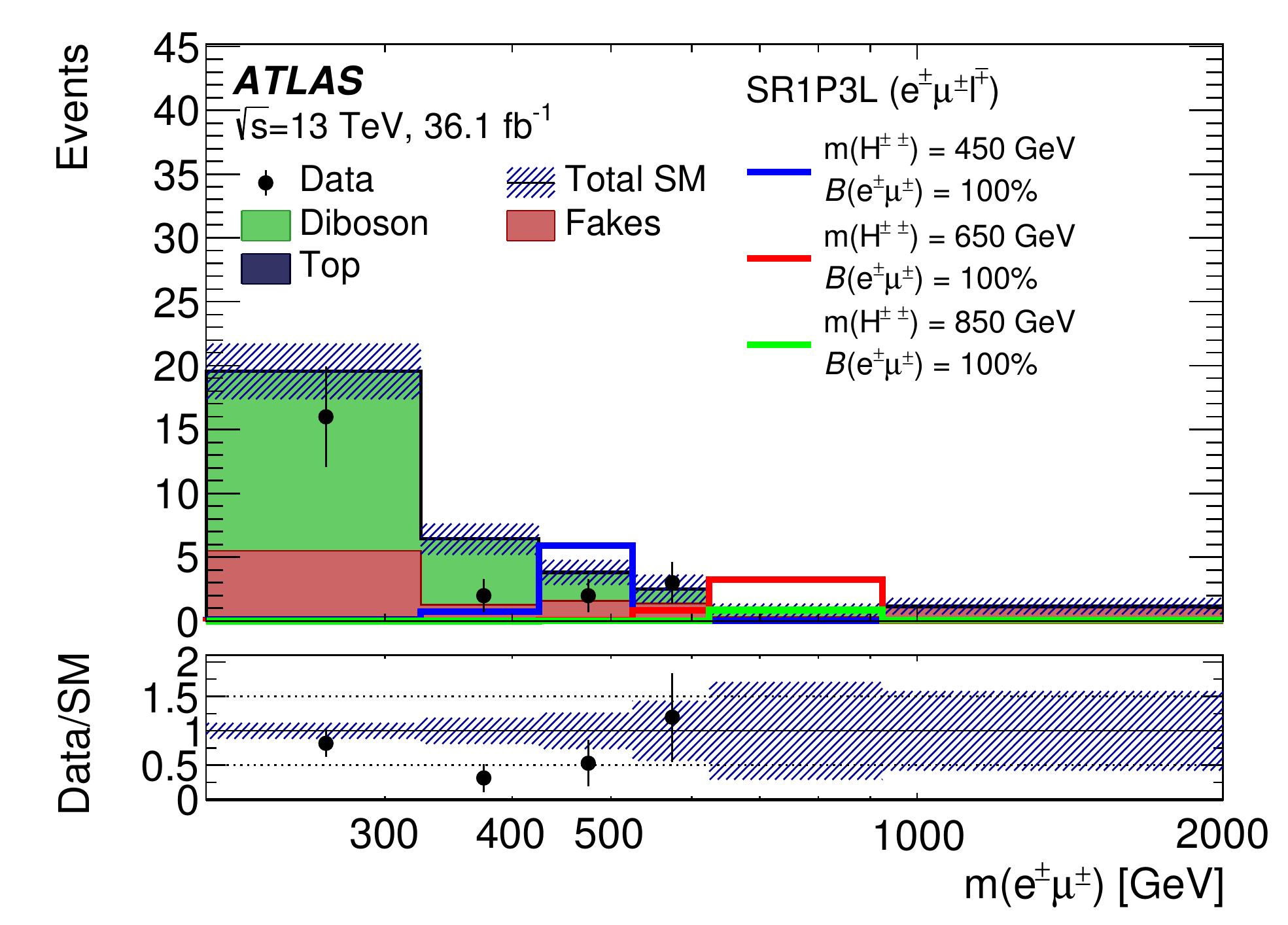}
\label{fig:histo_SR1P3Lem}
}
\subfloat[]{
\includegraphics[width= 0.45\textwidth]{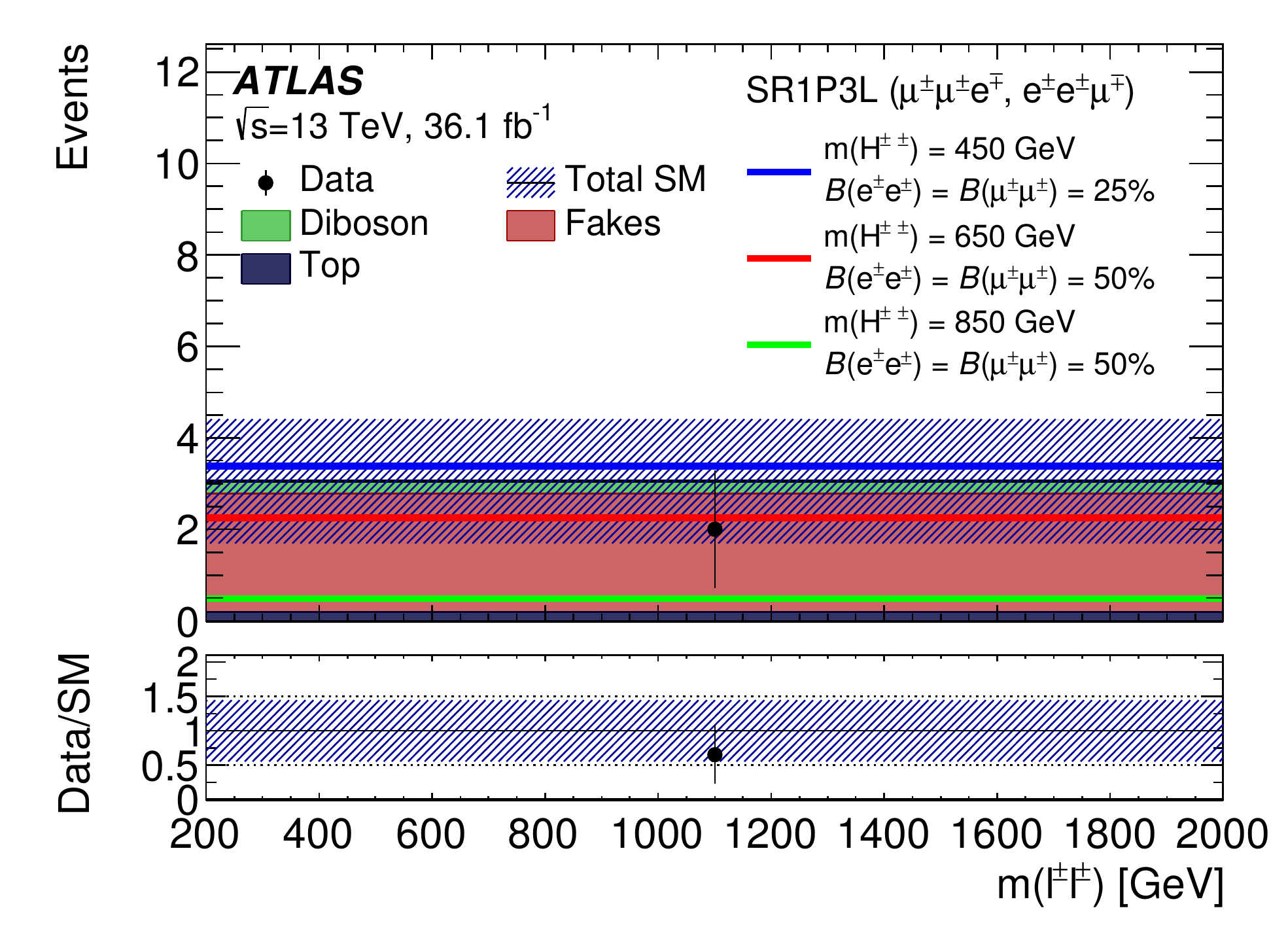}
\label{fig:histo_SR1P3L_mix}
}
\caption{Distributions of $m(\ell^{\pm}\ell^{\pm})$ in three-lepton signal regions, namely 
\protect\subref{fig:histo_SR1P3Lee} the three-electron SR (SR1P3L),
\protect\subref{fig:histo_SR1P3Lmm} the three-muon SR (SR1P3L),
\protect\subref{fig:histo_SR1P3Lem} the SR1P3L with an electron--muon same-charge pair ($e^{\pm}\mu^{\pm}\ell^{\mp}$), and
\protect\subref{fig:histo_SR1P3L_mix} the SR1P3L with a same-flavour same-charge pair ($e^{\pm}e^{\pm}\mu^{\mp}$ or $\mu^{\pm}\mu^{\pm}e^{\mp}$). 
The hatched bands include all systematic uncertainties post-fit with the correlations between various sources 
taken into account. The solid coloured lines correspond to signal samples, normalised using the theory cross-section, 
with the $H^{\pm \pm}$ mass and decay modes marked in the legend.
}
\label{fig:representativeSRs2}
\end{figure}

The observed and expected yields in all control, validation, and signal regions used 
in the analysis are presented in Figure~\ref{fig:allAnalysisRegions} and summarised 
in~\cref{tab:controlRegions,tab:validationRegions,tab:signalRegions}.
No significant excess is observed in any of the signal regions. Correlations between various 
sources of uncertainty are evaluated and used to estimate the total uncertainty in the 
SM background prediction. Two- and four-lepton signal regions are presented in Figure~\ref{fig:representativeSRs} 
and three-lepton signal regions are presented in Figure~\ref{fig:representativeSRs2}. In the four-lepton 
signal region only one data event is observed. It is an $e^{+}\mu^{+}e^{-}\mu^{-}$ event 
with invariant masses of \SI{228}{\GeV} and \SI{207}{\GeV} for the same-charge lepton pairs.

\begin{table}[t]
\caption{The number of predicted background events in control regions after the fit, compared to the data. 
Uncertainties correspond to the total uncertainties in the predicted event yields, and are smaller for the 
total than the sum of the components in quadrature due to correlations between these components. Due to 
rounding the totals can differ from the sums of components. Background processes with a negligible yield 
are marked with the en dash (--).}
\label{tab:controlRegions}
\begin{center}
{\small
\setlength{\tabcolsep}{6pt}
\begin{tabular}{lS[table-format=6.0] @{\,}@{$\pm$}@{\,} S[table-format=3.2]S[table-format=3.2] @{\,}@{$\pm$}@{\,} S[table-format=2.2]S[table-format=4.2] @{\,}@{$\pm$}@{\,} S[table-format=2.2]S[table-format=3.2] @{\,}@{$\pm$}@{\,} S[table-format=2.2]S[table-format=3.2] @{\,}@{$\pm$}@{\,} S[table-format=2.2]}
\toprule
           & \multicolumn{2}{c}{OCCR}            & \multicolumn{2}{c}{DBCR}            & \multicolumn{2}{c}{DBCR}            & \multicolumn{2}{c}{DBCR}            & \multicolumn{2}{c}{4LCR}              \\
           & \multicolumn{2}{c}{$e^{\pm}e^{\mp}$}            & \multicolumn{2}{c}{$e^{\pm}e^{\pm}e^{\mp}$}            & \multicolumn{2}{c}{$e^{\pm}\mu^{\pm}\ell^{\mp}$}            & \multicolumn{2}{c}{$\mu^{\pm}\mu^{\pm}\mu^{\mp}$}            & \multicolumn{2}{c}{$\ell^{\pm}\ell^{\pm}\ell^{\mp}\ell^{\mp}$}              \\
\midrule
Observed events          & \multicolumn{2}{c}{\num{184569}}              & \multicolumn{2}{c}{\num{576}}              & \multicolumn{2}{c}{\num{1025}}              & \multicolumn{2}{c}{\num{797}}              & \multicolumn{2}{c}{\num{140}}                    \\
\midrule
Total background         & 184570 & 430  & 574 & 24  & 1025 & 32  & 797 & 28  & 140 & 12      \\
\midrule
        Drell--Yan         & 169980 & 990  & \multicolumn{2}{c}{--}  & \multicolumn{2}{c}{--}  & \multicolumn{2}{c}{--}  & \multicolumn{2}{c}{--}      \\
        Diboson         & 5060 & 900  & 449 & 28  &909 & 35  & 775 & 29  & 138 & 12      \\
        Fakes         & 2340 & 300  & 123 & 15  & 113 & 14  & 19.9 & 6.5  & 1.31 & 0.16      \\
        Top         & 7200 & 250  & 1.58 & 0.06  & 2.90 & 0.11  & 2.04 & 0.08  & 0.37 & 0.01   \\
 \bottomrule
 \end{tabular}
}
\end{center}
\end{table}

\begin{table}[t]
\caption{The number of predicted background events in two-lepton and four-lepton validation regions (top) and 
three-lepton validation regions (bottom) after the fit, compared to the data. 
Uncertainties correspond to the total uncertainties in the predicted event yields, and are smaller for the 
total than the sum of the components in quadrature due to correlations between these components.
Due to rounding the totals can differ from the sums of components. Background 
processes with a negligible yield are marked with the en dash (--).}
\label{tab:validationRegions}
\begin{center}
{\small
\setlength{\tabcolsep}{6pt}
\begin{tabular}{lS[table-format=4.1] @{\,}@{$\pm$}@{\,} S[table-format=3.1]S[table-format=4.1] @{\,}@{$\pm$}@{\,} S[table-format=2.2]S[table-format=3.2] @{\,}@{$\pm$}@{\,} S[table-format=2.2]S[table-format=1.3] @{\,}@{$\pm$}@{\,} S[table-format=1.3]}
\toprule
           & \multicolumn{2}{c}{SCVR}            & \multicolumn{2}{c}{SCVR}            & \multicolumn{2}{c}{SCVR}            & \multicolumn{2}{c}{4LVR}              \\
           & \multicolumn{2}{c}{$e^{\pm}e^{\pm}$}            & \multicolumn{2}{c}{$e^{\pm}\mu^{\pm}$}            & \multicolumn{2}{c}{$\mu^{\pm}\mu^{\pm}$}            & \multicolumn{2}{c}{$\ell^{\pm}\ell^{\pm}\ell^{\mp}\ell^{\mp}$}              \\
\midrule
Observed events          & \multicolumn{2}{c}{\num{3237}}              & \multicolumn{2}{c}{\num{1162}}              & \multicolumn{2}{c}{\num{1006}}              & \multicolumn{2}{c}{\num{3}}                    \\
\midrule
Total background         & 3330 & 210  & 1119 & 51  & 975 & 50  & 4.62 & 0.40      \\
\midrule
        Drell--Yan         & 2300 & 190  & \multicolumn{2}{c}{--}  & \multicolumn{2}{c}{--}  & \multicolumn{2}{c}{--}      \\
        Diboson         & 319 & 25  & 547 & 23  & 719 & 30  & 4.59 & 0.4     \\
        Fakes         & 640 & 65  & 502 & 54  & 249 & 47  & \multicolumn{2}{c}{--}      \\
        Top         & 71.5 & 6.8  & 70.5 & 2.6  & 6.93 & 0.27  & 0.033 & 0.001      \\
\bottomrule
\end{tabular}
\\[5mm]
\begin{tabular}{lS[table-format=3.2] @{\,}@{$\pm$}@{\,} S[table-format=1.2]S[table-format=3.2] @{\,}@{$\pm$}@{\,} S[table-format=1.2]S[table-format=3.2] @{\,}@{$\pm$}@{\,} S[table-format=1.2]S[table-format=3.2] @{\,}@{$\pm$}@{\,} S[table-format=1.2]}
\toprule
           & \multicolumn{2}{c}{3LVR}            & \multicolumn{2}{c}{3LVR}            & \multicolumn{2}{c}{3LVR}            & \multicolumn{2}{c}{3LVR}              \\
           & \multicolumn{2}{c}{$e^{\pm}e^{\pm}e^{\mp}$}            & \multicolumn{2}{c}{$e^{\pm}\mu^{\pm}\ell^{\mp}$}            & \multicolumn{2}{c}{$\mu^{\pm}\mu^{\pm}\mu^{\mp}$}            & \multicolumn{2}{c}{$\mu^{\pm}\mu^{\pm}e^{\mp}, e^{\pm}e^{\pm}\mu^{\mp}$}              \\
\midrule
Observed events          & \multicolumn{2}{c}{\num{108}}              & \multicolumn{2}{c}{\num{180}}              & \multicolumn{2}{c}{\num{126}}              & \multicolumn{2}{c}{\num{16}}                    \\
\midrule
Total background         & 88.1 & 5.8  & 192.9 & 9.9  & 107.0 & 5.1  & 27.0 & 3.9      \\
\midrule
        Diboson         & 64.4 & 5.8  & 147.3 & 9.0  & 100.9 & 5.0  & 4.72 & 0.79      \\
        Fakes         & 23.3 & 3.0  & 43.9 & 4.9  & 5.3 & 1.2  & 21.3 & 3.4      \\
        Top         & 0.50 & 0.03  & 1.73 & 0.09  & 0.82 & 0.05  & 1.01 & 0.15      \\
 \bottomrule
 \end{tabular}
}
\end{center}
\end{table}

\begin{table}[t]
\caption{The number of predicted background events in two-lepton and four-lepton signal regions (top)
and three-lepton signal regions (bottom) after the fit, compared to the data. 
Uncertainties correspond to the total uncertainties in the predicted event yields, and are smaller for the 
total than the sum of the components in quadrature due to correlations between these components.
Due to rounding the totals can differ from the sums of components. Background processes with a negligible 
yield are marked with the en dash (--).}
\label{tab:signalRegions}
\begin{center}
{\small
\setlength{\tabcolsep}{6pt}
\begin{tabular}{lS[table-format=3.2] @{\,}@{$\pm$}@{\,} S[table-format=2.2]S[table-format=3.2] @{\,}@{$\pm$}@{\,} S[table-format=1.2]S[table-format=2.2] @{\,}@{$\pm$}@{\,} S[table-format=1.2]S[table-format=1.3] @{\,}@{$\pm$}@{\,} S[table-format=1.3]}
\toprule
           & \multicolumn{2}{c}{SR1P2L}            & \multicolumn{2}{c}{SR1P2L}            & \multicolumn{2}{c}{SR1P2L}            & \multicolumn{2}{c}{SR2P4L}              \\
           & \multicolumn{2}{c}{$e^{\pm}e^{\pm}$}            & \multicolumn{2}{c}{$e^{\pm}\mu^{\pm}$}            & \multicolumn{2}{c}{$\mu^{\pm}\mu^{\pm}$}      & \multicolumn{2}{c}{$\ell^{\pm}\ell^{\pm}\ell^{\mp}\ell^{\mp}$}              \\
\midrule
Observed events          & \multicolumn{2}{c}{\num{132}}              & \multicolumn{2}{c}{\num{106}}              & \multicolumn{2}{c}{\num{26}}              & \multicolumn{2}{c}{\num{1}}                    \\
\midrule
Total background         & 160 & 14  & 97.1 & 7.7  & 22.6 & 2.0  & 0.33 & 0.23      \\
\midrule
        Drell--Yan         & 70 & 10  & \multicolumn{2}{c}{--}  & \multicolumn{2}{c}{--}  & \multicolumn{2}{c}{--}      \\
        Diboson         & 30.5 & 3.0  &  40.4 & 4.5  & 20.3 & 1.8  & 0.11 & 0.06      \\
        Fakes         & 52.2 & 5.0  & 53.1 & 5.8  & 1.94 & 0.47  & 0.22 & 0.19      \\
        Top         & 7.20 & 0.97  & 3.62 & 0.53  & 0.42 & 0.03  &0.007 & 0.002      \\
\bottomrule
\end{tabular}
\\[5mm]
\begin{tabular}{lS[table-format=2.2] @{\,}@{$\pm$}@{\,} S[table-format=1.2]S[table-format=2.2] @{\,}@{$\pm$}@{\,} S[table-format=1.2]S[table-format=2.2] @{\,}@{$\pm$}@{\,} S[table-format=1.2]S[table-format=3.2] @{\,}@{$\pm$}@{\,} S[table-format=1.2]}
\toprule
           & \multicolumn{2}{c}{SR1P3L}            & \multicolumn{2}{c}{SR1P3L}            & \multicolumn{2}{c}{SR1P3L}            & \multicolumn{2}{c}{SR1P3L}              \\
           & \multicolumn{2}{c}{$e^{\pm}e^{\pm}e^{\mp}$}            & \multicolumn{2}{c}{$e^{\pm}\mu^{\pm}\ell^{\mp}$}            & \multicolumn{2}{c}{$\mu^{\pm}\mu^{\pm}\mu^{\mp}$}            & \multicolumn{2}{c}{$\mu^{\pm}\mu^{\pm}e^{\mp}, e^{\pm}e^{\pm}\mu^{\mp}$}              \\
\midrule
Observed events          & \multicolumn{2}{c}{\num{11}}              & \multicolumn{2}{c}{\num{23}}              & \multicolumn{2}{c}{\num{13}}              & \multicolumn{2}{c}{\num{2}}                    \\
\midrule
Total background         & 13.0 & 1.6  & 34.2 & 3.6  & 13.2 & 1.3  & 3.1 & 1.4      \\
\midrule
        Diboson         & 9.5 & 1.3  & 23.1 & 2.9  & 13.1 & 1.3  & 0.27 & 0.14      \\
        Fakes         & 3.3 & 0.67  & 10.7 & 1.7  & \multicolumn{2}{c}{--}  & 2.6 & 1.2      \\
        Top         & 0.14 & 0.02  & 0.45 & 0.04  & 0.12 & 0.01  & 0.19 & 0.08      \\
 \bottomrule
 \end{tabular}
}
\end{center}
\end{table}

\begin{figure}[htpb]
\centering
\subfloat[]{
\includegraphics[width=0.45\textwidth]{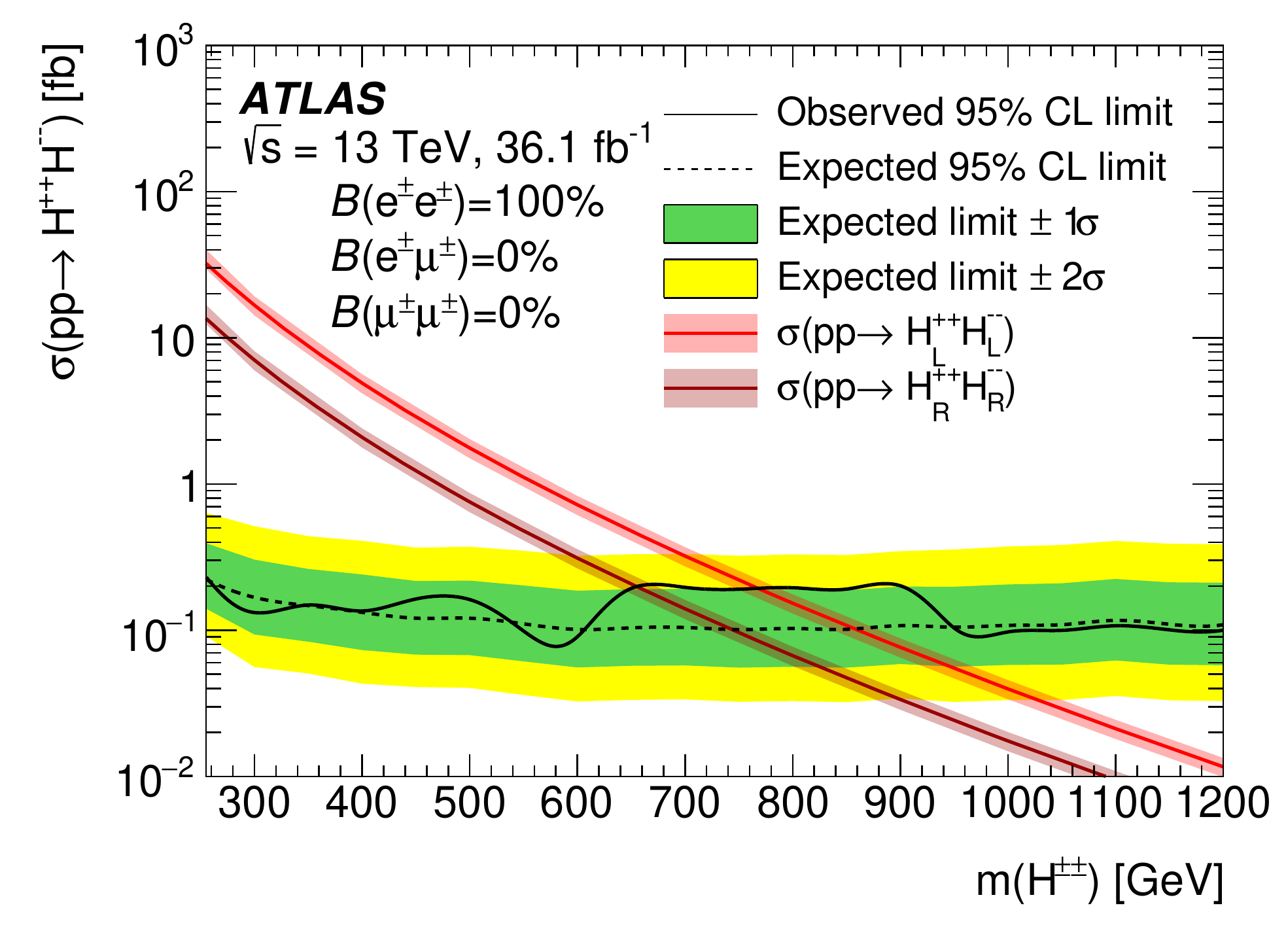}
\label{fig:ee100em0mm0}
}
\subfloat[]{
\includegraphics[width=0.45\textwidth]{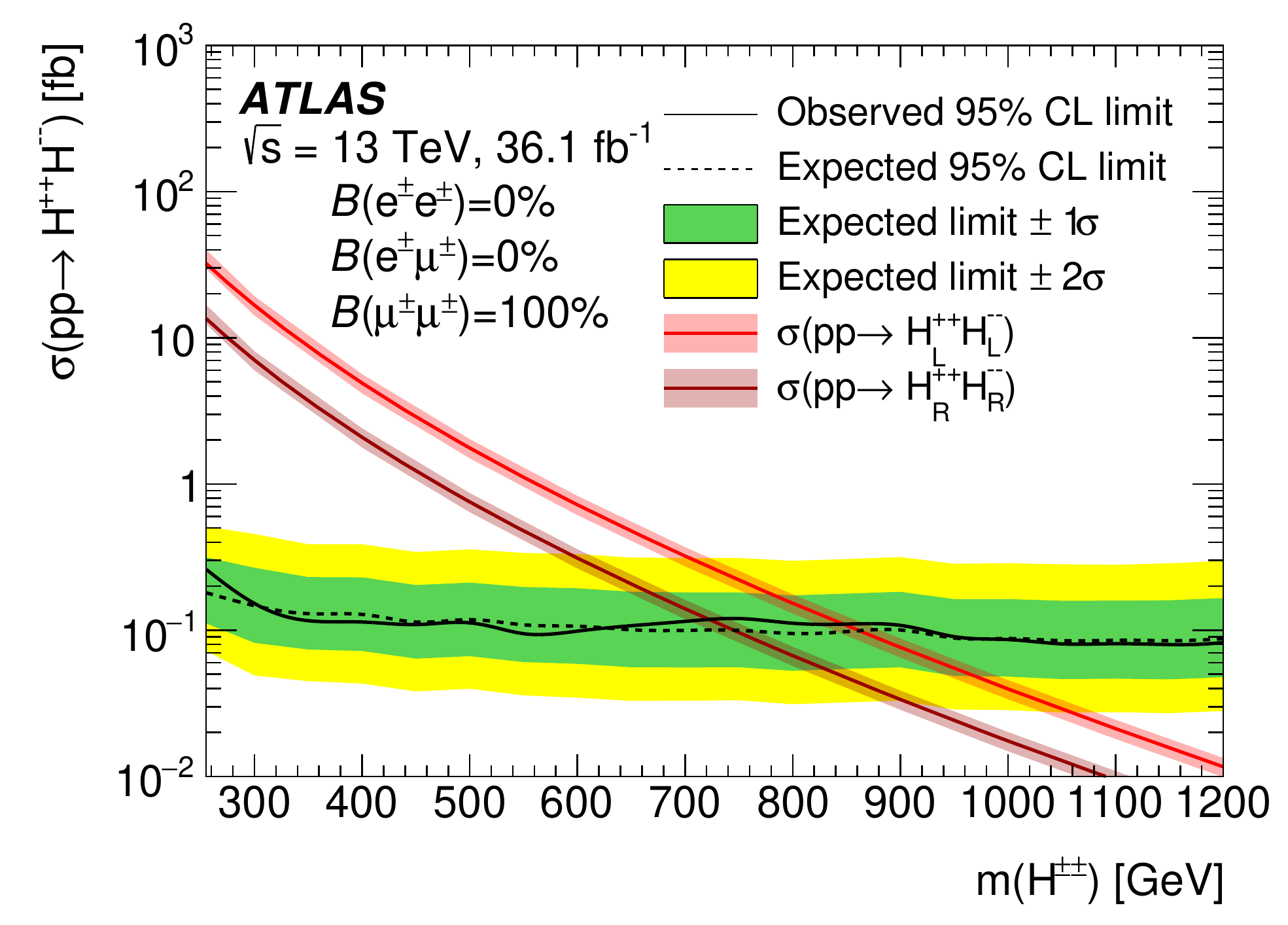}
\label{fig:ee0em0mm100}
}
\\
\subfloat[]{
\includegraphics[width= 0.45\textwidth]{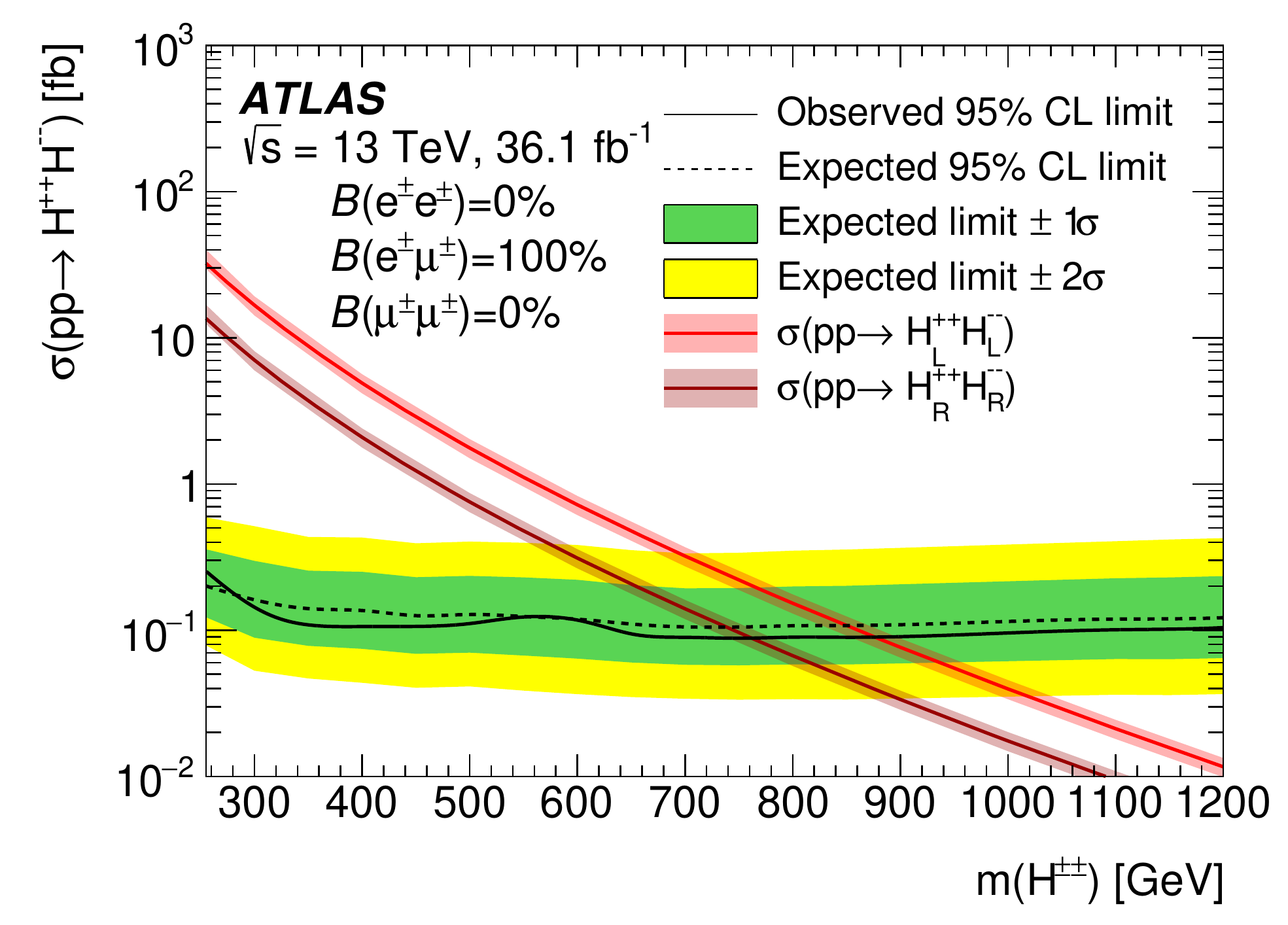}
\label{fig:ee0em100mm0}
}
\subfloat[]{
\includegraphics[width= 0.45\textwidth]{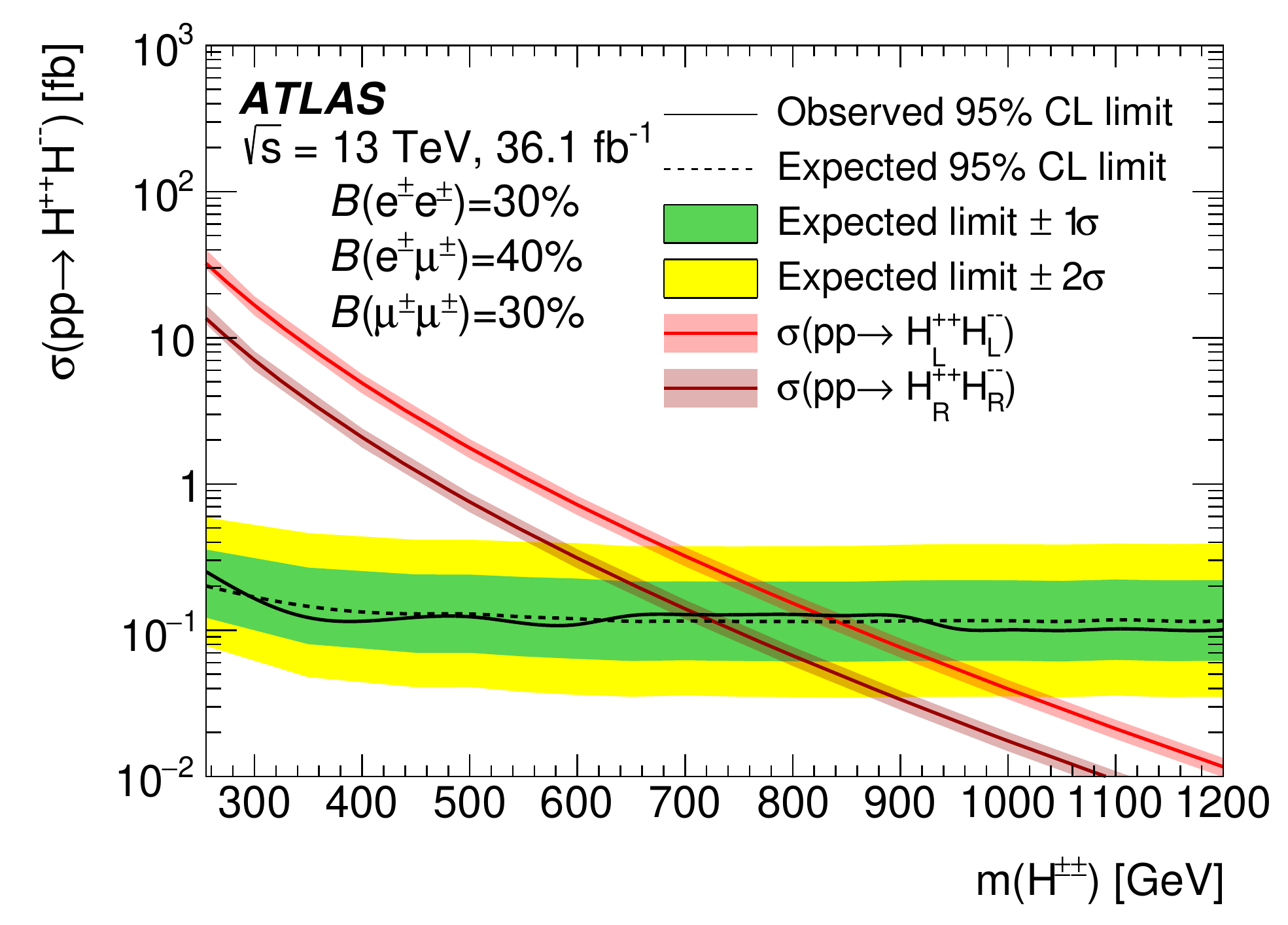}
\label{fig:ee30em40mm30}
}
\caption{Upper limit on the cross-section for $pp\rightarrow H^{++}H^{--}$ for several branching ratio values presented in the form $B(ee)/B(e\mu)/B(\mu\mu)$:
\protect\subref{fig:ee100em0mm0} $100\%/0\%/0\%$,
\protect\subref{fig:ee0em0mm100} $0\%/0\%/100\%$,
\protect\subref{fig:ee0em100mm0} $0\%/100\%/0\%$, and
\protect\subref{fig:ee30em40mm30} $30\%/40\%/30\%$.
The theoretical uncertainty in the cross-section for $pp\rightarrow H^{++}H^{--}$ is presented with the shaded band around the
central value.
}
\label{fig:exclusionCombinedMain}
\end{figure}

The likelihood fit to the two-, three-, and four-lepton control and signal regions was designed
to fully exploit the pair production of the $H^{\pm\pm}$ boson with its boosted topology
and lepton multiplicity. For $B(H^{\pm \pm}\rightarrow \ell^{\pm}\ell^{\pm})=100\%$ the production
cross-section is excluded down to \SI{0.1}{\fb}, corresponding to 3--4 signal events, which is the 
theoretical limit of a $95\%$ CL exclusion. Some representative cross-section upper limits as a function 
of the $H^{\pm \pm}$ boson mass are presented in Figure~\ref{fig:exclusionCombinedMain}, for different 
combinations of the branching ratios for decay into light-lepton pairs.

\begin{figure}[htpb]
\centering
\subfloat[]{
\includegraphics[width=0.47\textwidth]{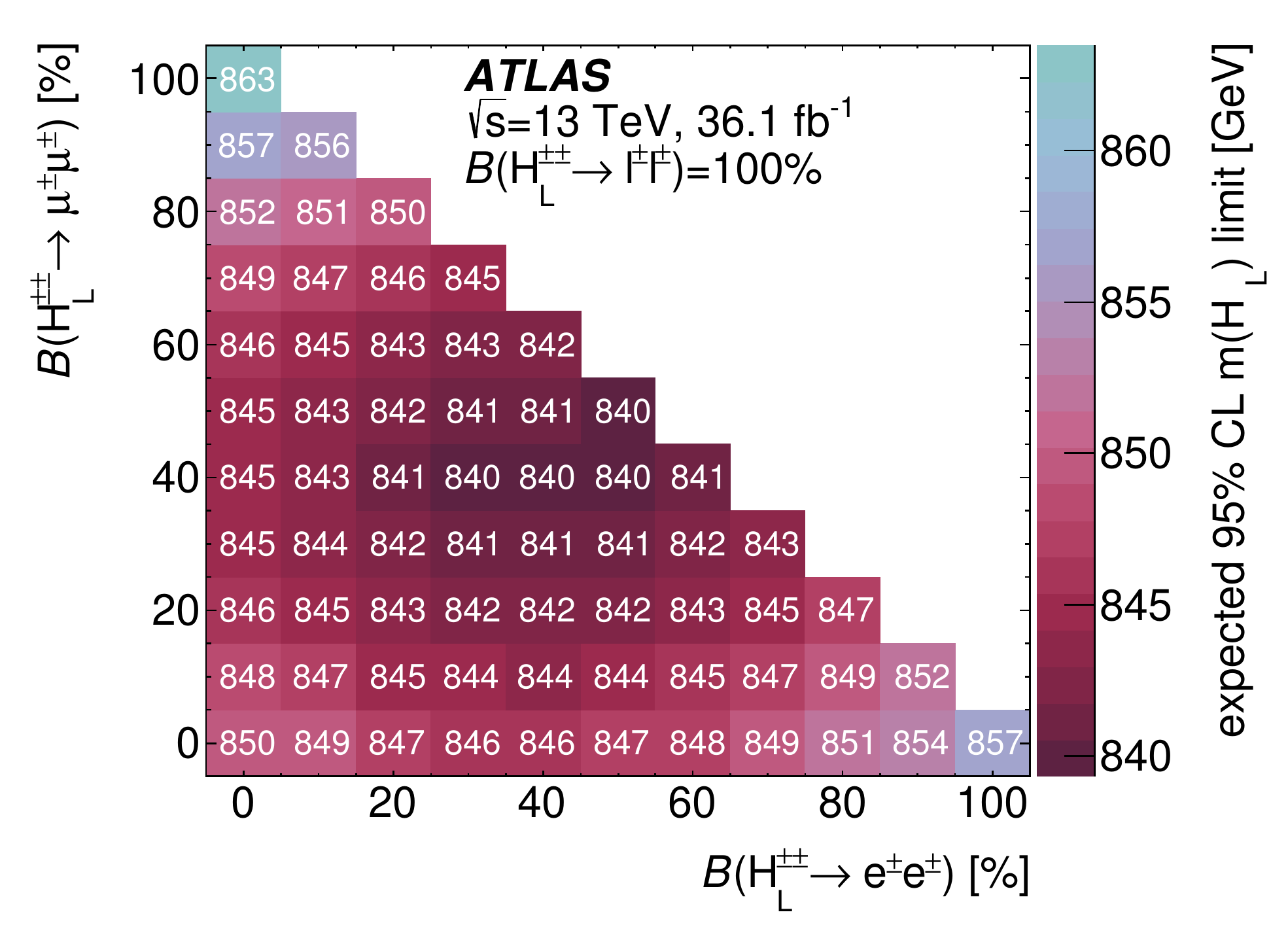}
\label{fig:combinedLimit2D_HL}
}
\subfloat[]{
\includegraphics[width=0.47\textwidth]{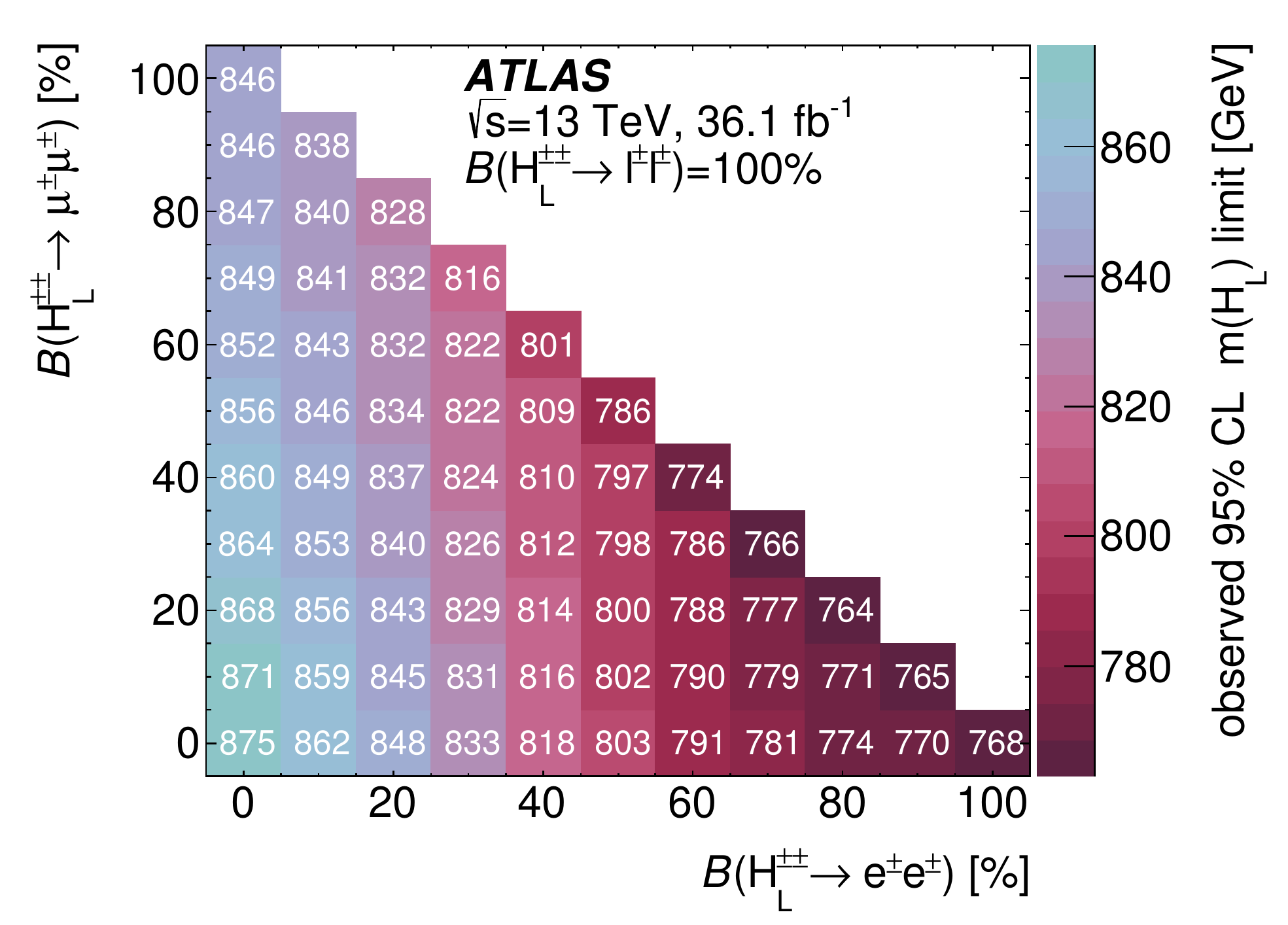}
\label{fig:combinedLimit2D_HLobs_BRsum100}
}
\caption{The \protect\subref{fig:combinedLimit2D_HL} expected and \protect\subref{fig:combinedLimit2D_HLobs_BRsum100} observed 
lower limits on the $H_{L}^{\pm\pm}$ boson mass for all branching ratio combinations that sum to $100\%$.}
\label{fig:expectedLimsCombinedFitHL}
\end{figure}

\begin{figure}[htpb]
\centering
\subfloat[]{
\includegraphics[width=0.47\textwidth]{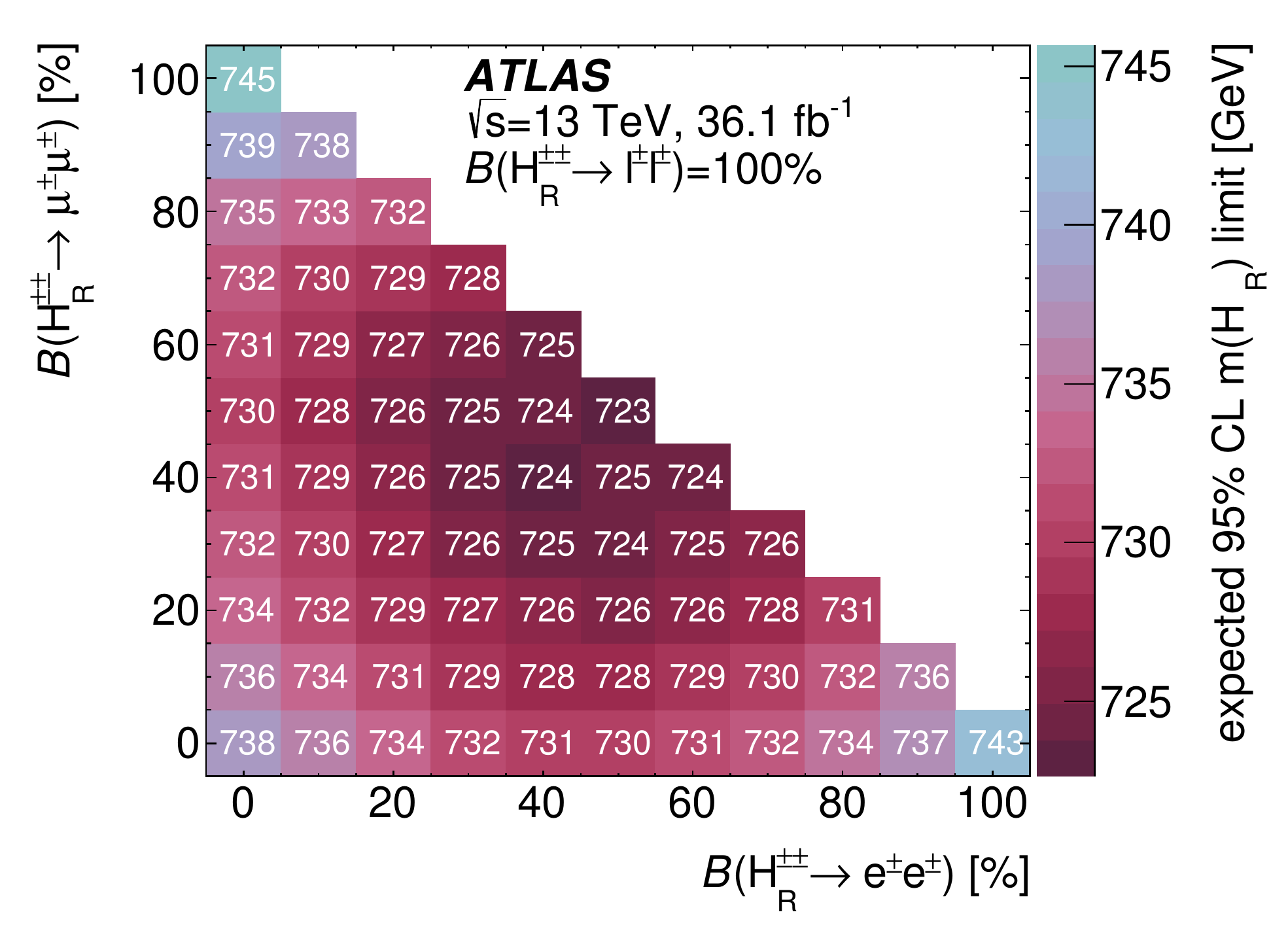}
\label{fig:combinedLimit2D_HR_BRsum100}
}
\subfloat[]{
\includegraphics[width=0.47\textwidth]{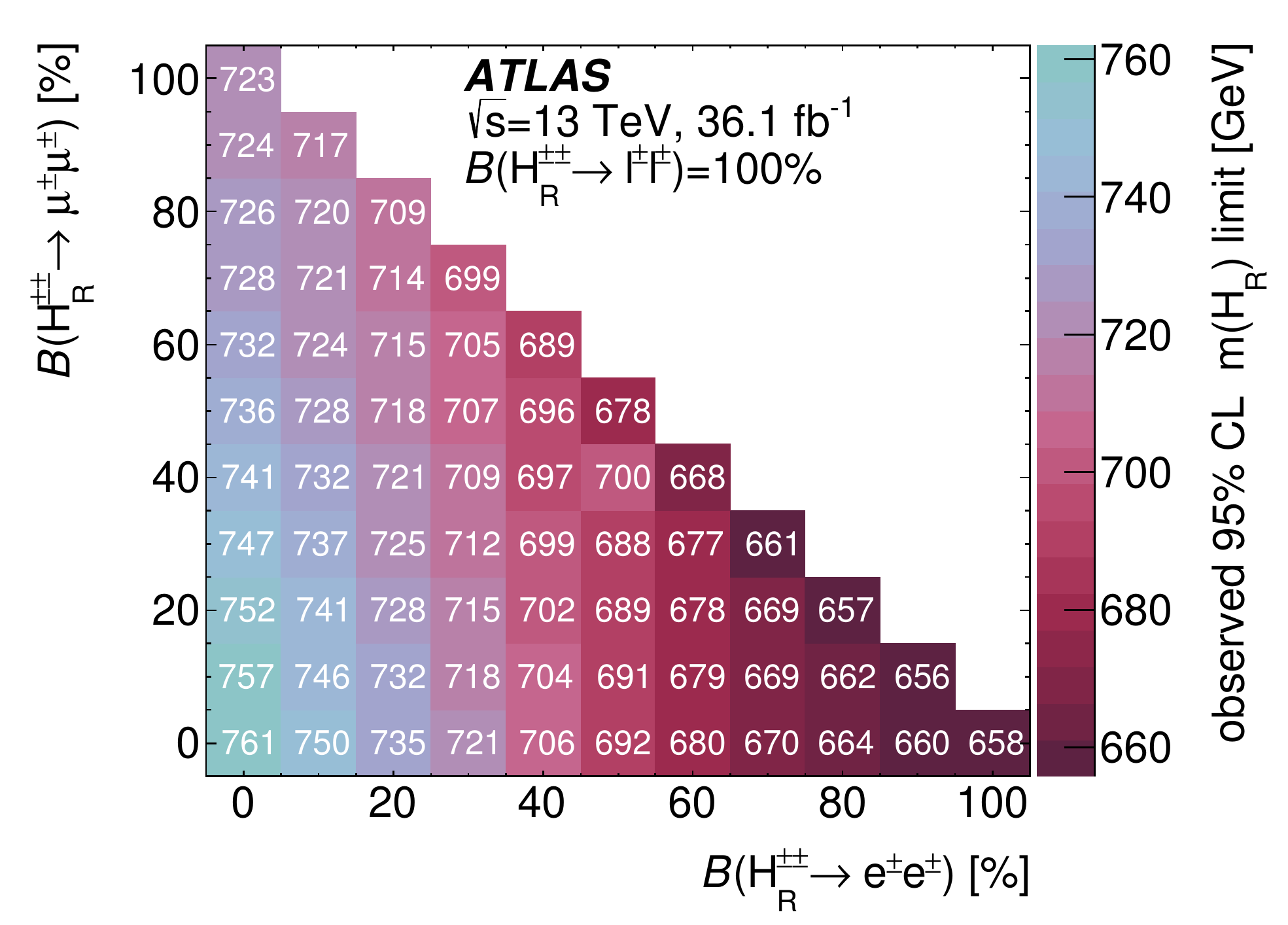}
\label{fig:combinedLimit2D_HRobs_BRsum100}
}
\caption{The \protect\subref{fig:combinedLimit2D_HR_BRsum100} expected and \protect\subref{fig:combinedLimit2D_HRobs_BRsum100} observed
lower limits on the $H_{R}^{\pm\pm}$ boson mass for all branching ratio combinations that sum to $100\%$.}
\label{fig:expectedLimsCombinedFitHR}
\end{figure}

\begin{figure}[htpb]
\centering
\subfloat[]{
\includegraphics[width=0.45\textwidth]{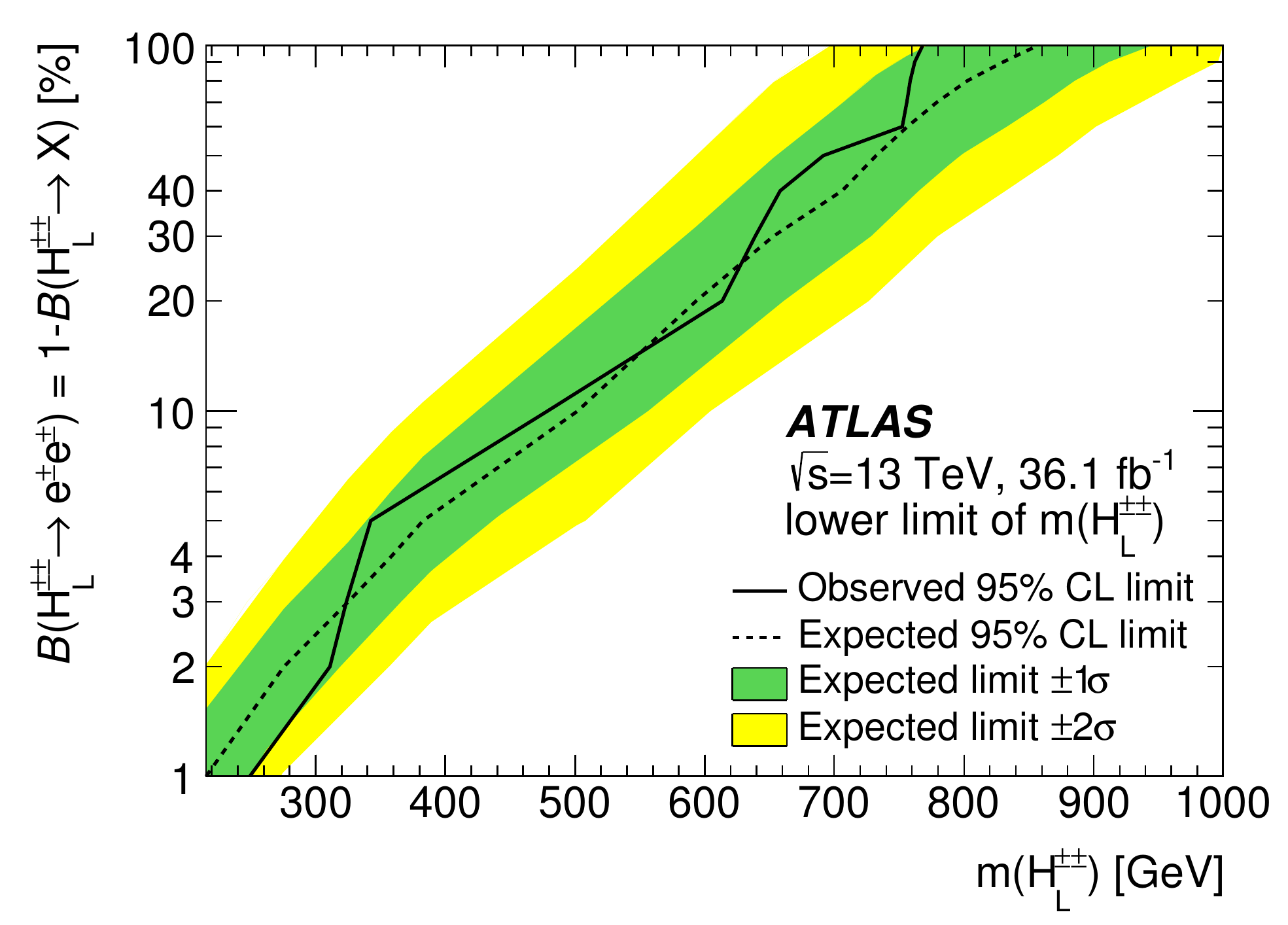}
\label{fig:HL_exp_1Dee}
}
\subfloat[]{
\includegraphics[width=0.45\textwidth]{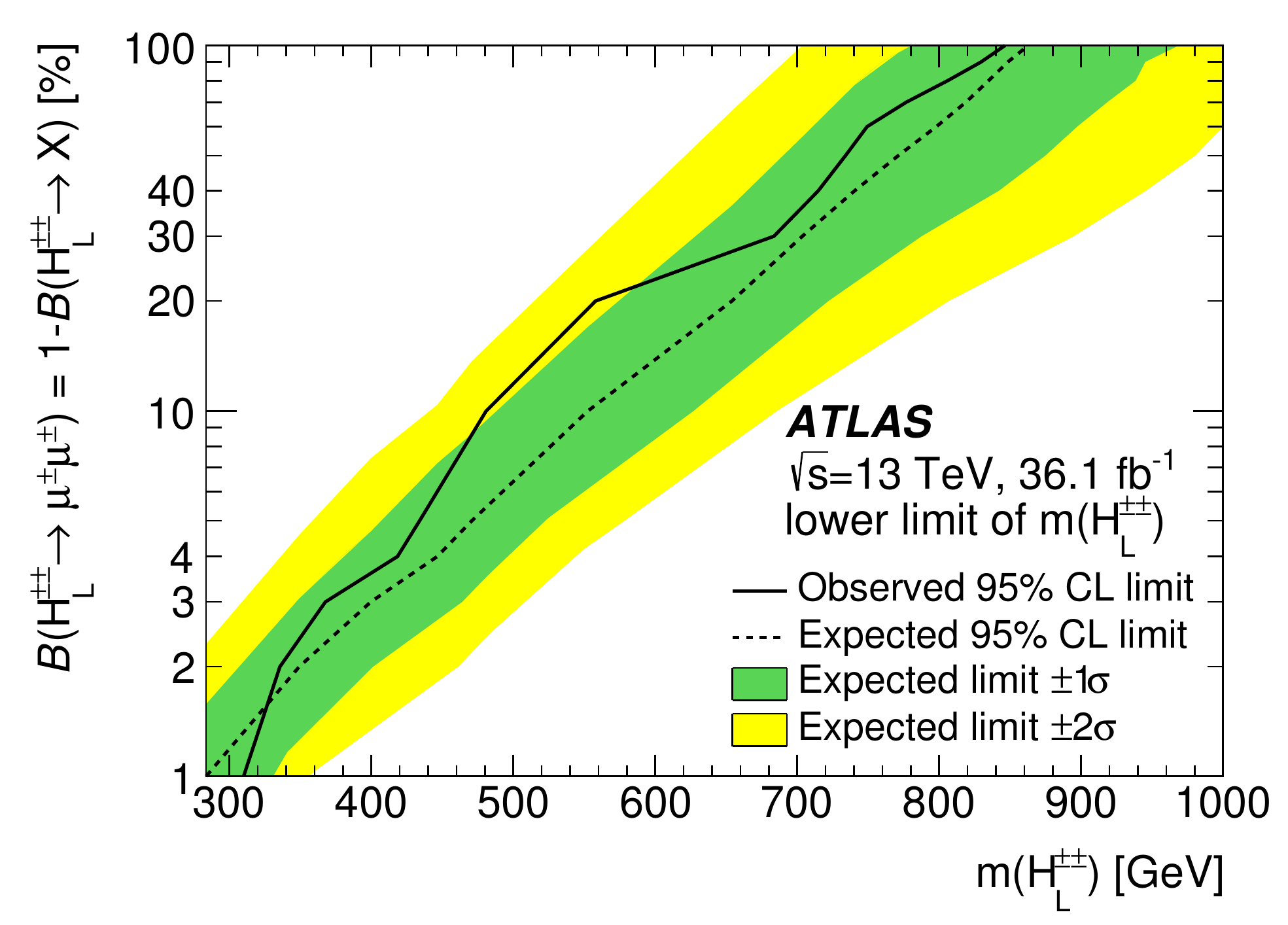}
\label{fig:HL_exp_1Dmm}
}
\\
\subfloat[]{
\includegraphics[width= 0.45\textwidth]{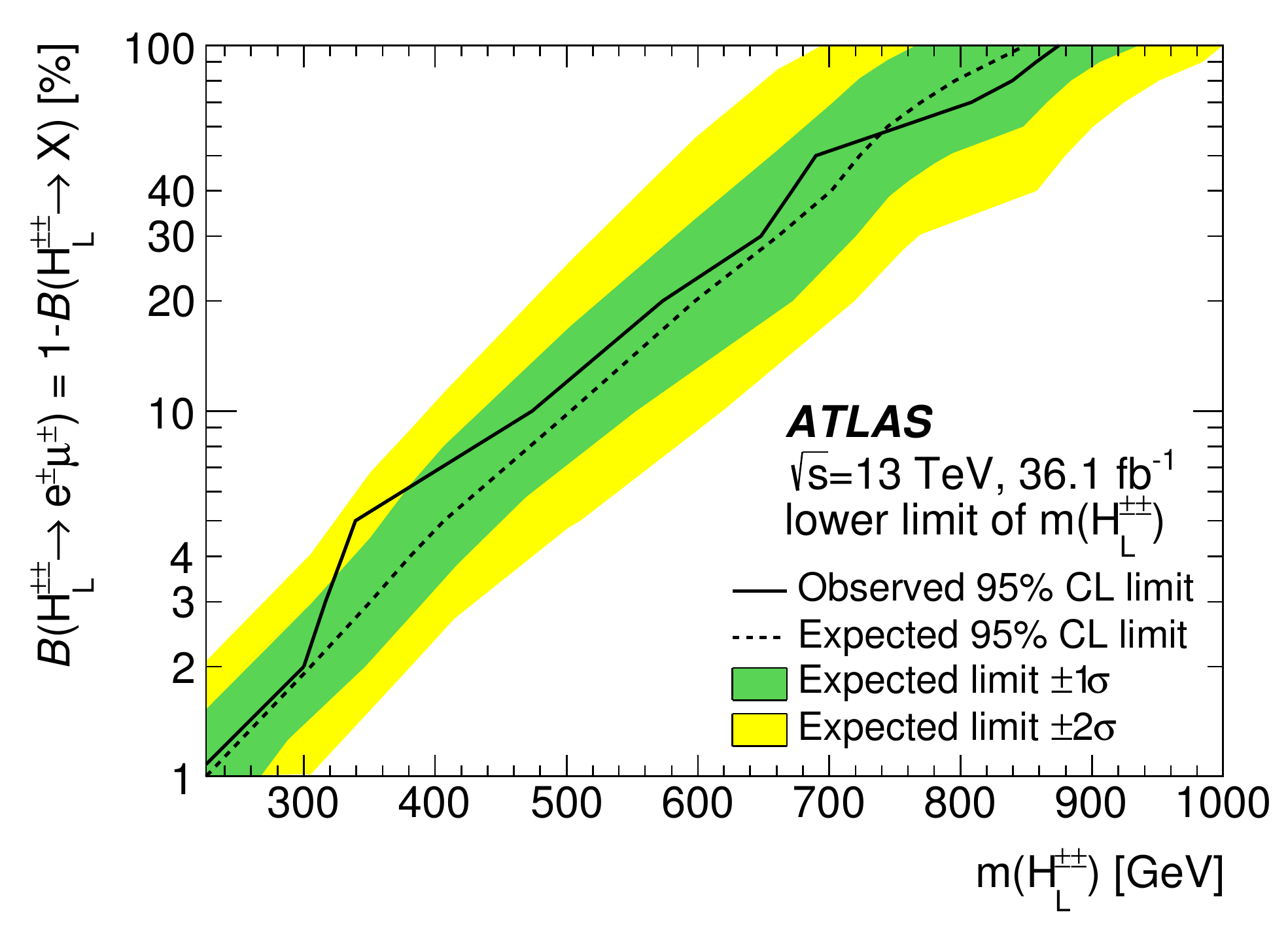}
\label{fig:HL_exp_1Dem}
}
\subfloat[]{
\includegraphics[width=0.45\textwidth]{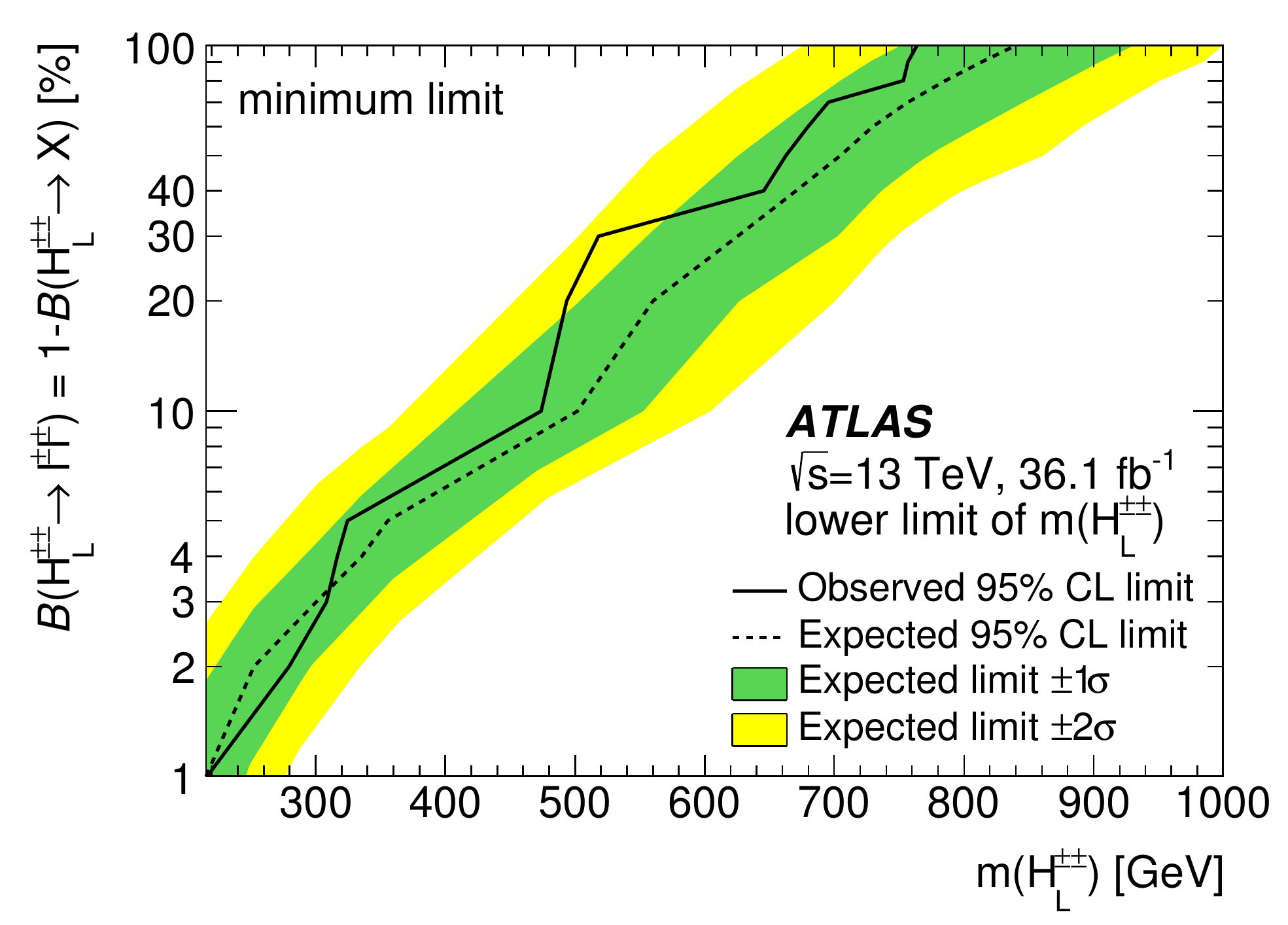}
\label{fig:HL_exp_1D_min}
}
\caption{Lower limit on the $H_{L}^{\pm \pm}$ boson mass as a function of the branching ratio 
$B(H_{L}^{\pm \pm}\rightarrow \ell^{\pm}\ell^{\pm})$. Several cases are presented: 
\protect\subref{fig:HL_exp_1Dee} $H_{L}^{\pm \pm}$ decays only into electrons and "$X$",
\protect\subref{fig:HL_exp_1Dmm} $H_{L}^{\pm \pm}$ decays only into muons and "$X$", and
\protect\subref{fig:HL_exp_1Dem} $H_{L}^{\pm \pm}$ decays only into electron--muon pairs and "$X$", with "$X$" not entering any of the signal regions.
Plot \protect\subref{fig:HL_exp_1D_min} shows the minimum observed and expected limit as a function of $B(H_{L}^{\pm \pm}\rightarrow \ell^{\pm}\ell^{\pm})$.
}
\label{fig:combinedFitAllBrHL}
\end{figure}

\begin{figure}[htpb]
\centering
\subfloat[]{
\includegraphics[width=0.45\textwidth]{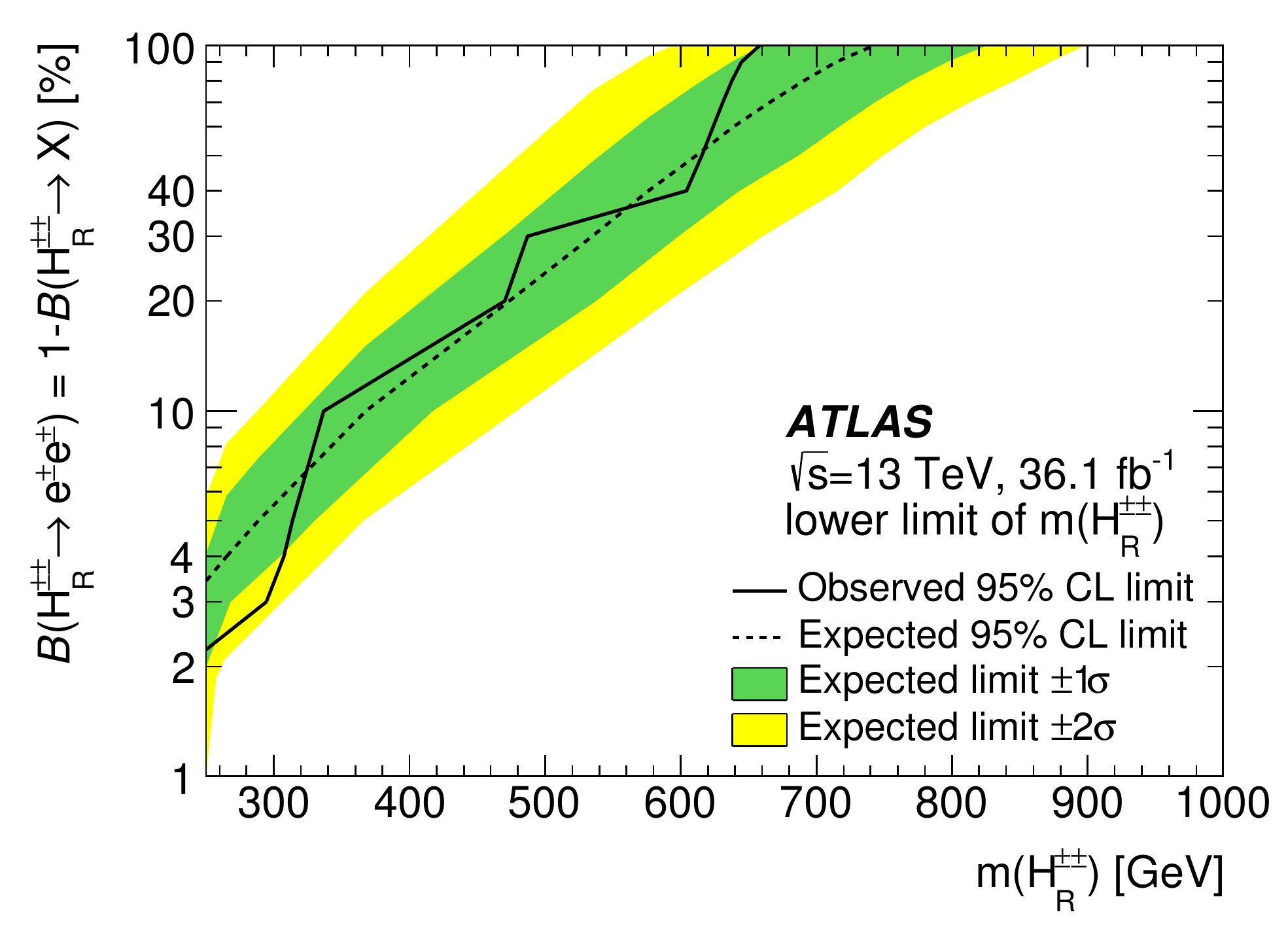}
\label{fig:HR_exp_1Dee}
}
\subfloat[]{
\includegraphics[width=0.45\textwidth]{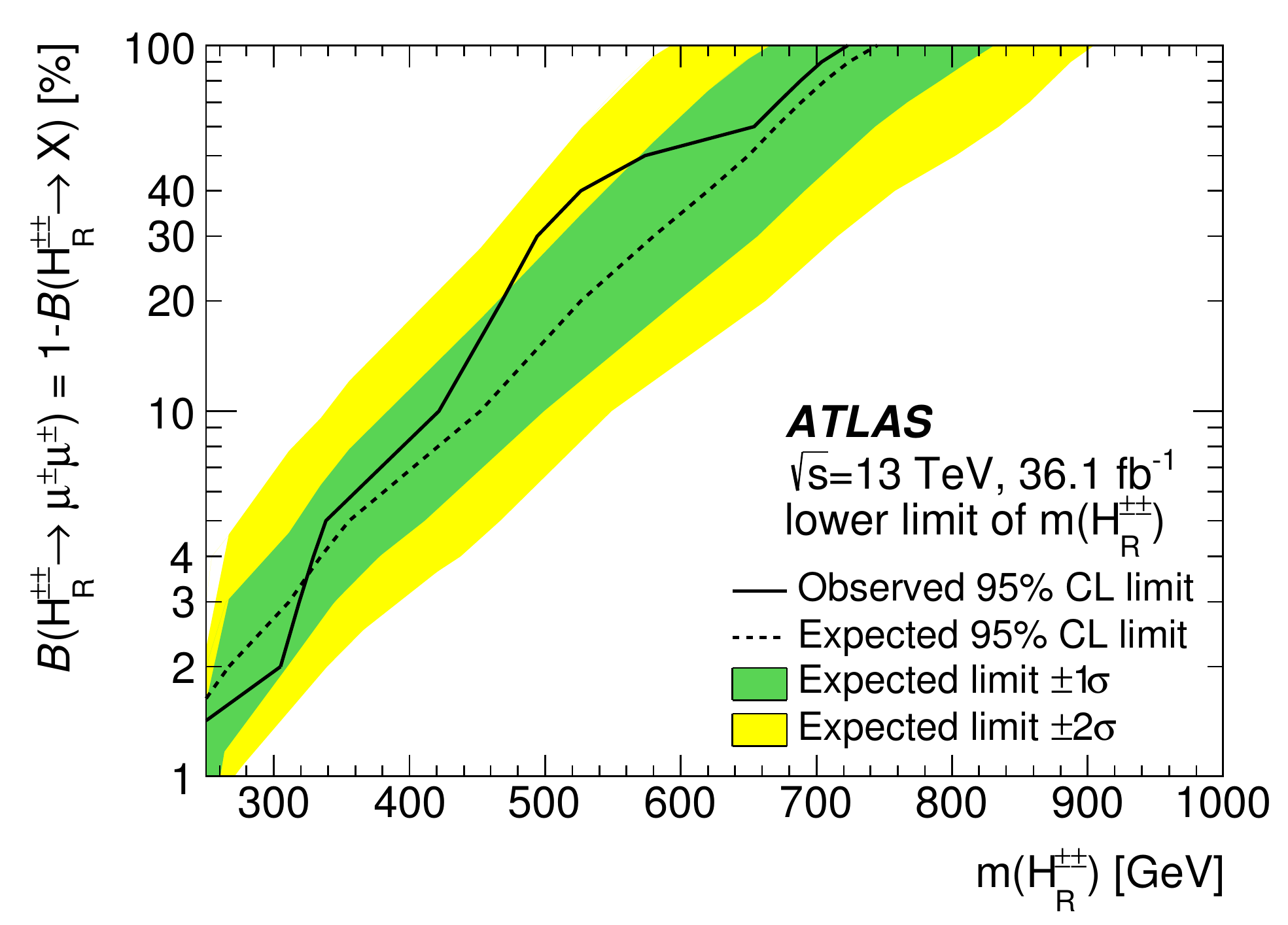}
\label{fig:HR_exp_1Dmm}
}
\\
\subfloat[]{
\includegraphics[width= 0.45\textwidth]{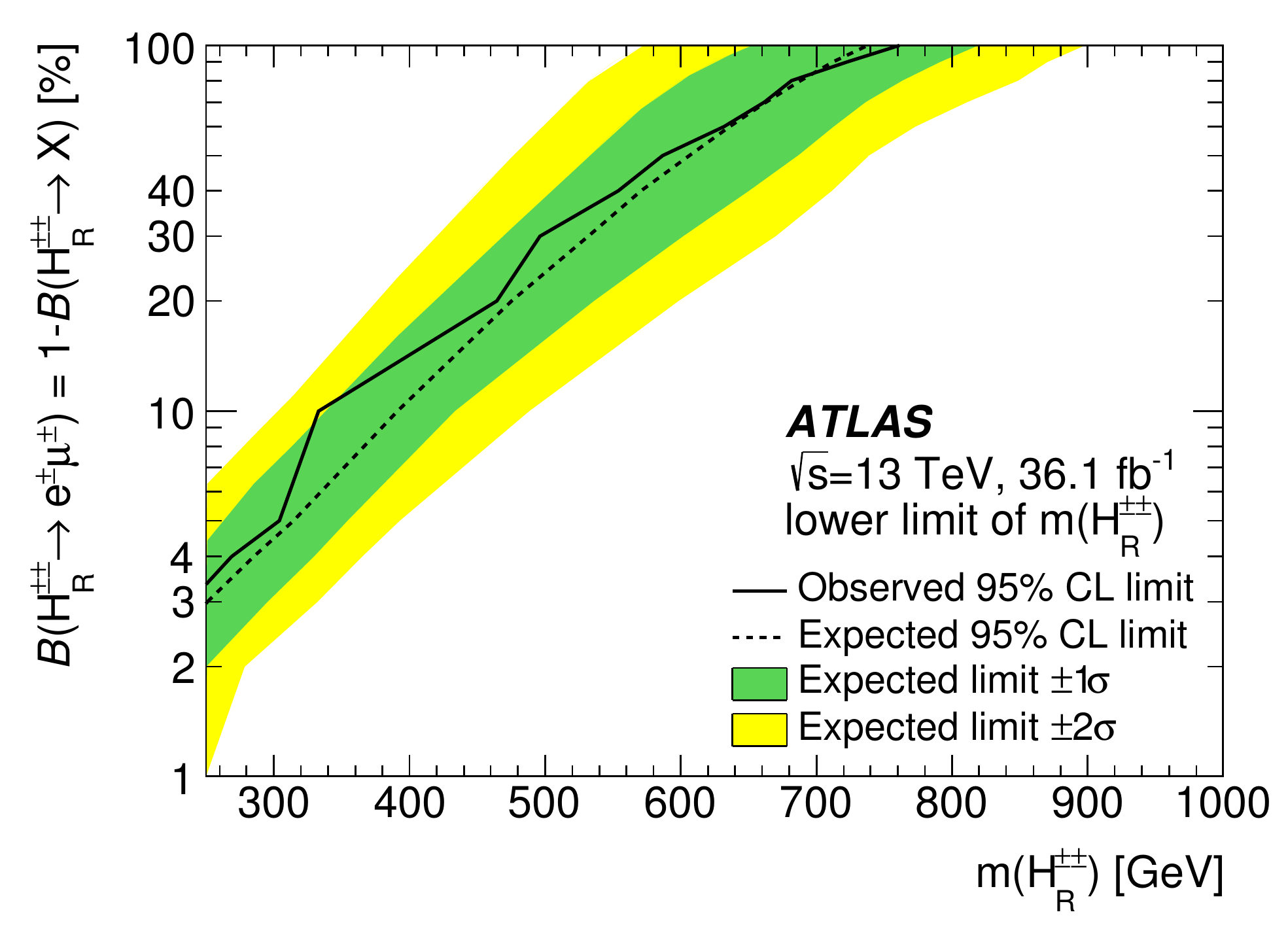}
\label{fig:HR_exp_1Dem}
}
\subfloat[]{
\includegraphics[width=0.45\textwidth]{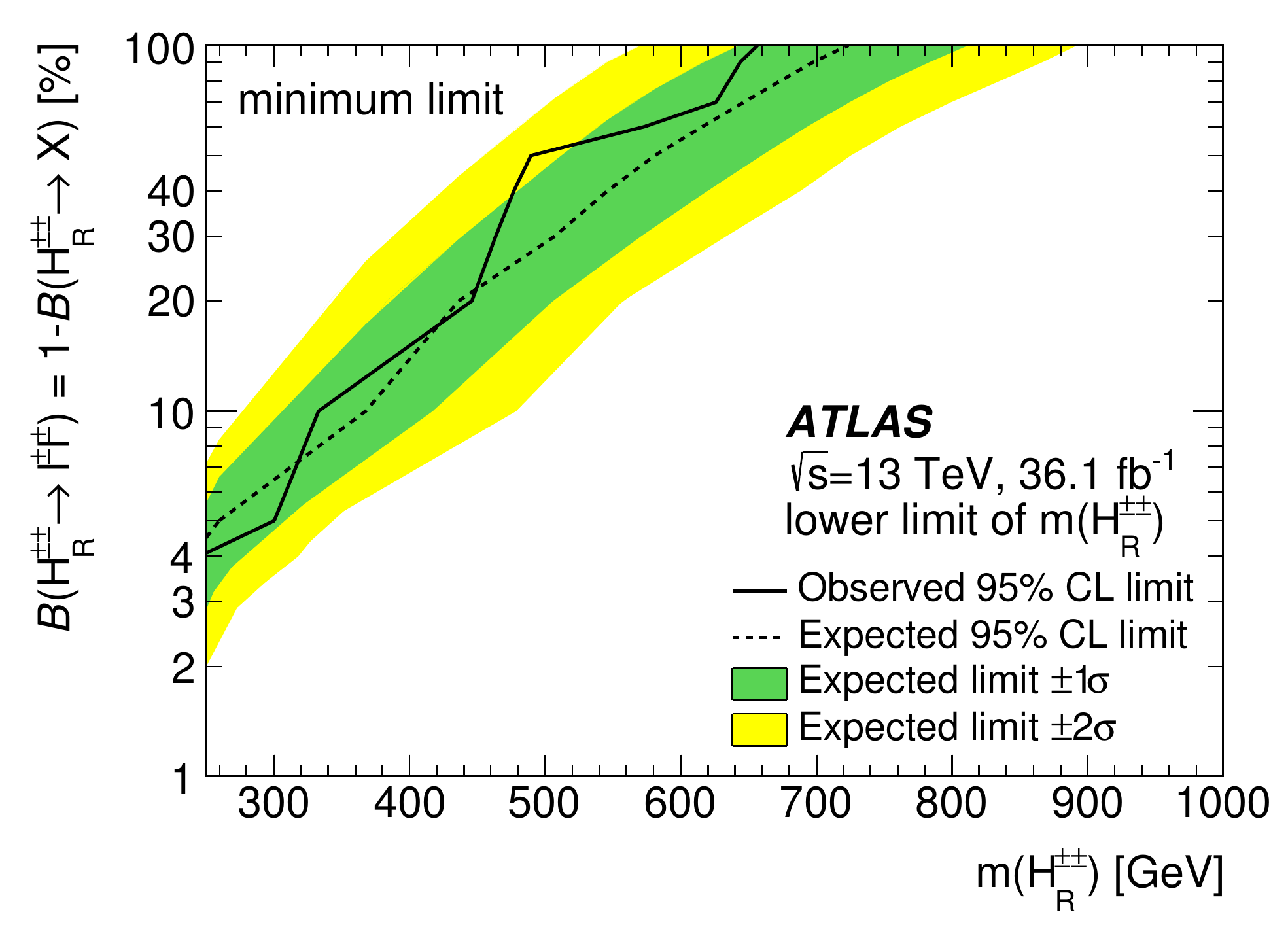}
\label{fig:HR_exp_1D_min}
}
\caption{Lower limit on the $H_{R}^{\pm \pm}$ boson mass as a function of the branching ratio $B(H_{R}^{\pm \pm}\rightarrow \ell^{\pm}\ell^{\pm})$. Several cases are presented: 
\protect\subref{fig:HL_exp_1Dee} $H_{R}^{\pm \pm}$ decays only into electrons and "$X$",
\protect\subref{fig:HR_exp_1Dmm} $H_{R}^{\pm \pm}$ decays only into muons and "$X$", and
\protect\subref{fig:HR_exp_1Dem} $H_{R}^{\pm \pm}$ decays only into electron--muon pairs and "$X$", with "$X$" not entering any of the signal regions.
Plot \protect\subref{fig:HR_exp_1D_min} shows the minimum observed and expected limit as a function of $B(H_{R}^{\pm \pm}\rightarrow \ell^{\pm}\ell^{\pm})$.
}
\label{fig:combinedFitAllBrHR}
\end{figure}

The final result of the fit is a lower limit on the two-dimensional grid of the $H^{\pm\pm}$ boson mass for 
any combination of light lepton branching ratios that sum to a certain value. The fit was 
performed for values of $B(H^{\pm \pm}\rightarrow \ell^{\pm}\ell^{\pm})$ from $1\%$ to 
$5\%$ in $1\%$ intervals, and from $10\%$ to $100\%$ in $10\%$ intervals. Expected limits 
for $B(H^{\pm \pm}\rightarrow \ell^{\pm}\ell^{\pm})=100\%$ are presented in 
Figure~\ref{fig:expectedLimsCombinedFitHL} for $H_{L}^{\pm\pm}$ and in 
Figure~\ref{fig:expectedLimsCombinedFitHR} for $H_{R}^{\pm\pm}$.
Results of the fit are presented in Figures~\ref{fig:combinedFitAllBrHL} and~\ref{fig:combinedFitAllBrHR} 
for $H_L^{\pm\pm}$ and $H_R^{\pm\pm}$, respectively. Here, three specific decay scenarios to
only $e^{\pm}e^{\pm}$, $\mu^{\pm}\mu^{\pm}$, and $e^{\pm}\mu^{\pm}$, are considered and the minimum 
limit for each value of $B(H^{\pm\pm} \rightarrow \ell^{\pm}\ell^{\pm})$ is given. The minimum limit 
is obtained by taking, for each value of $B(H^{\pm\pm} \rightarrow \ell^{\pm}\ell^{\pm})$, 
the least stringent limit for any combination of branching ratios that sum to 
$B(H^{\pm\pm} \rightarrow \ell^{\pm}\ell^{\pm})$. The lower mass limits for these four 
cases are similar, which indicates that the analysis is almost equally sensitive to each 
decay channel.

The observed lower mass limits vary from \SI{770}{\GeV} to \SI{870}{\GeV} for $H_{L}^{\pm\pm}$ with 
$B(H^{\pm \pm}\rightarrow \ell^{\pm}\ell^{\pm}) = 100\%$ and are above \SI{450}{\GeV} 
for $B(H^{\pm \pm}\rightarrow \ell^{\pm}\ell^{\pm}) \geq 10\%$. For $H_{R}^{\pm\pm}$ the lower 
mass limits vary from \SI{660}{\GeV} to \SI{760}{\GeV} for $B(H^{\pm\pm}\rightarrow \ell^{\pm}\ell^{\pm})=100\%$ 
and are above \SI{320}{\GeV} for $B(H^{\pm\pm}\rightarrow \ell^{\pm}\ell^{\pm}) \geq 10\%$.

\FloatBarrier

\section{Conclusion}
\label{sec:conclusion}

The ATLAS detector at the Large Hadron Collider was used to search for doubly charged 
Higgs bosons in the same-charge dilepton invariant mass spectrum at high values, using
$e^{\pm}e^{\pm}$, $e^{\pm}\mu^{\pm}$ and $\mu^{\pm}\mu^{\pm}$ final states as well as
final states with three or four leptons (electrons and/or muons). The 
search was performed with $\SI{36.1}{\ifb}$ of data from proton–proton collisions at 
$\sqrt{s}=\SI{13}{\TeV}$, recorded during the 2015 and 2016 data-taking periods.
No significant excess above the Standard Model prediction was found.
As a result of the search, lower limits are set on the mass of doubly-charged Higgs bosons. 
These vary between \SI{770}{\GeV} and \SI{870}{\GeV} for the $H_{L}^{\pm\pm}$ mass
and for $B(H^{\pm\pm}\rightarrow \ell^{\pm}\ell^{\pm})=100\%$
and above \SI{450}{\GeV} for $B(H^{\pm\pm}\rightarrow \ell^{\pm}\ell^{\pm}) \geq 10\%$ for any combination
of partial branching ratios. The observed lower limits on the $H_{R}^{\pm\pm}$ mass vary from
\SI{660}{\GeV} to \SI{760}{\GeV} for $B(H^{\pm\pm}\rightarrow \ell^{\pm}\ell^{\pm})=100\%$ and are above
\SI{320}{\GeV} for $B(H^{\pm\pm}\rightarrow \ell^{\pm}\ell^{\pm}) \geq 10\%$. The observed limits
are consistent with the expected limits.
The lower limits on the $H_L^{\pm\pm}$ and $H_R^{\pm\pm}$ masses obtained in this search,
under the assumption $B(H^{\pm\pm}\rightarrow \ell^{\pm}\ell^{\pm})=100\%$,
are \SI{300}{\GeV} higher than those from the previous ATLAS analysis~\cite{EXOT-2012-25}
and \SI{450}{\GeV} higher than those from the CMS analysis~\cite{CMS-HIG-12-005}.

\section*{Acknowledgements}

We thank CERN for the very successful operation of the LHC, as well as the
support staff from our institutions without whom ATLAS could not be
operated efficiently.

We acknowledge the support of ANPCyT, Argentina; YerPhI, Armenia; ARC, Australia; BMWFW and FWF, Austria; ANAS, Azerbaijan; SSTC, Belarus; CNPq and FAPESP, Brazil; NSERC, NRC and CFI, Canada; CERN; CONICYT, Chile; CAS, MOST and NSFC, China; COLCIENCIAS, Colombia; MSMT CR, MPO CR and VSC CR, Czech Republic; DNRF and DNSRC, Denmark; IN2P3-CNRS, CEA-DRF/IRFU, France; SRNSFG, Georgia; BMBF, HGF, and MPG, Germany; GSRT, Greece; RGC, Hong Kong SAR, China; ISF, I-CORE and Benoziyo Center, Israel; INFN, Italy; MEXT and JSPS, Japan; CNRST, Morocco; NWO, Netherlands; RCN, Norway; MNiSW and NCN, Poland; FCT, Portugal; MNE/IFA, Romania; MES of Russia and NRC KI, Russian Federation; JINR; MESTD, Serbia; MSSR, Slovakia; ARRS and MIZ\v{S}, Slovenia; DST/NRF, South Africa; MINECO, Spain; SRC and Wallenberg Foundation, Sweden; SERI, SNSF and Cantons of Bern and Geneva, Switzerland; MOST, Taiwan; TAEK, Turkey; STFC, United Kingdom; DOE and NSF, United States of America. In addition, individual groups and members have received support from BCKDF, the Canada Council, CANARIE, CRC, Compute Canada, FQRNT, and the Ontario Innovation Trust, Canada; EPLANET, ERC, ERDF, FP7, Horizon 2020 and Marie Sk{\l}odowska-Curie Actions, European Union; Investissements d'Avenir Labex and Idex, ANR, R{\'e}gion Auvergne and Fondation Partager le Savoir, France; DFG and AvH Foundation, Germany; Herakleitos, Thales and Aristeia programmes co-financed by EU-ESF and the Greek NSRF; BSF, GIF and Minerva, Israel; BRF, Norway; CERCA Programme Generalitat de Catalunya, Generalitat Valenciana, Spain; the Royal Society and Leverhulme Trust, United Kingdom.

The crucial computing support from all WLCG partners is acknowledged gratefully, in particular from CERN, the ATLAS Tier-1 facilities at TRIUMF (Canada), NDGF (Denmark, Norway, Sweden), CC-IN2P3 (France), KIT/GridKA (Germany), INFN-CNAF (Italy), NL-T1 (Netherlands), PIC (Spain), ASGC (Taiwan), RAL (UK) and BNL (USA), the Tier-2 facilities worldwide and large non-WLCG resource providers. Major contributors of computing resources are listed in Ref.~\cite{ATL-GEN-PUB-2016-002}.

\printbibliography

\newpage

\begin{flushleft}
{\Large The ATLAS Collaboration}

\bigskip

M.~Aaboud$^\textrm{\scriptsize 137d}$,
G.~Aad$^\textrm{\scriptsize 88}$,
B.~Abbott$^\textrm{\scriptsize 115}$,
O.~Abdinov$^\textrm{\scriptsize 12}$$^{,*}$,
B.~Abeloos$^\textrm{\scriptsize 119}$,
S.H.~Abidi$^\textrm{\scriptsize 161}$,
O.S.~AbouZeid$^\textrm{\scriptsize 139}$,
N.L.~Abraham$^\textrm{\scriptsize 151}$,
H.~Abramowicz$^\textrm{\scriptsize 155}$,
H.~Abreu$^\textrm{\scriptsize 154}$
% ...
\bigskip
\\
% ...
$^{12}$ Institute of Physics, Azerbaijan Academy of Sciences, Baku, Azerbaijan\\
% ...
$^{88}$ CPPM, Aix-Marseille Universit{\'e} and CNRS/IN2P3, Marseille, France\\
% ...
$^{115}$ Homer L. Dodge Department of Physics and Astronomy, University of Oklahoma, Norman OK, United States of America\\
% ...
$^{119}$ LAL, Univ. Paris-Sud, CNRS/IN2P3, Universit{\'e} Paris-Saclay, Orsay, France\\
% ...
$^{137}$ $^{(a)}$ Facult{\'e} des Sciences Ain Chock, R{\'e}seau Universitaire de Physique des Hautes Energies - Universit{\'e} Hassan II, Casablanca; $^{(b)}$ Centre National de l'Energie des Sciences Techniques Nucleaires, Rabat; $^{(c)}$ Facult{\'e} des Sciences Semlalia, Universit{\'e} Cadi Ayyad, LPHEA-Marrakech; $^{(d)}$ Facult{\'e} des Sciences, Universit{\'e} Mohamed Premier and LPTPM, Oujda; $^{(e)}$ Facult{\'e} des sciences, Universit{\'e} Mohammed V, Rabat, Morocco\\
% ...
$^{139}$ Santa Cruz Institute for Particle Physics, University of California Santa Cruz, Santa Cruz CA, United States of America\\
% ...
$^{151}$ Department of Physics and Astronomy, University of Sussex, Brighton, United Kingdom\\
% ...
$^{154}$ Department of Physics, Technion: Israel Institute of Technology, Haifa, Israel\\
$^{155}$ Raymond and Beverly Sackler School of Physics and Astronomy, Tel Aviv University, Tel Aviv, Israel\\
% ...
$^{161}$ Department of Physics, University of Toronto, Toronto ON, Canada\\
% ...
$^{a}$ Also at Department of Physics, King's College London, London, United Kingdom\\
$^{b}$ Also at Institute of Physics, Azerbaijan Academy of Sciences, Baku, Azerbaijan\\
$^{c}$ Also at Novosibirsk State University, Novosibirsk, Russia\\
$^{d}$ Also at TRIUMF, Vancouver BC, Canada\\
$^{e}$ Also at Department of Physics {\&} Astronomy, University of Louisville, Louisville, KY, United States of America\\
% ...
$^{*}$ Deceased
\end{flushleft}

\end{document}